\documentclass[12pt,titlepage]{book}

\author{Pedro D. Alvarez}
\date{July, 2009}
\usepackage{amsfonts}
\usepackage{amsmath}
\usepackage{amssymb}
\usepackage{graphicx}
\usepackage{latexsym}
\usepackage[latin1]{inputenc}
\usepackage[spanish]{babel}
\usepackage{appendix}
\usepackage{makeidx}
\usepackage{url}
\usepackage{fancyhdr}

\usepackage[
pdfkeywords={Newton-Hooke, Exotic Newton-Hooke symmetry, Nonlinear realization, Duality, Wave equations,Projective phase, Noncommutative plane, Conformal symmetry, W Symmetry, Extended Supersymmetry,Non-Commutative Geometry, Exotic Galilei, Landau Problem, conformal},%
pdfsubject={Nonrelativistic symmetries in the noncommutative plane},%
pdftitle={Exotic Newton-Hooke group, noncommutative plane and superconformal symmetry},%
pdfauthor={Pedro D. Alvarez}%
]{hyperref}

\usepackage{booktabs}
\usepackage{multirow}

\include{def}

\begin{document}
\frontmatter
\pdfbookmark[0]{Title}{Portada}

\begin{center}
\vspace{-35mm}{\Large UNIVERSIDAD DE SANTIAGO DE CHILE}

{\large \vspace{1cm} FACULTAD DE CIENCIA}

{\large \vspace{0.5cm} DEPARTAMENTO DE FÍSICA}{\Large
\vspace{3cm}}

{\LARGE \textbf{Grupo de Newton-Hooke exótico, plano no conmutativo y simetría superconforme}}{\huge
\bigskip }

{\Large \vspace{1cm} Tesis para optar el grado de}{\LARGE \bigskip }

{\Large Doctor en Ciencias con Mención en Física\vspace{2cm}}%
{\LARGE \textbf{\bigskip }}

{\LARGE Pedro Álvarez Núñez \vspace{1cm}}

\begin{tabular}{ll}
{\Large Profesor guía:}{\LARGE \qquad \medskip } & {\Large Dr.
Mikhail Plyushchay}
\end{tabular}%
{\Large \vspace{2.5cm}\vspace{2.5cm}}

{\large SANTIAGO -- CHILE}

{\large Julio 2009}

\end{center}


\chapter*{Resumen}


En este trabajo de tesis hemos estudiado algunos sistemas con simetrías exóticas, las cuales son una peculiaridad de las 2+1 dimensiones de espacio-tiempo. Codificadas dentro de la estructura exotica aparecen no conmutatividad de coordenadas, y una estructura de fases.

Este tipo de sistemas ha despertado interés en diversas areas de física en forma paralela. Entre ellas se destacan: teoría de representaciones proyectivas de grupos, física de anyones, algunos sistemas de materia condensada, por ejemplo el efecto de Hall cuántico, teorías de gauge y de gravitación planar, teoría de campos no conmutativa,  geometría no conmutativa y mecánica cuántica no conmutativa.

En esta tesis discutiremos sistemas con simetrías exóticas no relativistas, enfocándonos en tres problemas concretos:

\begin{itemize}

\item La simetría de Newton-Hooke exótica;

\item La relación entre el sistema de Newton-Hooke exótico y el problema de Landau no conmutativo;

\item Las simetrías del problema de Landau no conmutativo, su extensión conforme y supersimétrica.

\end{itemize}

El grupo de Newton-Hooke exótico corresponde al límite no relativista de los grupos de de Sitter, y contiene como caso particular al de Galileo exótico. Para la simetría de Newton-Hooke exótica hemos construido una acción de partícula libre y hemos hecho un estudio completo de las propiedades clásicas y cuánticas. Este sistema está íntimamente relacionado con el problema de Landau no conmutativo, que estudiamos aparte. Nosotros mostramos que en el problema de Landau no conmutativo la inclusión de los grados de libertad de spin lleva a la  integración natural del grupo de Newton-Hooke exótico con la simetría conforme y la supersimetría.

\chapter*{}
\begin{figure}[h]
\begin{center}
  \includegraphics[width=1\textwidth]{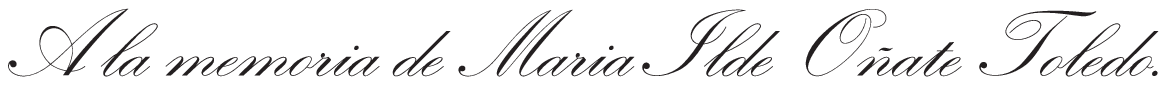}\\
     \end{center}
\end{figure} 

\urlstyle{rm}
\fancyhead{}
  \fancyhead[LO]{\slshape \rightmark}
  \fancyhead[RO,LE]{\thepage}
  \fancyhead[RE]{\slshape \leftmark}
  \fancyfoot{}
  \pagestyle{fancy}
  \renewcommand{\chaptermark}[1]{\markboth{\chaptername \ \thechapter \ \ #1}{}}
  \renewcommand{\sectionmark}[1]{\markright{\thesection \ \ #1}}

\tableofcontents
\chapter{Prefacio}

En este trabajo de tesis hemos investigado sistemas planares con simetrías
no relativistas exóticas. A nivel de la álgebra de simetría, en 2+1 dimensiones, aparece la posibilidad de considerar una segunda extensión central, esto gracias a una cohomología de Eilenberg-Chevalley de las 2-formas no trivial.

El primer sistema que analizamos es la partícula con simetría de Newton-Hooke exótica. La simetría de Newton-Hooke aparece como el límite no relativista de los grupos de de Sitter (dS), de forma análoga a como la simetría de Galileo es el límite no relativista del grupo de Poincaré. Dependiendo de si se toma dicho límite sobre AdS o dS, se llega al grupo de Newton-Hooke oscilatorio o al hiperbólico. La estructura exótica aparece para el caso especial de 2+1 dimensiones, lo cual corresponde a una partícula planar. Esta estructura exótica proporciona algunas propiedades físicas no observadas en otras dimensiones, como por ejemplo, permite asociar de forma natural coordenadas no conmutativas con la segunda extensión central de la álgebra. Además en el caso trigonométrico (oscilatorio) de la simetría exótica de Newton-Hooke, aparece una estructura de fases que depende de los valores de los parámetros del sistema. Existen tres fases, subcrítica, crítica y supercrítica. La fase subcrítica reproduce, tomando el limite plano, el sistema con simetría de Galileo exótica. En la fase crítica se reducen los grados de libertad físicos del sistema, esta fase separa las otras dos. En la fase supercrítica el sistema posee algunas propiedades inusuales como, por ejemplo, un espectro de energía no acotado por abajo. Esta estructura de fases es análoga a la que aparece en el sistema del problema de Landau no conmutativo.

Posteriormente, estudiamos las simetrías de un oscilador anisótropo exótico. En este sistema la anisotropía y la estructura exótica se complementan de tal forma que el sistema posee simetría rotacional. La anisotropía está cuantificada por un parámetro adimensional. Este sistema uniparamétrico interpola de forma continua los sistemas de Newton-Hooke exótico (caso isótropo) y el problema de Landau no conmutativo (caso maximalmente anisótropo). Coordenadas no conmutativas, la estructura de fases y las peculiaridades de la fase supercrítica son propiedades genéricas del caso anisótropo.

Finalmente, estudiamos las simetrías del problema de Landau no conmutativo. Analizamos las extensiones conformes y la extensión supersimétrica, que está relacionada con grados de libertad de spin. La estructura del Hamiltoniano es tal que la energía es acotada de abajo para cualquier fase. En la fase crítica los niveles de energía se separan infinitamente, dejando al sistema congelado en el estado básico, el cual gracias a la supersimetría tiene energía cero. De este modo sale de forma natural una justificación para la llamada ``substitución de Peierls'', usada en el efecto cuántico de Hall.

\bigskip Este trabajo de tesis está basado en los artículos \cite{AGKP,NH-NCL,Alvarez:2009nz}.

\bigskip\noindent{\it {\large Descripción por capítulo}}

\bigskip El capítulo \ref{motiv} está reservado a una introducción elemental en el tema de las simetrías exóticas y la no conmutatividad.

Luego, en el capítulo \ref{preambulo} haremos una breve discusión de algunos de los conceptos básicos involucrados en los grupos cinemáticos de simetría, la simetría de Newton-Hooke, las simetrías exóticas y conceptos afines. Fijaremos algunas notaciones y pondremos en contexto los grupos cinemáticos y las álgebras correspondientes.

Los tres capítulos siguientes \ref{enh chapter}, \ref{aho chapter} y \ref{senclp chapter} incluyen los principales resultados de esta tesis. El capítulo \ref{enh chapter} trata sobre la simetría de Newton-Hooke exótica. Construimos el Lagrangiano para la partícula con esta simetría y estudiamos sus aspectos clásicos y cuánticos. Luego, en el capítulo \ref{aho chapter} analizamos la relación existente entre las simetrías de Newton-Hooke exótica y la del problema de Landau no conmutativo. Después, en el capítulo \ref{senclp chapter} discutimos las simetrías del problema de Landau no conmutativo y su extensión conforme y supersimétrica.

Finalmente, está incluido un apéndice donde se indican algunos aspectos del grupo $SO(4)$ inhomogéneo.

\mainmatter

\renewcommand{\tablename}{Tabla}

\chapter{Introducción}\label{motiv}

Cuando miramos el mundo en 2+1 dimensiones cosas peculiares ocurren. Por lo general en física, cuando disminuimos las dimensiones, o grados de libertad, los sistemas se simplifican. La excepción a esta regla son los sistemas planares, por un lado son más simples que sus análogos 3+1 dimensionales, pero a la vez, en muchos casos, poseen suficientes sutilezas, que al final, conducen a una estructura más rica que la de sus análogos 3+1 dimensionales. ¿Qué es tan especial con 2+1 dimensiones?, o quizás lo especial sea nuestra propia realidad 3+1 dimensional... Lamentablemente las respuestas a estas preguntas no se encuentran al alcance de esta tesis, pero el estudio de algunos sistemas planares con simetrías exóticas presentados aquí es útil en esta dirección.

Modelos con características exóticas, tales como anyones, coordenadas no conmutativas o paraestadística han aparecido de forma casi paralela desde diferentes areas de física, sin embargo de una u otra forma parecen tener ciertos rasgos en común. A continuación daremos un breve resumen de algunos sistemas exóticos.

Podemos empezar mencionando un caso particular de las 2+1 dimensiones. En el plano hay sólo un generador de rotaciones, con lo cual tenemos un grupo abeliano y no se obtiene una cuantización del spin. En este contexto aparecen los anyones \cite{Leinaas:1977fm,Goldin:1981sm,Wilczek:1981du,Wilczek:1982wy,Semenoff:1988jr,Plyushchay:1990cv,Jackiw:1990ka,Pl1,Pl2,mariano1}. El nombre proviene de que su estadística (anyonica) interpola entre la bosónica y la fermiónica.

Existe otro acercamiento para llegar a estadísticas no usuales.Las partículas con estadística bosónica (o fermiónica) deben satisfacer reglas de conmutación (o anticonmutación). Es posible postular desde el inicio que tenemos campos que satisfacen relaciones que combinan las de conmutación con anticonmutación, de este modo llegamos a partículas con paraestadística \cite{Green:1952kp,Messiah:1900zz,Greenberg:1964pe,Greenberg:1963kk,Greenberg:1989ty,Polychronakos:1999sx}. Es posible modelar el grado de mezcla de la estadística bosónica y fermiónica mediante un parámetro $q$, en este caso nos referiremos a las correspondientes partículas por ``$q-$ones'' \cite{Greenberg:1991ec}. Lo observado hasta ahora en la naturaleza son partículas con estadística bosónica o fermiónica, entonces haciendo el parámetro de paraestadística $q$ pequeño podemos estudiar una violación leve a la estadística usual, que podría ser observable en sistemas experimentales gracias al refinamiento de las técnicas de laboratorio. Una ventaja del acercamiento desde la paraestadística es, que a priori, no requiere tratar con un sistema 2+1 dimensional.

Por otro lado, podemos mencionar las extensiones centrales de las álgebras de simetría cinemáticas. Consideremos el caso más discutido, que es el relativo a la simetría de Galileo. Matemáticamente las extensiones centrales se deben a una cohomología no trivial de las álgebras. Para cualquier dimensión, y en particular en el caso de 3+1, Galileo acepta una extensión central, la cual es asociada a la masa de la partícula \cite{LL1963,Bacry:1968zf}. Pero en el caso particular de 2+1 dimensiones y sólo en este, la álgebra de Galileo acepta una segunda extensión central \cite{LL}, de aquí el nombre extensión exótica. La naturaleza física de la estructura exótica era más bien desconocida. La relación entre anyones y este tipo de simetrías exóticas fue indicada por primera vez por Jackiw y Nair \cite{JackN}, quienes mostraron cómo es posible obtener boosts de Galileo no conmutativos (que caracterizan a la álgebra de Galileo exótica), a partir de un límite no relativista especial de un anyon.

Esto nos lleva a la clase de sistemas correspondientes con coordenadas no conmutativas. La primera instancia en que aparecieron en la literatura es gracias a Snyder \cite{Snyder:1946qz}, quien exploró esta posibilidad en búsqueda de algún mecanismo que sirviera para aminorar los problemas de autoenergías divergentes \footnote{Para un breve relato de como Snyder llegó esta idea, que originalmente se debe a Heisenberg, ver \cite{Jackiw:2001dj}.}. Después de Snyder, pasaron cerca de cincuenta años para que el estudio de teorías no conmutativas volviera a despertar interés, esta vez desde diversas áreas tales como, geometría no conmutativa \cite{Connes:1997cr}, teorías de gauge y D-branas \cite{Seiberg:1999vs}, gravedad cuántica \cite{Szabo:2006wx,Harikumar:2006xf}, modelo estándar no conmutativo \cite{Chaichian:2001py,Carroll:2001ws,CCGM} y el estudio de las simetrías de teorías de campos no conmutativas \cite{NCS1,NCS2,NCS3,NCS4,Luk,Az}.

Como límite de bajas energías aparece la mecánica cuántica no conmutativa
\cite{Bigatti:1999iz,Gamboa:2000yq,Chaichian:2000si,Nair:2000ct,Nair:2000ii,Chaichian:2000hy,Gamboa:2001fg,Morariu:2001dv,Gamboa:2001qa,Falomir:2002ih,Horv}.
Mucho interés existe en buscar sistemas tales que la no conmutatividad produzca efectos no
triviales y a la vez medibles. Como algunos ejemplos de sistemas sensibles a la no conmutatividad
podemos mencionar el efecto de Aharonov-Bohm en el plano no conmutativo
\cite{Chaichian:2000hy,Falomir:2002ih} y el problema de Landau no conmutativo
\cite{Gamboa:2001fg,Horvathy:2002wc,Hellerman:2001rj}. En particular, es posible ver que el
problema de Landau y la simetría de Galileo exótica reproducen las condiciones ideales para
justificar la ``substitución de Peierls'', usada en la descripción del efecto de Hall cuántico
\cite{Dunne:1992ew,Duval:2000xr,Duval:2001hu,DH2}. Señalamos aquí también que los aspectos de spin
fraccionario y no conmutatividad de coordenadas en 2+1 dimensiones están relacionados íntimamente
\cite{JackN,Schonfeld:1980kb,anyonnc}.

Los sistemas no conmutativos de mecánica cuántica poseen simetrías no relativistas de tipo de Galileo. Es conocido que las representaciones de Galileo usadas en física son representaciones proyectivas, que corresponden a que en la álgebra de Galileo aparece una extensión central, la cual es asociada a la masa \cite{LL1963,Bacry:1968zf,LL,bargmann,azcarragabook}. En el caso especial de 2+1 dimensiones la álgebra de Galileo admite una segunda extensión central, por esto llamada exótica. Mucho interés atrajo el estudio de las consecuencias físicas de la extensión exótica de la álgebra de Galileo  \cite{LL,JackN,Duval:2000xr,Duval:2001hu,BGO,Grigore:1993fz,Bose:1994sj,Bose:1994si,Brihaye:1995nv,Hagen:2002pg,Horvathy:2002vt,LSZ,OP}, la cual mostró ser responsable de la aparición de coordenadas no conmutativas. También mostró tener aplicación en el efecto de Hall cuántico \cite{Duval:2000xr,Duval:2001hu,Dunne:1989hv,Dunne:1992ew}.

Otra simetría no relativista es la de Newton-Hooke. Los grupos de Newton-Hooke aparecen como los modelos cosmológicos no relativistas más simples, en el sentido de que se obtienen como límites no relativistas de los espacios con curvatura constante de de Sitter. Sorprendentemente Newton-Hooke exótico ha sido estudiado sólo en unos pocos trabajos \cite{olmo nh}.

El estudio de las extensiones centrales, y en particular de las exóticas, de las álgebras no relativistas aparece motivado también desde el contexto de la correspondencia AdS/CFT no relativista \cite{Maldacena:1997re,Gubser:1998bc,Witten:1998qj,Aharony:1999ti,LeivaPl,Son:2008ye,CJPAdS}. Este enfoque a la conjetura de Maldacena ofrece nuevas posibilidades, ya que estos sistemas son en principio, accesibles experimentalmente. Se ha indicado que el sistema de los fermiones fríos en la unitaridad \cite{Son:2008ye,Balasubramanian:2008dm,Nishida:2007pj} es un buen candidato a estudiar, ya que posee simetría de Schrödinger, que es una versión no relativista de la simetría conforme \cite{NRConf}. La álgebra de Schrödinger contiene como subálgebra a Galileo centralmente extendido, y la extensión central ha resultado ser una pieza clave para la identificación de geometrías asociadas \cite{Duval:1984cj,Duval:2008jg}. La determinación de geometrías es necesaria para la posterior búsqueda de teorías de gravedad duales. En este contexto el estudio de las extensiones exóticas de Newton-Hooke es relevante, ya que pueden estar relacionadas a un conjunto más rico de geometrías.

Finalmente, no podemos dejar de mencionar el caso especial de relatividad general en 2+1 dimensiones. La teoría no posee grados de libertad propagantes, pero aún así ofrece soluciones no triviales como el agujero de BTZ \cite{Banados:1992wn}, el cual acepta una descripción en términos de teorías de gauge en 2+1 dimensiones mediante dos campos de Chern-Simons \cite{AchTown,Witten}.

\renewcommand{\tablename}{Tabla}
\chapter{Simetrías cinemáticas}\label{preambulo}

Las simetrías están detrás de las propiedades que caracterizan un
sistema físico. Una clase importante son las simetrías del
espacio-tiempo, llamadas cinemáticas.

Los grupos cinemáticos incluyen transformaciones de rotaciones,
traslaciones en el tiempo y espacio, y ``transformaciones
inerciales'', que expresan la isotropía del espacio, la
homogeneidad y la correspondencia de marcos de referencia
inerciales respectivamente.

Las leyes de la física deben ser equivalentes para cualquier
observador, y al ser las transformaciones cinemáticas las que
conectan distintos observadores, lo usual es pedir que las
ecuaciones sean covariantes bajo el grupo dado. De este modo uno
puede decir que el grupo cinemático es el grupo de relatividad de
la naturaleza.

En este contexto es necesario decir que la palabra relatividad la
usamos en el sentido de covariancia, y no en el sentido de tiempo
relativo o absoluto. De hecho, como veremos, matemáticamente el
espacio donde vivimos bien puede ser relativista, o no, e incluso
puede tener curvatura.

De este modo la existencia de este conjunto de simetrías no fija
la estructura precisa del grupo. Aún falta decir de que forma las
leyes de la física transforman y si existe alguna curvatura del
espacio-tiempo. Estos dos aspectos están descritos por la
estructura del grupo. Dentro de los grupos de relatividad más
discutidos en física están Galileo y Poincaré, una pregunta
natural es, ¿existen otras posibilidades?, la respuesta es
conocida gracias a Bacry y Lévy-Leblond \cite{Bacry:1968zf}.

Dentro de los grupos cinemáticos que se obtienen, hay cuatro grupos no relativistas. Uno de estos es el llamado ``Newton-Hooke'', el cual tiene el sentido de límite no relativista de un espacio de de Sitter, de forma análoga a como la simetría de Galileo es el límite no relativista de la de Poincaré. Esto será explicado en detalle en este capítulo. Para las simetrías no relativistas ha sido observado que, las representaciones con una interpretación física apropiada, son las llamadas representaciones proyectivas, las cuales tienen asociadas las álgebras de simetría con extensiones centrales \cite{LL1963}. Las extensiones centrales de Galileo y de Newton-Hooke aparecen para cualquier dimensión de espacio tiempo, y está asociada a la masa de la partícula.

En la sección \ref{pos kin sec} revisaremos los diferentes grupos cinemáticos, su clasificación y su interpretación física. Al final de la misma, comentaremos sobre las representaciones proyectivas. Después, en la sección \ref{NH sec}, profundizaremos en el grupo de Newton-Hooke. Luego, en la sección \ref{ex gal sec}, discutiremos algunos aspectos básicos de la simetría de Galileo exótica.

\section{Posibles cinemáticas}\label{pos kin sec}

La aproximación al problema es desde el punto de vista de las
álgebras de Lie. Llamaremos a los generadores por $H$, $P_i$,
$J_i$ y $K_i$ (i=1,2,3), que producen traslaciones en tiempo,
espacio, rotaciones y boosts en el eje i. Estudiaremos el caso de 3+1 dimensiones, pero muchas de las conclusiones son válidas para cualquier dimensión mayor a dos.

Es posible determinar todos los grupos cinemáticos bajo tres
suposiciones físicamente razonables:\\
(\small{i}) La \emph{primera} es isotropía del
espacio, significando que $H$ es un escalar y $P_i$, $J_i$ y $K_i$
transformen como vectores
\begin{eqnarray}\label{isotropic}
&\left[\mathbf{J},H\right]=0,&\\
&\left[\mathbf{J},\mathbf{P}\right]=\mathbf{P}, \qquad
\left[\mathbf{J},\mathbf{K}\right]=\mathbf{K}, \qquad
\left[\mathbf{J},\mathbf{J}\right]=\mathbf{J}, \qquad &
\end{eqnarray}
donde hemos usado las notaciones $[\mathbf{A},\mathbf{B}]=\mathbf{C}$ como abreviaciones para $[A_i,B_j]=\epsilon_{ijk}C_k$.\\

\noindent (\small{ii}) La \emph{segunda} es que paridad $\Pi$ e inversión del tiempo $\Theta$
sean automorfismos de la álgebra
\begin{eqnarray}
\Pi&=&\{H\rightarrow H, P_i\rightarrow -P_i, J_i\rightarrow J_i, K_i\rightarrow -K_i\},\\
\Theta&=&\{H\rightarrow -H, P_i\rightarrow P_i, J_i\rightarrow
J_i, K_i\rightarrow -K_i\}.
\end{eqnarray}
Esta exigencia no es obligatoria, pero sirve para reducir el
número de diferentes álgebras que obtendremos\footnote{Podemos notar que la acción combinada $\Gamma = \Pi \Theta$, la
cual puede reemplazar a $\Pi$
\begin{equation}
\Gamma=\{H\rightarrow -H, P_i\rightarrow -P_i, J_i\rightarrow J_i,
K_i\rightarrow K_i\},
\end{equation}
es intercambiada con $\Theta$ cuando $P_i$ y $K_i$ son
intercambiados.}. De hecho después
podemos pasar a las combinaciones lineales
\begin{eqnarray}
&P'_i=\alpha P_i+\beta K_i +\gamma J_i,&\\
&K'_i=\alpha' P_i+\beta' K_i +\gamma' J_i,&
\end{eqnarray}
donde $\alpha$, $\beta$, $\gamma$, $\alpha'$, $\beta'$ y $\gamma'$
son reales y $\alpha \beta'-\alpha' \beta \ne 0$. Con estas
combinaciones levantamos la segunda condición y accedemos a más
grupos \cite{BacNuy}.\\

\noindent (\small{iii}) Como \emph{tercera} suposición establecemos que las transformaciones inerciales
sean un subgrupo no compacto. Esto es totalmente razonable, de no
hacerla podemos tener un boost lo suficientemente grande tal que
sea proporcional a la unidad, es decir, que sea equivalente a no
hacer nada, lo cual es físicamente inaceptable. Notar que la
condición análoga para traslaciones no se hace, de hecho el
espacio bien puede ser compacto.

Como resultado aparecen ocho álgebras diferentes, correspondientes a
once cinemáticas distintas:
\begin{description}
\item[R1.] Las dos álgebras de de Sitter (dS), AdS $\cong so(3,2)$ y dS $\cong so(4,1)$;
\item[R2.] La álgebra de Lie de Poincaré (P);
\item[R3.] Dos álgebras de Lie de ``Para-Poincaré'' (P'). Una isomorfa a la álgebra de Poincaré ordinaria\footnote{El isomorfismo se ve haciendo $P\rightarrow -K, K\rightarrow P.$}, pero físicamente distinta. La otra es isomorfa a la del grupo $SO(4)$ inhomogéneo, o también llamado grupo Euclideano E(4), ver apéndice \ref{Inhso(4)};
\item[R4.] La álgebra de Carroll (C);
\item[A1.] Las dos álgebras de Newton-Hooke (NH);
\item[A2.] La álgebra de Lie de Galileo (G).;
\item[A3.] La álgebra de ``Para-Galileo'' (G'), isomorfa a la de Galileo usual bajo la identificación $P\leftrightarrow K$;
\item[A4.] La álgebra ``Estática'' (St).
\end{description}

Una característica en común a todas estas álgebras, es que
contienen las relaciones (\ref{isotropic}), el resto de
conmutadores están dados en la tabla \ref{possible kinematics tabla}. Es importante comentar que aparte de las algebras de de Sitter, Poincaré y Galileo, el
resto no son muy discutidas en la literatura.

\begin{table}[h]
\begin{center}
\scalebox{0.75}[0.75]{
\begin{tabular}{c cc c cc c cc c c c}
\toprule
& \multicolumn{2}{c}{R1 (dS)} & R2 (P) & \multicolumn{2}{c}{R3 (P')} &R4 (C) & \multicolumn{2}{c}{A1 (NH)}& A2 (G) & A3 (G') & A4 (St) \\ \addlinespace
\cmidrule{2-12}
& SO(4,1) & SO(3,2) & ISO(3,1) & Inh.SO(4) & Para-Poinc. & Carroll & NH$_+$ & NH$_-$ & Galileo & Para-Gal. & Estático \\ \addlinespace
\midrule
{\lbrack H,\textbf{P}]} & \textbf{K} & -\textbf{K}& 0 & \textbf{K} & -\textbf{K} & 0 & \textbf{K} & -\textbf{K} & 0 & \textbf{K}
& 0 \\ \addlinespace
{\lbrack H,\textbf{K}]} & \textbf{P} & \textbf{P} & \textbf{P} & 0 & 0 & 0 & \textbf{P} & \textbf{P} & \textbf{P} & 0 &
0 \\ \addlinespace
{\lbrack \textbf{P},\textbf{P}]} & \textbf{J} & -\textbf{J} & 0 & \textbf{J} & -\textbf{J} & 0 & 0 & 0 & 0 & 0
& 0 \\ \addlinespace
{\lbrack \textbf{K},\textbf{K}]} & -\textbf{J} & -\textbf{J} & -\textbf{J} & 0 & 0 & 0 & 0 & 0 & 0 & 0
& 0 \\ \addlinespace
{\lbrack \textbf{P},\textbf{K}]} & H & H & H & H & H & H & (I) & (I) & (I)
& (I) & (I) \\ \addlinespace \bottomrule
\end{tabular}
}\label{possible kinematics tabla}
\caption{\small{ Paréntesis característicos para los diferentes grupos cinemáticos. La notación $[\mathbf{P},\mathbf{K}]=H$ quiere decir $[P_i,K_j]=H \delta_{ij}$. El término $(I)$ corresponde a posibles extensiones centrales, se anula para las álgebras no extendidas. A pesar de que algunos grupos son isomorfos, su interpretación física cambia, tales casos son: $P'\cong P$, $G'\cong G$.}}
\end{center}
\end{table}

Podemos ver que los grupos naturalmente quedan divididos en dos
familias, ``R'' de tiempo relativo y ``A'' de tiempo absoluto,
caracterizadas por el conmutador $[P_i,K_j]=\delta_{ij}H$ para
las álgebras R, mientras que $[P_i,K_j]=(I)$ para las álgebras A, donde $(I)$ significa que el conmutador es cero o una carga central. Todas las álgebras tipo A admiten, para \emph{cualquier} dimensión, cargas centrales $I$ \cite{LL1963,azcarragabook,Mukunda:1974dr}, mientras que las de clase R no admiten extensiones centrales no triviales \cite{bargmann,AzHer}
.

Es posible obtener una interpretación física de estos espacios
mediante el método de las contracciones \cite{Segal,IW1,Saletan}, el cual  permite
reemplazar simetrías cinemáticas ``exactas'' por las ``aproximadas''. El proceso de contracción siempre produce cierta abelianización de la álgebra, por lo que en este marco las álgebras de de Sitter son las únicas ``exactas''\footnote{No se pueden obtener, vía una contracción, de ninguna otra de la tabla \ref{possible kinematics tabla}, aunque si mediante un proceso llamado ``expansiones'' de la álgebra, el cual tiene la naturaleza opuesta a las contracciones.}.
La contracción siempre está definida con respecto a un subgrupo
dado de la álgebra y debido a que los únicos subgrupos comunes a
todas las álgebras son los generados por $\{J_i,H\}$, $\{J_i,P_j\}$, $\{J_i,K_j\}$ y
$\{J_i\}$, es posible hacer cuatro tipos de contracciones.

\bigskip\noindent{\it 1. Contracción de velocidad-espacio}

\bigskip Está definida por

\begin{equation}\label{con-speed-space}
\mathbf{K} \rightarrow \epsilon \mathbf{K}, \qquad \mathbf{P} \rightarrow
\epsilon \mathbf{P}, \qquad \epsilon \rightarrow 0.
\end{equation}

\begin{figure}[h]
\begin{center}
  \includegraphics[width=0.55\textwidth]{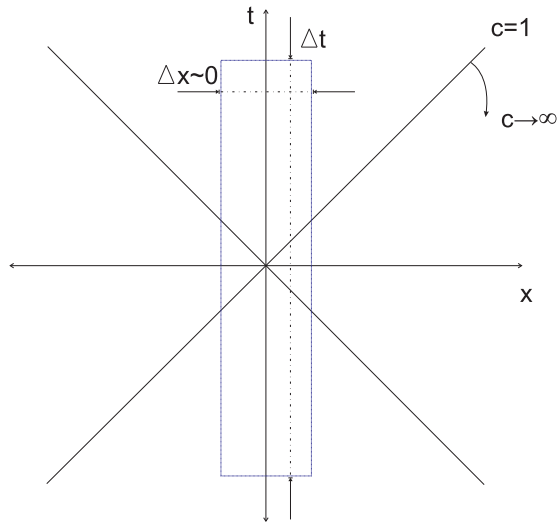}\\
  \caption{Contracción velocidad-espacio, $c$ es la velocidad de la luz.}\label{speed-space}
     \end{center}
\end{figure}

Esta contracción es respecto al subgrupo generado por $\{J_i,H\}$. Físicamente
significa considerar velocidades pequeñas e intervalos tipo tiempo
grandes (ver Fig. \ref{speed-space}).

 Corresponde a pasar de tiempo relativo a absoluto, y mapea
uno a uno las álgebras de los tipos R a las A,

\begin{center}
R1 $\rightarrow$ A1, \qquad R2 $\rightarrow$ A2, \qquad R3 $\rightarrow$
A3, \qquad R4 $\rightarrow$ A4.
\end{center}

Esta contracción lleva de Poincaré a Galileo, y lo que nos interesará más adelante, de las álgebras de de Sitter a las de Newton-Hooke. Más precisamente, según sea el caso, de AdS o dS, resultan grupos de Newton-Hooke de tipo trigonométrico o hiperbólico, respectivamente.

\bigskip\noindent{\it 2. Contracción de velocidad-tiempo}

\bigskip Debemos hacer

\begin{equation}
\mathbf{K} \rightarrow \epsilon \mathbf{K}, \qquad H \rightarrow
\epsilon H, \qquad \epsilon \rightarrow 0.
\end{equation}

\begin{figure}[h]
\begin{center}
  \includegraphics[width=0.55\textwidth]{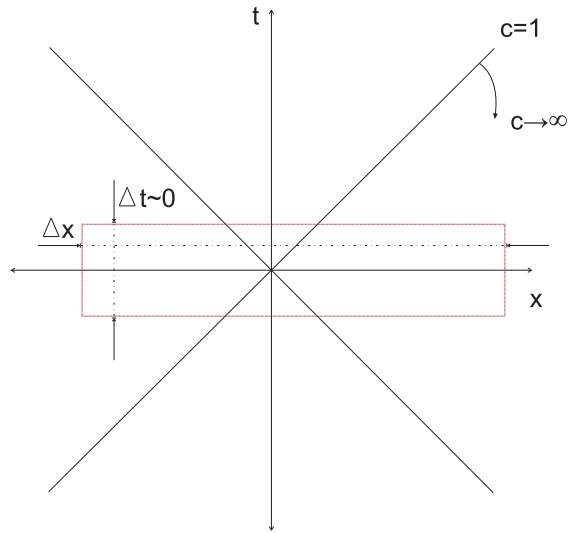}\\
  \caption{Contracción velocidad-tiempo.}\label{speed-time}
     \end{center}
\end{figure}

Esta contracción es respecto al subgrupo euclideano tridimensional, generado por $\{J_i,P_j\}$. El grupo contraído se puede interpretar como considerar bajas velocidades e intervalos tipo \emph{espacio} grandes (ver Fig. \ref{speed-time}).
  De acuerdo con la definición de esta contracción obtenemos grupos que describen intervalos que conectan eventos sin conexión causal, por lo tanto el interés físico es reducido. El mapeo de esta contracción va de la siguiente forma
\begin{center}
R1 $\rightarrow$ R3, \qquad R2 $\rightarrow$ R4, \qquad A1 $\rightarrow$
A3, \qquad A2 $\rightarrow$ A4.
\end{center}
Es útil mencionar que obtenemos intervalos tipo espacio absoluto. Esta contracción lleva de Poincaré al espacio de Carroll.

\bigskip\noindent{\it 3. Contracción de espacio-tiempo}

\begin{figure}[h]
\begin{center}
  \includegraphics[width=0.55\textwidth]{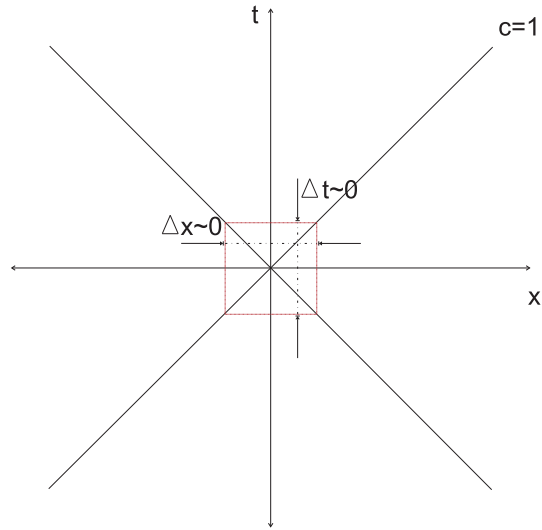}\\
  \caption{Contracción espacio-tiempo.}\label{space-time}
     \end{center}
\end{figure}

\bigskip Se contrae respecto al grupo de rotaciones y boosts (grupo de Lorentz), generado por $\{J_i,K_j\}$,

\begin{equation}
\mathbf{P} \rightarrow \epsilon \mathbf{P}, \qquad H \rightarrow
\epsilon H, \qquad \epsilon \rightarrow 0.
\end{equation}
Físicamente corresponde a considerar pequeñas unidades de espacio y tiempo (ver Fig. \ref{space-time}). Los grupos que aparecen describen propiedades locales y, a diferencia de los casos anteriores, para velocidades arbitrariamente grandes.

Los espacios resultantes se pueden entender como el límite ``plano'' de los grupos originales, esta contracción produce el siguiente mapeo
\begin{center}
R1 $\rightarrow$ R2, \qquad R3 $\rightarrow$ R4, \qquad A1 $\rightarrow$
A2, \qquad A3 $\rightarrow$ A4.
\end{center}
De acuerdo la interpretación es natural llamar a R1, R3, A1 y A3 como grupos ``cosmológicos'' y ``locales'' a R2, R4, A2 y A4. Un resumen con las tres contracciones se puede ver en la tabla \ref{cont tabl}.

\begin{figure}[h]
\begin{center}
  \includegraphics[width=0.55\textwidth]{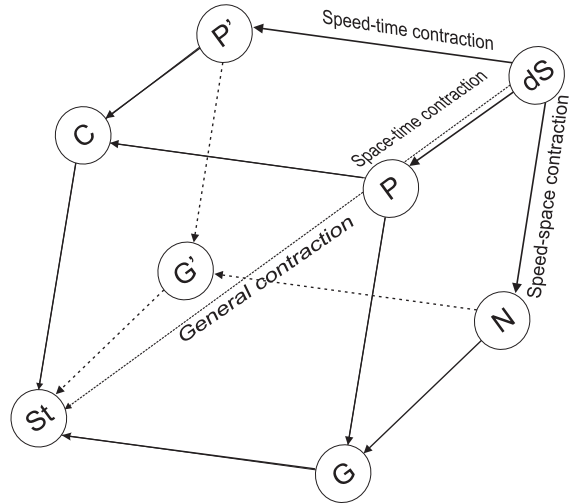}\\
  \caption{Los ocho grupos cinemáticos pueden ser situados en los vértices de un cubo tridimensional, luego las tres contracciones velocidad-tiempo, velocidad-espacio y espacio tiempo actúan por las aristas, y la contracción general actúa por la diagonal interna.}\label{cubo}
     \end{center}
\end{figure}
\begin{table}[h]
\begin{center}
\scalebox{0.8}[0.8]{
\begin{tabular}{l|lccccc}
{\small Tipo de contracción} & \multicolumn{6}{|c}{\small Tipos de grupos
} \\ \hline\hline
\multirow{3}{*}[3 mm]{\small 1. Velocidad - Espacio} & {\small Tiempo relativo}
& {\small :} & {\small (dS)} & {\small (P)} & {\small (P')} & {\small (C)}
\\
\multicolumn{1}{l|}{} & {\small Tiempo absoluto} & {\small :} & {\small (NH)}
& {\small (G)} & {\small (G')} & {\small (St)} \\ \hline
\multirow{3}{*}[3 mm]{\small 2. Velocidad - Tiempo} & {\small Espacio relativo}
& {\small :} & {\small (dS)} & {\small (NH)} & {\small (P)} & {\small (G)}
\\
\multicolumn{1}{l|}{} & {\small Espacio absoluto} & {\small :} & {\small (P')%
} & {\small (G')} & {\small (C)} & {\small (St)} \\ \hline
\multirow{3}{*}[3 mm]{\small 3. Espacio - Tiempo} & {\small Cosmol\'{o}gicos}
& {\small :} & {\small (dS)} & {\small (NH)} & {\small (P')} & {\small (G')}
\\
\multicolumn{1}{l|}{} & {\small Locales} & {\small :} & {\small (P)} &
{\small (G)} & {\small (C)} & {\small (St)}%
\end{tabular}
}\caption{Resumen con las contracciones tipo 1,2 y 3, y la relativa interpretación física de los grupos. Cada contracción de la columna izquierda lleva de la primera a la segunda línea de la columna derecha.}\label{cont tabl}
\end{center}
\end{table}

\bigskip\noindent{\it 4. Contracción general}

\bigskip En este caso el único subgrupo que permanece invariable es el de rotaciones,
\begin{equation}
\mathbf{P} \rightarrow \epsilon \mathbf{P}, \qquad H \rightarrow
\epsilon H, \qquad \mathbf{K} \rightarrow \epsilon \mathbf{K}, \qquad \epsilon \rightarrow 0.
\end{equation}
Esta contracción combina las características de las precedentes y, consecuentemente, partiendo de cualquier grupo llegamos al grupo Estático.

Un hecho destacable corresponde a que, sin la necesidad de hipótesis adicionales, cada uno de los grupos mencionados pueden ser considerados como un grupo de transformaciones actuando en coordenadas de espacio tiempo. Claro que esto resulta de las suposiciones ya hechas, de donde se deduce que las rotaciones y las transformaciones de boosts forman un subgrupo para cualquiera de los grupos cinemáticos, ver tabla \ref{possible kinematics tabla}. Luego para cada grupo cinemático, existe un espacio inhomogéneo que corresponde al grupo cuociente entre el grupo en sí mismo y el subgrupo seis dimensional de las rotaciones más los boosts. De cualquier forma es necesario aclarar que la sola existencia de este espacio inhomogéneo no ofrece garantías de que el grupo actúe efectivamente, es decir, que cada generador produzca transformaciones no triviales.

Gracias a que el grupo estático es la versión más abeliana de todos los grupos cinemáticos, podemos ver fácilmente el espacio inhomogéneo asociado, mediante la regla de composición,
\begin{equation}
(b',\mathbf{a}',\mathbf{u}',R')(b,\mathbf{a},\mathbf{u},R)=(b'+b,\mathbf{a}'+R\mathbf{a},\mathbf{u}'+R\mathbf{u},R'R),\label{ac st grp}
\end{equation}
donde
\begin{equation}
(b,\mathbf{a},\mathbf{u},R)=e^{b H}e^{\mathbf{a}\cdot \mathbf{P}}e^{\mathbf{u}\cdot \mathbf{K}}e^{\mathbf{n}\cdot \mathbf{J}},
\end{equation}
entonces todos los sistemas están en reposo para cualquier marco de referencia inercial y las transformaciones de boosts no producen ningún tipo de movimiento. A partir de (\ref{ac st grp}) vemos que los boosts actúan trivialmente.

De la figura \ref{cubo}, es claro que los grupos de Para-Galileo (G') o el Estático (St) se pueden obtener mediante la aplicación combinada de las contracciones tipo 1 y 2,
\begin{equation}
\mathbf{P}\rightarrow \epsilon \mathbf{P}, \qquad H\rightarrow \epsilon H, \qquad \mathbf{K} \rightarrow \epsilon^2 \mathbf{K}, \qquad \epsilon \rightarrow 0,
\end{equation}
sobre las álgebras de de Sitter (dS) o Poincaré (P), respectivamente. Entonces, pueden describir un modelo con intervalos tipo-espacio y tipo-tiempo reducidos, pero con velocidades totalmente despreciables.

\bigskip\noindent{\it Sobre las extensiones centrales}

\bigskip Cada una de las álgebras tipo A aceptan una extensión central en el conmutador de traslaciones y boosts, $[\mathbf{P},\mathbf{K}]=(I)$. De acuerdo con estas extensiones aparecen representaciones proyectivas de los grupos correspondientes,
\be
U_r U_s=\Omega(r,s)U_{rs},
\ee
donde $U_r$ es un elemento del grupo en la representación dada y $\Omega(r,s)$ toma valores complejos de norma unidad, de tal forma de preservar las amplitudes de probabilidad cuánticas. $\Omega$ es llamado ``factor local''. Dado que $U_e=Id$, tenemos
\be
\Omega(r,e)=1=\Omega(e,s).\label{Omega=1}
\ee
Además, en virtud de la asociatividad del producto $U_r(U_s U_t)=(U_r U_s)U_t$, $\Omega(r,s)$ debe satisfacer
\be
\Omega(r,s)\Omega(rs,t)=\Omega(s,t)\Omega(r,st).\label{omega factor}
\ee
A cualquier función que satisfaga (\ref{Omega=1}) y (\ref{omega factor}) se le llama factor local.
De la libertad de elección de representativos del grupo, $U'_r=\phi(r) U_r$, $|\phi(r)|=1$, tenemos que $\Omega'(r,s)$ dado por
\be
\Omega'(r,s)=\frac{\phi(r)\phi(s)}{\phi(rs)}\Omega(r,s),
\ee
es equivalente a $\Omega(r,s)$. Una forma alternativa es trabajar con ``exponentes locales'' $\omega(r,s)$, relacionados por
\be
\Omega(r,s)=\exp{i\omega(r,s)},
\ee
análogamente, llamaremos exponente local a cualquier función que satisfaga
\bea
&\omega(r,e)=0=\omega(e,s),&\label{omega=0} \ (\mathrm{mod} \ 2\pi)\\
&\omega(r,s)+\omega(rs,t)=\omega(s,t)+\omega(r,st),&\label{omega cociclo}
\eea
podemos ver que trabajando con exponentes locales traspasamos la notación multiplicativa a una aditiva.
Dos exponentes locales serán equivalentes cuando estén relacionados vía
\be
\omega'(r,s)=\omega(r,s)+\Delta_{cob}(r,s), \qquad \Delta_{cob}(r,s)=\z(r)+\z(s)-\z(rs).
\ee

La condición (\ref{omega cociclo}) es la definición de los dos-cociclos. Los dos-cociclos de la forma $\Delta_{cob}(r,s)$ son dos-cociclos triviales o dos-cobordes, ya que satisfacen la condición de los dos-cocilos (\ref{omega cociclo}) por construcción. Naturalmente podemos definir los dos-cociclos y dos-cobordes en notación aditiva o multiplicativa. En este lenguaje dos dos-cociclos son equivalentes cuando difieren en un dos-coborde. Luego es posible construir grupos de cohomologías $H^2(G,U(1))$ con las clases de equivalencia de los dos-cociclos \cite{azcarragabook}.

La importancia de las representaciones proyectivas en el grupo de Galileo ha sido observada de la siguiente forma. En el caso de representaciones reales (exponentes locales triviales), aparecen problemas al intentar definir coordenadas localizables \cite{Bacry:1968zf,IW2,Wightman1}. En el caso del grupo de Galileo, Inönü y Wigner \cite{IW2} mostraron que bajo ninguna condición las autofunciones pueden ser interpretadas como funciones de onda de partículas físicas. Hamermesh \cite{Hamer}, haciendo el converso, afirmó que el operador de posición sólo se puede construir en el caso de representaciones proyectivas no triviales. Es posible ver que las soluciones de la ecuación de Schrödinger para la partícula libre transforman de acuerdo a las representaciones proyectivas del grupo de Galileo. Por estas razones, algunas veces, se refieren a las representaciones proyectivas del grupo de Galileo por ``representaciones físicas''.

Es interesante contrastar el hecho de que las únicas representaciones físicas de Galileo son proyectivas, con el teorema que dice que \emph{para representaciones finitas del grupo cualquier factor local es equivalente a $1$} \cite{bargmann} (ver también \cite{Weyl}, pag. 183-184). Por otro lado recordemos que hay un resultado general de teoría de grupos de Lie, que dice que cualquier representación finita de un grupo no compacto es necesariamente no unitaria.

Volveremos al tema de las representaciones proyectivas y las extensiones centrales de Galileo en la sección \ref{ex gal sec}, donde mencionaremos otros dos aspectos muy relevantes a nivel físico: la introducción de una regla de superselección asociada a la masa, y la íntima relación entre posibles geometrías en el contexto de la correspondencia AdS/CFT con las extensiones centrales de Galileo.

\section{Grupos de Newton-Hooke}\label{NH sec}

En esta sección haremos una breve introducción a uno de los grupos que nos interesará cuando estudiemos las extensiones exóticas, este es el grupo de Newton-Hooke\footnote{Para una historia del nombre de este grupo ver \cite{Hoogland:1978qs}}.

Como ya se mencionó, los grupos de Newton-Hooke aparecen cuando hacemos una contracción de velocidad-espacio sobre los grupos de de Sitter, de hecho, aquella contracción es la única que lleva de Poincaré a Galileo, por lo que corresponde al límite no relativista usual.

En esta sección aprovecharemos de mostrar una forma alternativa a (\ref{con-speed-space}) para realizar la contracción. Partiremos escribiendo la álgebra de adS (anti-de Sitter) en 3+1 dimensiones en su forma covariante
\begin{eqnarray}\label{adS}
&\left[M_{\mu\nu},M_{\rho\sigma}\right]=\left(\eta_{\mu\rho}M_{\nu\sigma}-\eta_{\mu\sigma}M_{\nu\rho}+\eta_{\nu\sigma}M_{\mu\rho}-\eta_{\nu\rho
}M_{\mu\sigma}\right),&\label{AdS1} \\
&\left[M_{\mu\nu},P_{\rho}\right]=\left(\eta_{\mu\rho}P_\nu-\eta_{\nu\rho}P_\mu\right),&\label{AdS2} \\
&\left[P_{\mu},P_{\nu}\right]=-\frac{1}{R^2}M_{\mu\nu}.&\label{AdS3}
\end{eqnarray}
donde $\eta_{\mu \nu}=diag(-,+,+,+)$. Este espacio naturalmente posee curvatura no trivial\footnote{Un hecho anecdótico comentado por Dyson \cite{Dyson:1972sd}, es que Bacry y Levy-Leblond fueron capaces de predecir un espacio físico con curvatura no trivial sin usar relatividad general ni geometría diferencial. A pesar de que el paper aparece muchos años después del desarrollo de esas teorías, su razonamiento es totalmente independiente.}. Es posible ver que $R$ tiene el sentido de radio de AdS y el escalar de Ricci toma el valor $Ric=-12/R^2$. Como paso intermedio, es útil reescribirla descomponiendo en partes espaciales y temporales
\begin{eqnarray}\label{adS noncov}
&\left[K_i,K_j\right]=-\epsilon_{ijk}J_k, \qquad \left[J_i,K_j\right]=\epsilon_{ijk}K_k, \qquad \left[J_i,J_j\right]=\epsilon_{ijk}J_k,& \\
&\left[K_i,H\right]=P_i, \qquad \left[K_i,P_j\right]= H \delta_{ij},&\\
&\left[J_i,H\right]=0, \qquad \left[J_i,P_j\right]=\epsilon_{ijk}P_k,&\\
&\left[H,P_i\right]=\frac{1}{R^2}K_i, \qquad \left[P_i,P_j\right]=-\frac{1}{R^2}\epsilon_{ijk}J_k,&
\end{eqnarray}
donde
\begin{equation}
H=P^0=-P_0, \qquad K_i=-M_{0i}, \qquad J_i=\frac{1}{2}\epsilon_{ijk}M_{jk},
\end{equation}
aquí $i=1,2,3$ y $\epsilon_{ijk}$ es el tensor totalmente antisimétrico de tres componentes con la normalización $\epsilon_{123}=1$. La contracción esta vez la tomaremos haciendo
\begin{equation}
K_i \rightarrow \epsilon K_i, \qquad H \rightarrow \frac{1}{\epsilon} H, \qquad R \rightarrow \epsilon R,
\end{equation}
notar que de esta forma también es necesario incluir en el rescalamiento al radio $R$. Como resultado obtenemos la álgebra de Newton-Hooke $NH_-$ no extendida \cite{Bacry:1968zf,BacNuy,AldBar,Mariano,Gao,GibPat,BGK}
\begin{eqnarray}\label{NH 3+1}
&\left[K_i,K_j\right]=0, \qquad \left[K_i,J_j\right]=\epsilon_{ijk}K_k, \qquad \left[J_i,J_j\right]=\epsilon_{ijk}J_k,& \label{NH 3+1_1}\\
&\left[K_i,H\right]=P_i, \qquad \left[K_i,P_j\right]=0,&\\
&\left[J_i,H\right]=0, \qquad \left[J_i,P_j\right]=\epsilon_{ijk}P_k,&\\
&\left[H,P_i\right]=\frac{1}{R^2}K_i, \qquad \left[P_i,P_j\right]=0.&\label{NH 3+1_4}
\end{eqnarray}
Podemos ver que las subálgebras asociadas con la isotropía del espacio permanecen inalteradas. En cambio del conmutador de boosts con traslaciones vemos que tenemos un grupo asociado a un espacio de tiempo absoluto. La última línea de (\ref{NH 3+1}) la interpretamos como que la curvatura en las direcciones $\overrightarrow{ij}$ se ha ``aplanado'' pero aún sobrevive cierta curvatura en las direcciones $\overrightarrow{0i}$.


Para determinar la ley de composición del grupo debemos escoger una parametrización. Una posibilidad es hacer

\begin{equation}
(t,\mathbf{x},\mathbf{v},R_\mathbf{n}(\theta))=e^{t H}e^{\mathbf{x}\cdot\mathbf{P}}e^{\mathbf{v}\cdot\mathbf{K}}e^{\theta\mathbf{n}\cdot\mathbf{J}},
\end{equation}
donde $R_\mathbf{n}(\theta)$ es una rotación en torno a un eje $\mathbf{n}$ en un ángulo $\theta$. Luego obtenemos\footnote{Para el cálculo es necesario usar $e^{L}\ A\ e^{-L}=A+\left[ L,A\right] +\frac{1}{2!}\left[ L,\left[ L,A\right]
\right] +\frac{1}{3!}\left[ L,\left[ L,\left[ L,A\right] \right] \right] +...$}
\begin{eqnarray}
&(t',\mathbf{x}',\mathbf{v}',R_{\mathbf{n}'}(\theta'))(t,\mathbf{x},\mathbf{v},R_\mathbf{n}(\theta)) =(t'+t,\cos{\frac{t}{R}} \mathbf{x}'-R\sin{\frac{t}{R}} \mathbf{v}'+R'\mathbf{x},&\\
&\cos{\frac{t}{R}} \mathbf{v}'-\frac{1}{R}\sin{\frac{t}{R}} \mathbf{x}'+R'\mathbf{v},R_{\mathbf{n}'}(\theta')R_\mathbf{n}(\theta)).&\label{nh trans 3+1}
\end{eqnarray}

Tomaremos el espacio-tiempo como el coset
\begin{equation}
(t,\mathbf{x})=(t,\mathbf{x},\mathbf{v},R_\mathbf{n}(\theta))/(0,0,\mathbf{v},R_\mathbf{n}(\theta)),
\end{equation}
entonces las transformaciones de Newton-Hooke están dadas por
\begin{eqnarray}
&\mathbf{x}'=R_\mathbf{n}(\theta)\mathbf{x}-R\mathbf{v}\sin{\frac{t}{R}}+\mathbf{a}\cos{\frac{t}{R}},&\label{trans inh NH 3+1 1}\\
&t'=t+b.&\label{trans inh NH 3+1 2}
\end{eqnarray}
Estas transformaciones incluyen una dependencia no lineal en el tiempo, esto es porque los boost ya no son movimientos con velocidad constante, lo cual es heredado de las propiedades de AdS, y a su vez es una gran diferencia con Poincaré y Galileo. Podemos ver que en el límite plano $R \rightarrow \infty$ recuperamos las transformaciones del grupo de Galileo. Ya sea para el grupo de Newton-Hooke o el de Galileo, de (\ref{nh trans 3+1}) vemos la complejidad de la ley de composición de estos grupos versus la de Poincaré.

No es difícil ver que los campos vectoriales que producen las transformaciones de $NH_-$ (\ref{trans inh NH 3+1 1},\ref{trans inh NH 3+1 2}) están dados por
\begin{eqnarray}
{\cal X}_{J_i}&=&\epsilon_{ijk}x_j\frac{\partial}{\partial x_k},\\
{\cal X}_{H}&=&-\frac{\partial}{\partial t},\\
{\cal X}_{P_i}&=& \cos{\frac{t}{R}}\frac{\partial}{\partial x_i},\\
{\cal X}_{K_i}&=&R \sin{\frac{t}{R}}\frac{\partial}{\partial x_i},
\end{eqnarray}
los cuales satisfacen la álgebra $NH_-$ (\ref{NH 3+1_1}-\ref{NH 3+1_4}).

Otra manera de ver este sistema es notando que las transformaciones (\ref{trans inh NH 3+1 1},\ref{trans inh NH 3+1 2}) son simplemente las simetrías de un oscilador armónico isótropo con frecuencia de oscilación $1/R$
\be
\ddot x_i(t)+\frac{1}{R^2}x_i(t)=0,
\ee
las cuales se derivan del lagrangiano usual del oscilador armónico
\be
L=\frac{m}{2}\dot x^2_i-\frac{m}{2 R^2}x^2_i.\label{Lag oa}
\ee
Usando el teorema de Noether podemos obtener las integrales de movimiento asociadas a las transformaciones de Newton-Hooke (\ref{trans inh NH 3+1 1},\ref{trans inh NH 3+1 2}). Bajo rotaciones y traslaciones en el tiempo el Lagrangiano (\ref{Lag oa}) es invariante, mientras que para traslaciones espaciales y boosts es cuasi-invariante \cite{levyleblond69,Marmo:1987rv}, es decir aparece un término de derivada total. Con esto obtenemos las expresiones
\begin{eqnarray}
J_i&=& \epsilon_{ijk}x_j p_k,\\
H&=& \frac{m}{2} p^2_i+\frac{m}{2 R^2}x^2_i,\label{H nh}\\
P_i&=& \cos{t/R} \ p_i+\frac{m}{R} \sin{t/R} \ x_i,\\
K_i&=&m \cos{t/R} \ x_i-R\sin{t/R} \ p_i.
\end{eqnarray}
donde $p_i=\partial L/\partial\dot x_i$ es el momento canónico conjugado a $x_i$. El momento angular y el Hamiltoniano son los usuales para un oscilador armónico. El generador de traslaciones de Newton-Hooke es $P_i$ y el de boosts $K_i$. Estos generadores forman, con respecto a los paréntesis de Poisson, la álgebra de Newton-Hooke centralmente extendida por la masa $m$
\begin{eqnarray}
&\left\{J,H\right\}=0, \qquad \left\{J,P_i\right\}=\epsilon_{ij}P_j, \qquad \left\{J,K_i\right\}=\epsilon_{ij}K_j,& \\
&\left\{H,P_i\right\}=\frac{1}{R^2} K_i, \qquad \left\{H,K_i\right\}= - P_i,& \\
&\left\{K_i,P_j\right\}=m \delta_{ij}.&
\end{eqnarray}
Notar que $P_i$ y $K_i$ no conmutan con $H$, pero son integrales de movimiento en el sentido de $dP_i/dt=\{P_i,H\}+\partial P_i/\partial t=0$, y análogamente para $K_i$. En el límite plano recuperamos las expresiones usuales para los generadores de traslaciones y boosts Galileanos.

De (\ref{H nh}) es muy intrigante notar que la energía cinética de la partícula está cuantizada, lo cual evidentemente está relacionado a la ``compacidad'' en el tiempo del espacio-tiempo de Newton-Hooke (o al menos de la parte espacial). De cualquier forma la separación entre niveles esta dada por
\be
\Delta E=\frac{\hbar}{R},
\ee
la cual puede ser extremadamente pequeña si $R$ es del orden del tiempo de vida del universo.

\section{Grupo de Galileo Exótico}\label{ex gal sec}

Hemos comentado que las álgebras tipo A poseen extensiones centrales, y que de hecho son fundamentales para describir sistemas físicos que realizan aquellas simetrías. De este modo, aparece de forma natural el estudio de las extensiones exóticas. El nombre proviene de que para el caso de 2+1 dimensiones (y sólo para este) las álgebras tipo A admiten dos extensiones centrales adicionales \cite{BGO,Grigore:1993fz,Bose:1994sj,Bose:1994si,Brihaye:1995nv,Hagen:2002pg}.

Comenzaremos escribiendo las transformaciones de Galileo en 2+1 dimensiones
\begin{eqnarray}\label{Gal trans 2+1}
\mathbf{x}'&=&R(\theta)\mathbf{x}+\mathbf{v}t+\mathbf{u},\\
t'&=&t+\tau,
\end{eqnarray}
donde \footnote{Una descripción alternativa es considerando numeros complejos $(t,\mathbf{x})=(t,x_1+i x_2)$, $R(\theta)=e^{i\theta}$, $\mathbf{u}=u_1+i u_2$ y $\mathbf{v}=v_1+i v_2$.}
\begin{equation}
R(\theta)=\left(\begin{array}{cc}\cos{\theta} & -\sin{\theta}\\ \sin{\theta} & \cos{\theta} \end{array}\right), \qquad \mathbf{u}=\left(\begin{array}{c} u_1\\u_2\end{array}\right), \qquad \mathbf{v}=\left(\begin{array}{c} v_1\\v_2\end{array}\right).
\end{equation}
No es difícil realizar la composición de dos transformaciones
\begin{equation}
\mathbf{x} \overset{(\tau,\mathbf{u},\mathbf{v},\theta)}\longrightarrow \mathbf{x}' \overset{(\tau',\mathbf{u}',\mathbf{v}',\theta')}\longrightarrow \mathbf{x}'',
\end{equation}
de donde se puede ver la ley de composición del grupo ${G}$ (Galileo)
\begin{equation}
(\tau',\mathbf{u}',\mathbf{v}',\theta')(\tau,\mathbf{u},\mathbf{v},\theta)=(\tau+\tau',R(\theta')\mathbf{u}+\mathbf{v}'\tau+\mathbf{u}',R(\theta')\mathbf{v}+\mathbf{v}',\theta+\theta').
\end{equation}

Por simplicidad podemos considerar la siguiente realización finita
\begin{equation}
g(\tau,\mathbf{u},\mathbf{v},\theta)=
\left(\begin{array}{ccc}
R(\theta)&\mathbf{v}&\mathbf{u}\\
0&1&\tau\\
0&0&1\\
\end{array}\right),
\end{equation}
luego la acción sobre el espacio homogéneo está dada matricialmente
\begin{equation}
\left(\begin{array}{ccc}
\mathbf{x}'\\
t'\\
1\\
\end{array}\right)=
\left(\begin{array}{ccc}
R(\theta)&\mathbf{v}&\mathbf{u}\\
0&1&\tau\\
0&0&1\\
\end{array}\right)\left(\begin{array}{ccc}
\mathbf{x}\\
t\\
1\\
\end{array}\right).
\end{equation}

Los elementos neutro e inverso están dados respectivamente por
\begin{equation}
e=
\left(\begin{array}{ccc}
R(0)&0&0\\
0&1&0\\
0&0&1\\
\end{array}\right), \qquad
g^{-1}(\tau,\mathbf{u},\mathbf{v},\theta)=
\left(\begin{array}{ccc}
R^{-1}&-R^{-1}\mathbf{v}&-R^{-1}(\mathbf{u}-\mathbf{v}\tau)\\
0&1&-\tau\\
0&0&1\\
\end{array}\right),
\end{equation}
donde $R^{-1}=R(-\theta)$.

Como subgrupos tenemos:
\begin{enumerate}
\item Traslaciones de espacio-tiempo $T$
\begin{equation}
T=\left\{\left(\begin{array}{ccc}
1&0&\mathbf{u}\\
0&1&\tau\\
0&0&1\\
\end{array}\right)\right\},
\end{equation}
el cual es posible ver que es normal
\begin{equation}
\left(\begin{array}{ccc}
R'&\mathbf{v}'&\mathbf{u}'\\
0&1&\tau'\\
0&0&1\\
\end{array}\right)\left(\begin{array}{ccc}
1&0&\mathbf{u}\\
0&1&\tau\\
0&0&1\\
\end{array}\right)\left(\begin{array}{ccc}
R'&\mathbf{v}'&\mathbf{u}'\\
0&1&\tau'\\
0&0&1\\
\end{array}\right)^{-1}=\left(\begin{array}{ccc}
1&0&R'\mathbf{u}+\mathbf{v}'\tau\\
0&1&\tau\\
0&0&1\\
\end{array}\right).
\end{equation}
Es claro que $T$ es isomorfo a $\mathbb{R}^3$, ya que es producto directo de los subgrupos de traslaciones temporales $T_t$ con las espaciales $T_x$,
\be
T = T_t \ \otimes \ T_x.
\ee
\item El sector homogéneo de Galileo ${G}_0$, definido como
\begin{equation}
{G}_0=\left\{\left(\begin{array}{ccc}
R&\mathbf{v}&0\\
0&1&0\\
0&0&1\\
\end{array}\right)\right\}.
\end{equation}
$G_0$ a diferencia de $T$ no es invariante. Entonces podemos escribir $G$ como el producto semidirecto de $T$ con $G_0$
\begin{equation}
G=T \ \otimes_s \ G_0.
\end{equation}
Podemos descomponer $G_0$ aún más, con los subgrupos de rotaciones $J$ y boosts $K$
\bea
J&=&\left\{\left(\begin{array}{ccc}
R&0&0\\
0&1&0\\
0&0&1\\
\end{array}\right)\right\},\\
K&=&\left\{\left(\begin{array}{ccc}
1&\mathbf{v}&0\\
0&1&0\\
0&0&1\\
\end{array}\right)\right\},
\eea

se puede ver que $K$ es subgrupo normal de $G_0$, entonces $G_0$ es el producto semidirecto de $K$ con $J$
\begin{equation}
G_0=K \ \otimes_s \ J,
\end{equation}
luego podemos descomponer $G$ de la forma
\begin{equation}
G=(T_t \ \otimes \ T_x) \ \otimes_s \ (K \ \otimes_s \ J).\label{descop G1}
\end{equation}
Notar que $G_0$ es isomorfo al grupo euclideano en dos dimensiones $G_0 \cong E_K(2)$. Por otra parte, también tenemos el subgrupo euclideano de $G$, dado por las rotaciones y traslaciones espaciales $E_{T_x}(2) = T_x \ \otimes_s \ J$, pero a diferencia de $E_K(2)$, $E_{T_x}(2)$ no es subgrupo normal de $G$.

\item Otro subgrupo normal de $G$ está dado por el producto directo del de traslaciones espaciales $T_x$ con el generado por los boosts $K$
    \be
    N=T_x \ \otimes \ K,
    \ee
$N$ es el grupo abeliano maximal.

\item También tenemos otro subgrupo de $G$ formado por el producto directo del de traslaciones temporales $T_t$ con el de rotaciones $J$
\be
    G_{ps}=T_t \ \otimes \ J,
    \ee
luego $N$ y $G_{ps}$ nos proporcionan una descomposición independiente a \ref{descop G1}
\begin{equation}
G=N \ \otimes_s \ G_{ps}=(T_x \ \otimes \ K) \ \otimes_s \ (T_t \ \otimes \ J).\label{descop G2}
\end{equation}

\end{enumerate}
Podemos descomponer un elemento $g$ de la forma
\begin{equation}
\left(\begin{array}{ccc}
R(\theta)&\mathbf{v}&\mathbf{u}\\
0&1&\tau\\
0&0&1\\
\end{array}\right)=
\left(\begin{array}{ccc}
1&0&0\\
0&1&\tau\\
0&0&1\\
\end{array}\right)
\left(\begin{array}{ccc}
1&0&\mathbf{u}\\
0&1&0\\
0&0&1\\
\end{array}\right)
\left(\begin{array}{ccc}
1&\mathbf{v}&0\\
0&1&0\\
0&0&1\\
\end{array}\right)
\left(\begin{array}{ccc}
R(\theta)&0&0\\
0&1&0\\
0&0&1\\
\end{array}\right),
\end{equation}
entonces es claro que podemos parametrizar exponencialmente
\begin{equation}
g(\tau,\mathbf{u},\mathbf{v},\theta)=e^{-i \tau H}e^{i \mathbf{u}\cdot \mathbf{P}}e^{i \mathbf{v}\cdot \mathbf{K}}e^{i \theta J},
\end{equation}
como resultado obtenemos la álgebra de Galileo (no extendida) en 2+1 dimensiones
\begin{eqnarray}
&\left[J,P_i\right]=i\epsilon_{ij}P_j, \qquad \left[J,K_i\right]=i \epsilon_{ij}K_j,&\\
&\left[K_i,H\right]=i P_i,&
\end{eqnarray}
todos los conmutadores restantes son triviales. Notar que esta álgebra es el limite plano, $R \rightarrow \infty$, del análogo 2+1 dimensional de (\ref{NH 3+1_1}-\ref{NH 3+1_4}).

Para estudiar las extensiones centrales de esta álgebra el procedimiento más inocente y directo corresponde a agregar elementos centrales $\mathbf{1}$ con coeficientes arbitrarios al lado derecho de todas las relaciones de conmutación (incluyendo las triviales) y luego de, (i) hacer re-definiciones de los $\mathbf{P}$'s y $\mathbf{K}$'s, y (ii) exigir que se satisfagan las identidades de Jacobi para cualquier generador llegamos a una familia de tres parámetros de extensiones centrales, las cuales aparecen en los conmutadores
\begin{eqnarray}
&\left[P_i,K_j\right]=-i m \delta_{ij},&\label{EGAL1}\\
&\left[K_i,K_j\right]=i \kappa \delta_{ij},\label{EGAL2}&\\
&\left[J,H\right]=i l \delta_{ij},&
\end{eqnarray}
el resto de conmutadores permanece sin cambio. En lo que sigue nos restringiremos al caso $l=0$. La razón de esto lo podemos ver de la siguiente forma: si $l\ne0$ entonces
\begin{equation}
e^{i\theta J}H e^{-i\theta J}=H+ l \theta,
\end{equation}
pero para tener resultados idénticos para $\theta=0$ y $\theta=2 \pi$ debemos hacer $l=0$ \cite{LSZ}, para más detalles ver Sección \ref{E-C coho}.

Podemos construir el grupo mediante exponenciación,
\be
g(z,w,\tau,\mathbf{u},\mathbf{v},\theta)=e^{i z m},e^{-i w \kappa},e^{-i \tau H},e^{i \mathbf{u} \cdot\mathbf{P}},e^{i \mathbf{v} \cdot\mathbf{K}},e^{i \theta J},
\ee
donde el signo en la segunda exponencial del lado derecho es puramente convencional. La composición de dos elementos produce
\bea
&g(z,w,\tau,\mathbf{u},\mathbf{v},\theta)g(z',w',\tau',\mathbf{u}',\mathbf{v}',\theta')=&\\
&g(z+z'+\frac{1}{2}\mathbf{v}^2\tau'+\mathbf{v}^T R(\theta)\mathbf{u}',w+w'+\frac{1}{2}\mathbf{v}^T R(\theta)\mathbf{v}',&\\&\tau+\tau',\mathbf{u}+R(\theta)\mathbf{u}'+\tau'\mathbf{v},\mathbf{v}+R(\theta)\mathbf{v}',\theta+\theta').\label{ex gal comp}&
\eea
Es posible chequear que la siguiente representación finita
\begin{equation}
g=\left(
\begin{array}{ccccc}

R&\mathbf{v}&0&\mathbf{u}&\half\varepsilon\mathbf{v}\\

0&1&0&\tau&0\\

\mathbf{v}^T R&\half\mathbf{v}^2&1&z&w\\
0&0&0&1&0\\
0&0&0&0&1
\end{array}
\right),
\label{BargmannMatrices}
\end{equation}
reproduce correctamente (\ref{ex gal comp}). Una sutileza a comentar es que (\ref{ex gal comp}) y también (\ref{BargmannMatrices}) corresponden de hecho a ``true'' representations, esto debido a que estamos considerando $m$ y $\kappa$ como generadores, y no números propiamente tal, esto corresponde a trabajar en el grupo extendido \cite{Duval:1984cj,Duval:2008jg}. Es pertinente remarcar esto debido al teorema que mencionamos al final de la Sección \ref{pos kin sec}: todos los factores de las representaciones de rayos de grupos finitos son equivalentes a la unidad.







\chapter{Simetría de Newton-Hooke exótica}\label{enh chapter}

En este capítulo estudiaremos la simetría de Newton-Hooke exótica. En la sección \ref{ENH algebra} obtenemos la álgebra doblemente extendida (centralmente), como contracción de AdS$_3$ y mostramos la existencia de dos bases de generadores que nos serán útiles para el estudio de distintos aspectos. Después, en la sección \ref{classicalLag}, derivamos el Lagrangiano para la partícula que realiza esta simetría y realizamos un estudio del punto de vista clásico de la dinámica, el análisis de vínculos y las simetrías. Luego, en la sección \ref{classicalLag chiral}, repetimos el análisis en términos de otras variables que además facilitan el estudio de algunas de las propiedades inusuales del sistema, debidas a la estructura exótica. En la sección \ref{reducedquant} realizamos la cuantización en espacio de fase reducido, y en la sección \ref{waveEquat} derivamos las ecuaciones de onda, es decir la cuantización en el cuadro de Schrödinger. En la sección \ref{Phaseproj} obtenemos las fases proyectivas con las cuales transforma la función de onda. Después, en la sección \ref{ext space sec}, estudiamos brevemente la simetría en espacio de fase extendido. En la sección \ref{map hyp case} comentamos sobre el paso al caso hiperbólico. Finalmente, en la sección \ref{conclus ENH}, damos algunas conclusiones para el capítulo.

\section{Álgebra exótica de Newton-Hooke}\label{ENH algebra}

Es posible obtener la álgebra exótica de NH$_-$ como contracción de la algebra de AdS${}_3$. Para realizar la contracción no relativista primero descompondremos (\ref{AdS1}-\ref{AdS3}) (su análogo cuántico en 2+1) en partes espaciales y temporales
\begin{equation}\label{P0M12}
    [P_0,M_{12}]=0,
\end{equation}
\begin{equation}\label{P0M0i}
    [P_0,M_{0i}]=iP_i,\qquad
    [P_0,P_i]=-i\frac{1}{R^2}M_{0i},
\end{equation}
\begin{equation}\label{M12Pi}
    [M_{12},P_i]=i\epsilon_{ij}P_j,\qquad
    [M_{12},M_{0i}]=i\epsilon_{ij}M_{0j},
\end{equation}
\begin{equation}\label{M0iPj}
    [M_{0i},P_j]=-i\delta_{ij}P_0,\qquad
    [M_{0i},M_{0j}]=-i\epsilon_{ij} M_{12},
\end{equation}
\begin{equation}\label{PiPj}
    [P_i,P_j]=-i\frac{1}{R^2}\epsilon_{ij}M_{12},
\end{equation}
donde $\mu=0,i$; $i,j=1,2$, $\epsilon_{12}=1$. Reemplazaremos los generadores y el radio $R$ de AdS${}_3$ por las cantidades reescaladas
\begin{equation}\label{contr}
    P^0=-P_0\rightarrow \omega Z +\frac{1}\omega H,\qquad
    M_{0i}\rightarrow  \omega K_i,\qquad
    M_{12}\rightarrow  \omega^2\tilde{Z}+J,\qquad
    R\rightarrow \omega R.
\end{equation}
Tomando el límite $\omega\rightarrow\infty$, obtenemos la álgebra exótica de Newton-Hooke $ENH_-$ con dos cargas centrales $Z$ y $\tilde{Z}$,
\begin{equation}\label{NH3ex}
    [H,J]=0,
\end{equation}
\begin{equation}\label{HPK}
    [H,K_i]=-iP_i,\qquad [H,P_i]=i\frac{1}{R^2}K_i,
\end{equation}
\begin{equation}\label{JKP}
    [J,P_i]=i\epsilon_{ij}P_j,\qquad
    [J,K_i]=i\epsilon_{ij}K_j,
\end{equation}
\begin{equation}\label{KPK}
    [K_i,P_j]=i\delta_{ij}Z,\qquad
    [K_i,K_j]=-i\epsilon_{ij}\tilde{Z},
\end{equation}
\begin{equation}\label{PP}
    [P_i,P_j]=-i\frac{1}{{R}^2}\epsilon_{ij}\tilde{Z}.
\end{equation}
Esta álgebra ha sido considerada antes en \cite{Mariano,Gao}. En el límite plano $R\rightarrow \infty$ reproduce la álgebra de Galileo exótica (\ref{EGAL1},\ref{EGAL2}) (con l=0)\cite{LL,BGO,Brihaye:1995nv}.

De la cohomología de Eilenberg-Chevalley de la álgebra de NH$_-$ no extendida ($Z=0$ y $\tilde{Z}=0$) es posible ver incluso que hay una tercera extensión en el conmutador $[H,J]=\tilde{\tilde{Z}}$, pero esta no es posible obtenerla como contracción de AdS${}_3$. Al incluir esta extensión se degeneran los Casimires asociados a la energía y el momento angular, por este motivo no la consideraremos.

Los Casimires cuadráticos de $ENH_-$ pueden ser obtenidos de los dos Casimires cuadráticos de AdS${}_3$
\begin{equation}\label{CasAdS1}
    C_1=-P_\mu
    P^\mu+\frac{1}{2R^2}M_{\mu\nu}M^{\mu\nu}=P_0^2-P_1^2-P_2^2+\frac{1}{R^2}
    (J_0^2-J_1^2-J_2^2),
\end{equation}
\begin{equation}\label{CasAdS2}
    C_2=-P_\mu J^\mu=P_0J_0-P_1J_1-P_2J_2,
\end{equation}
donde
\begin{equation}
    J_{\mu}=\frac{1}{2}\epsilon_{\mu\nu\lambda}M^{\nu\lambda}
\label{Jmu}
\end{equation}
con $\epsilon_{012}=+1$. Notar que nosotros podemos considerar $C_1$ y $\frac{1}{R}C_2$ como Casimires de la misma dimensión. Si tomamos la contracción (\ref{contr}), la parte finita de la expansión nos da
\begin{equation}\label{CasNH1}
    {\cal C}_1=2\left(ZH+\frac{1}{{R}^2}\tilde{Z}J\right)
    -P_i^2-\frac{1}{{R}^2}K_i^2,
\end{equation}
\begin{equation}\label{CasNH2}
    {\cal C}_2=-ZJ-\tilde{Z}H-\epsilon_{ij}P_iK_j.
\end{equation}
La base de generadores $\{H,J,P_i,K_i,Z,\tilde{Z}\}$ la llamaremos ``covariante'', debido a que respecto a esta base es supuesto que existe un espacio inhomogéneo cuyas transformaciones son las de NH estándar.
Es conveniente también representar $ENH_-$ en otra base, que llamaremos chiral. La existencia de esta base está basada en el siguiente isomorfismo de álgebras
\begin{equation}\label{Isomor}
    AdS_3\sim SO(2,2)\sim SO(2,1)\oplus SO(2,1)\sim
    AdS_2\oplus AdS_2.
\end{equation}
la primera relación se ve haciendo $M_{\mu 3}=RP_\mu$. De aquí es claro que sólo en 2+1 tenemos dos extensiones centrales independientes, ya que podemos incluir una extensión central por cada AdS${}_2$.
Definiendo
\begin{equation}\label{Jpmdef}
J^\pm_\mu=\frac{1}{2}(J_\mu\mp RP_\mu),
\end{equation}
obtenemos una forma equivalente de la álgebra AdS${}_3$,
\begin{equation}\label{JJAdS}
    [J^\pm_\mu,J^\pm_\nu]=i\epsilon_{\mu\nu}{}^\lambda
    J^\pm_\lambda,\qquad
    [J^+_\mu,J^-_\nu]=0,
\end{equation}
con los Casimires
\begin{equation}\label{Cas+-}
    C^\pm=\eta^{\mu\nu}J^\pm_\mu
    J^\pm_\nu={J^\pm_1}^2+{J^\pm_2}^2-{J^\pm_0}^2.
\end{equation}
Notar que $R$ está oculto en la definición (\ref{Jpmdef}), por esto la base chiral se degenera para el límite plano, con lo cual, cuando más adelante deseemos considerar este límite, deberemos usar la base covariante. En correspondencia con la contracción no relativista realizada en la base covariante, reemplazamos
\begin{equation}\label{contrchiral}
    J^\pm_0\rightarrow -\omega^2Z^\pm +{\cal J}^\pm,\qquad
    J^\pm_i\rightarrow \omega {\cal J}^\pm_i,
\end{equation}
y tomamos el límite $\omega\rightarrow \infty$. Como resultado obtenemos $ENH_-$ en la base chiral
\begin{equation}\label{NHchiral}
    [{\cal J}^\pm_i,{\cal J}^\pm_j]=i\epsilon_{ij}Z^\pm,\qquad
    [{\cal J}^\pm,{\cal J}^\pm_i]=i\epsilon_{ij}{\cal J}^\pm_j,\qquad
    [G^+_a,G^-_b]=0,
\end{equation}
donde $G^\pm_a=Z^\pm,{\cal J}^\pm,{\cal J}^\pm_i$. Los Casimires son
\begin{equation}\label{CasNHch}
    {\cal C}^\pm={{\cal J}_i^\pm}^2+2Z^\pm{\cal J}^\pm.
\end{equation}
En correspondencia con la descomposición (\ref{Isomor}), encontramos que la extensión central doble de $NH_-^3$ es suma directa de dos algebras centralmente extendidas $NH_-^2$, obtenidas por una contracción de sus respectivas partes AdS${}_2$.

Para más adelante será útil la relación entre generadores de ambas bases, chiral y covariante,
\begin{equation}\label{NHchNch}
    {\cal J}^\pm=\frac{1}{2}\left(J\pm  R H\right),\qquad
    {\cal J}^\pm_i=\frac{1}{2}\left(\epsilon_{ij}K_j\mp  RP_i\right),\qquad
    Z^\pm=-\frac{1}{2}\left(\tilde{Z}\pm  RZ\right).
\end{equation}


\section{Lagrangiano para el sistema con simetría de $ENH_-$}\label{classicalLag}

Construiremos el Lagrangiano para la partícula con simetría exótica de Newton-Hooke mediante el método de las realizaciones no lineales \cite{Coleman}, para esto debemos elegir un coset apropiado ${\cal G}/{\cal H}$. En nuestro caso ${\cal G}$ es el grupo $NH_-$ doblemente extendido y ${\cal H}$ es el grupo de rotaciones en dos dimensiones. Podemos parametrizar los elementos del coset de la forma
\begin{equation}
\label{coset1} g=e^{-iH x^0}\;e^{iP_ix^i}\;e^{iK_jv^j} \;e^{iZc}
\;e^{-i\tilde Z\tilde c}.
\end{equation}
Las coordenadas (de Goldstone) del coset dependen de $\tau$ que parametriza la linea de mundo de la partícula, mirar por ejemplo \cite{West,GomisKW}.


La uno-forma de Maurer-Cartan es
 \begin{equation}\label{MCar}
    \Omega=-ig^{-1}dg= -L_H H +L^i_P P_i+L^i_K K_i+L_J J
    -L_Z Z+ L_{\tilde Z} \tilde Z\, ,
\end{equation}
donde
\begin{eqnarray}
    L_H &=& dx^0,
    \qquad
    L^i_P = dx^i-v^i dx^0,
    \qquad
    L^i_K = dv^i+\frac{x^i}{R^2}\,dx^0,
    \qquad
    L_J = 0, \nn\\
    L_Z &=& -dc-v^idx^i+\frac{v_i^2}{2}dx^0+\frac{x_i^2}{2R^2}\,dx^0,
    \nn\\
    L^{\tilde Z} &=& -d\tilde c -\frac{1}{2}\epsilon_{ij}
    \left(v^idv^j+\frac{1}{R^2}x^i dx^j-
    \frac{2}{R^2}x^iv^jdx^0\right). \label{MCNH2}
\end{eqnarray}
La uno-forma (\ref{MCar}) satisface la ecuación de Maurer-Cartan $d\Omega+i\Omega\wedge\Omega=0$.

El lagrangiano se obtiene como pullback de una combinación lineal de las uno-formas no triviales invariantes bajo rotaciones, $L_H, L_Z,
L_{\tilde Z}$,
\begin{equation}\label{nhactionZZ}
{\cal L}=\mu\dot{x}^0+
    m\left(\dot c+v^i\dot x^i-\frac{v_i^2}{2}\dot{x}^0-\frac{x_i^2}{2R^2}\dot{x}^0\right)\\
    +\kappa\left(\dot{\tilde c}+\frac{1}{2}\epsilon_{ij}\left(v^i\dot v^j+\frac{1}{R^2}x^i \dot
    x^j- \frac{2}{R^2}x^iv^j\dot{x}^0\right)\right),
\end{equation}
donde $\mu$ y $m$ son constantes que tienen dimension de $R^{-1}$ mientras $\kappa$ es adimensional.
El Lagrangiano (\ref{nhactionZZ}) es invariante bajo reparametrizaciones $\tau \rightarrow f(\tau)$, fijando el gauge mediante $\tau=x^0\equiv t$ y descartando los términos de derivada total, obtenemos
\begin{equation}\label{nhaction}
{\cal L}_{nc}=
    m\left(v_i\dot x_i-\frac{v_i^2}{2}-\frac{x_i^2}{2R^2}\right)\\
    +\frac 12\kappa\epsilon_{ij}\left(v_i\dot v_j+\frac{1}{R^2}x_i \dot
    x_j- \frac{2}{R^2}x_iv_j\right).
\end{equation}

\subsection{Dinámica clásica}\label{classicaldyn}

La variación del Lagrangiano (\ref{nhaction}) en $v_i$ y $x_i$ produce las ecuaciones de movimiento
\begin{eqnarray}
&&m(\dot{x}_i-v_i)+{\kappa}\epsilon_{ij}\left(\dot{v}_j+R^{-2}x_j\right)=0,
\label{eom1}
\\
&&m\left(\dot{v}_i+R^{-2}
x_i\right)-R^{-2}\epsilon_{ij}(\dot{x}_j-v_j)=0, \label{eom2}
\end{eqnarray}
que equivalentemente, pueden ser presentadas en la forma
\begin{eqnarray}
&&(\dot{x}_i-v_i)+\rho
R\epsilon_{ij}\left(\dot{v}_j+R^{-2}x_j\right)=0, \label{eom1'}
\\
&&(\dot{x}_i-v_i)+\rho^{-1}
R\epsilon_{ij}\left(\dot{v}_j+R^{-2}x_j\right)=0, \label{eom2'}
\end{eqnarray}
donde
\begin{equation}  \label{rho}
\rho=\frac{\kappa}{mR}.
\end{equation}
Para $\rho^2 \ne 1$, estas ecuaciones implican
\begin{equation}\label{vdotx}
    \dot x_i-v_i=\dot{v}_i+R^{-2}x_i=0
    \qquad\to \qquad \ddot x_i+\frac{1}{R^2}x_i=0.
\end{equation}

Un punto importante que podemos ver de las ecuaciones (\ref{eom1},\ref{eom2}) es que $v_i=\dot{x}_i$ es generada por la variación del Lagrangiano (\ref{nhaction}) en una combinación lineal de $v_i$ y $x_i$, pero no en $v_i$ por si sólo. Si nosotros tratamos de substituir su ecuación en (\ref{nhaction}), como consecuencia obtenemos un Lagrangiano con altas derivadas en $x_i$, que produce ecuaciones para $x_i$ no equivalentes a (\ref{vdotx}) \cite{HeTe}. Como veremos más adelante, del análisis de vínculos, aparecen coordenadas no conmutativas, lo cual implica trabajar en un espacio configuracional extendido, análogo al espacio de fases en este caso.

El sistema luce como un oscilador isótropo en ambos casos $\rho^2<1$ y $\rho^2>1$, sin embargo, el último caso está caracterizado por algunas propiedades no usuales.

Para los valores críticos $\rho=\varepsilon$, $\varepsilon=\pm 1$, los cuales separan las fases subcrítica, $\rho^2<1$, y supercrítica, $\rho^2>1$, las ecuaciones (\ref{eom1'}), (\ref{eom2'}) se reducen a sólo una ecuación vectorial
\begin{equation}\label{eqrhocrit}
    (\dot{x}_i-v_i)+{\varepsilon}\,
    R\epsilon_{ij}(\dot{v}_j+R^{-2}x_j)=0.
\end{equation}
Esto refleja la aparición de una simetría de gauge para el caso crítico $\rho^2=1$,
\begin{equation}\label{gauge}
    \delta_\sigma x_i=\sigma_i(t),\quad
    \delta_\sigma v_i=
    {\varepsilon}R^{-1}\epsilon_{ij}\sigma_j(t).
\end{equation}
La formulación chiral es muy cómoda para la discusión del caso crítico, como ya lo veremos en la sección siguiente.

Para $\rho^2\neq 1$, la transformación de los parámetros
\begin{equation}\label{rhosym}
    \rho\rightarrow \rho^{-1}
\end{equation}
provoca un intercambio de las ecuaciones (\ref{eom1'}) y
(\ref{eom2'}). Más adelante veremos que esto es debido a una simetría de dualidad entre fases sub y supercrítica, la cual será más transparente de establecer en términos de la formulación chiral.

Para las simetrías de Newton-Hooke dedicaremos una subsección completa, después de analizar los vínculos que se deducen del Lagrangiano.

\subsection{Análisis de vínculos}

El sistema dado por (\ref{nhaction}) tiene dos pares de vínculos primarios
\begin{eqnarray}\label{constNon1}
    \Pi_i &=& \pi_i+\frac{\kappa}{2}\epsilon_{ij}v_j\approx 0,\\
    V_i &=& p_i-mv_i+\frac{\kappa}{2R^2}\epsilon_{ij}x_j\approx
    0,\label{constNon2}
\end{eqnarray}
donde $p_i$ y $\pi_i$ son los momentos canónicos conjugados a
$x^i$ y $v^i$. Los vínculos (\ref{constNon1}) satisfacen las relaciones
\begin{equation}\label{VPialg}
    \{\Pi_i,\Pi_j\}=\kappa\epsilon_{ij},\qquad
    \{V_i,V_j\}=\frac{\kappa}{R^2}\epsilon_{ij},\qquad
    \{\Pi_i,V_j\}=m\delta_{ij}.
\end{equation}
De donde se puede ver que el determinante de la matriz de los paréntesis de Poisson de los vínculos
$A_{ab}=\{\phi_a,\phi_b\}$, $\phi_a=(\Pi_i,V_i)$,
es
\begin{equation}\label{detA}
    \det A= m^4(1-\rho^2)^2.
\end{equation}
En el caso no crítico la matriz es no degenerada, y (\ref{constNon1}), (\ref{constNon2}) forman un conjunto de vínculos de segunda clase.

Mediante el conteo de los grados de libertad podemos ver que efectivamente estamos describiendo una partícula planar. Originalmente tenemos $2 \times 4=8$ variables de espacio de fase dadas por las $x_i, v_i$ y sus momentos canónicos conjugados $p_i$ y $\pi_i$, pero restando los cuatro vínculos de segunda clase (\ref{constNon1},\ref{constNon2}) nos quedamos con 4 grados de libertad que corresponden a los de una partícula planar.

Para $\rho^2=1$ la matriz se degenera, con lo cual aparecen vínculos de primera clase que analizaremos separadamente.

El Hamiltoniano canónico está dado por
\begin{equation}\label{HcanNC}
    H_{can}=\frac{m}{2}\left(v_i^2+\frac{1}{R^2}x_i^2\right)+\frac{\kappa}{R^2}\epsilon_{ij}
    x_iv_j.
\end{equation}
sumándole una combinación lineal de los vínculos primarios y aplicando el algoritmo de Dirac para el análisis de vínculos, para el caso no crítico llegamos a el Hamiltoniano total
\begin{equation}\label{HtotNC}
    H=p_iv_i-\frac{\pi_i x_i}{R^2}+\frac{m}{2}\left(
    \frac{x_i^2}{R^2}-v_i^2\right),
\end{equation}
\begin{equation}\label{NCconsistency}
    \{H,\Pi_i\}=V_i,\qquad
    \{H,V_i\}=-\frac{1}{R^2}\Pi_i.
\end{equation}

Las siguientes combinaciones lineales de las coordenadas del espacio de fase,
\begin{equation}\label{obserNC}
    {\cal P}_i=p_i-\frac{\kappa}{2R^2}\epsilon_{ij}x_j,\qquad
    {\cal X}_i=mx_i-\pi_i+\frac{1}{2}\kappa\epsilon_{ij}v_j,
\end{equation}
conmutan en el sentido de los paréntesis de Poisson con los vínculos (\ref{constNon1}) y (\ref{constNon2}), con lo cual son observables y satisfacen las relaciones
\begin{equation}\label{PPKK}
    \{{\cal P}_i,{\cal P}_j\}=-\frac{\kappa}{R^2}\epsilon_{ij},\qquad
    \{{\cal X}_i,{\cal P}_j\}=m\delta_{ij},\qquad
    \{{\cal X}_i,{\cal X}_j\}=-\kappa\epsilon_{ij}.
\end{equation}

Para $\kappa\neq 0$ los vínculos (\ref{constNon1}) forman un subconjunto de segunda clase. La reducción a la superficie que definen nos permite expresar $\pi_i$ en términos de $v_i$, resultando en los siguientes paréntesis no triviales
\begin{equation}\label{xpvv}
    \{x_i,p_j\}=\delta_{ij},\qquad
    \{v_i,v_j\}=-\frac{1}{\kappa}\epsilon_{ij}.
\end{equation}
En términos de estos paréntesis,
\begin{equation}\label{VV}
    \{V_i,V_j\}=-\frac{m^2}{\kappa}(1-\rho^2)\epsilon_{ij}.
\end{equation}
En correspondencia con (\ref{detA}), para $\rho^2\neq 1$
los vínculos (\ref{constNon2}) son de segunda clase. Para $\rho^2=1$ se vuelven de primera clase, y consecuentemente la dimensión del subespacio físico es menor en dos en comparación con el caso no crítico. Para $\rho^2\neq 1$, podemos reducirnos también a la superficie dada por (\ref{constNon2}), excluyendo $v_i$, obtenemos para $x_i$ y $p_i$
\begin{equation}\label{xxpx}
    \{x_i,x_j\}=\frac{\kappa}{m^2}\,\frac{1}{1-\rho^2}\epsilon_{ij},\qquad
    \{x_i,p_j\}=\frac{1-\frac{1}{2}\rho^2}{1-\rho^2}\delta_{ij},\qquad
    \{p_i,p_j\}=\frac{m^2}{4\kappa}\,\frac{\rho^4}{1-\rho^2}\epsilon_{ij}.
\end{equation}
La dinámica es generada por el Hamiltoniano en espacio de fase reducido
\begin{equation}\label{Hredxp}
    H^{*}=\frac{1}{2m}\left(p_i-\frac{\kappa}{2R^2}\epsilon_{ij}x_j\right)^2+
    \frac{m}{2R^2}(1-\rho^2)x_i^2
\end{equation}
que corresponde a la restricción del Hamiltoniano total (\ref{HtotNC}), a la superficie dada por los vínculos (\ref{constNon1}), (\ref{constNon2}).

De la forma explícita del Hamiltoniano (\ref{Hredxp}) es claro que las fases subcrítica, $\rho^2<1$, y supercrítica, $\rho^2>1$, tienen propiedades esencialmente diferentes: en el primer caso tiene el mismo signo que la masa $m$, mientras que el segundo puede tomar valores de ambos signos.

Notar que la estructura simpléctica del espacio de fase reducido (\ref{xxpx}) tiene una forma similar a la del problema de Landau en el plano no conmutativo \cite{Duval:2000xr,Duval:2001hu,Horvathy:2004fw}, ahondaremos en este tema en el siguiente capítulo.
Podemos ver que en el límite plano $R\rightarrow \infty$, la estructura simpléctica y el Hamiltoniano (\ref{Hredxp}) toman la forma de las para la partícula libre en el plano no conmutativo, la cual está descrita por la simetría de Galileo exótica \cite{PHMP}.

\subsection{Simetría de Newton-Hooke exótica}\label{classicalNH}

El Lagrangiano (\ref{nhaction}) es invariante bajo rotaciones y traslaciones en el tiempo,
\begin{eqnarray}\label{rotsym}
    &x'_i=x_i\cos \varphi +\epsilon_{ij}x_j\sin\varphi, \qquad
    v'_i=v_i\cos \varphi +\epsilon_{ij}v_j\sin\varphi,&\\
    &t'=t-\gamma.&\label{timetr}
\end{eqnarray}
mientras que es sólo cuasi invariante bajo las traslaciones y boosts de Newton-Hooke,
\begin{eqnarray}
&x_i^{\prime }=x_i+\alpha_i\cos R^{-1}t,\qquad v_i^{\prime
}=v_i-\alpha_iR^{-1}\sin
R^{-1}t,&  \label{trbo1} \\
&x_i^{\prime }=x_i+\beta_iR\sin R^{-1}t,\qquad v_i^{\prime
}=v_i+\beta_i\cos R^{-1}t,&  \label{trbo2}
\end{eqnarray}
podemos ver que ambas transformaciones están relacionadas haciendo una traslación en el tiempo, (\ref{trbo1}) se transforma en (\ref{trbo2}) haciendo el corrimiento $t\rightarrow t-\frac{\pi}{2}R$, acompañado con un cambio en los parámetros de transformación $\alpha_i\rightarrow R\beta_i$.

Estas transformaciones están generadas por los siguientes campos vectoriales
\begin{equation}
    {\cal X}_{P_i}=\cos
    R^{-1}t\frac{\partial}{\partial x_i}
    -R^{-1}\sin R^{-1}t \frac{\partial}{\partial v_i},\qquad
    {\cal X}_{K_i} =  R \sin R^{-1}t \frac{\partial}{\partial x_i}+\cos
    R^{-1}t\frac{\partial}{\partial v_i},\label{gen1}
\end{equation}
\begin{equation}
    {\cal X}_J=\epsilon_{ij}\left(x_j\frac{\partial}{\partial
    x_i}+v_j\frac{\partial}{\partial v_i}\right),\qquad
     {\cal X}_H = -\frac{\partial}{\partial t}.\label{gen2}
\end{equation}
Estos campos forman la álgebra de Newton-Hooke
\begin{equation}\label{alg01}
    [{\cal X}_H, {\cal X}_{K_i}]=-{\cal X}_{P_i},\qquad
    [{\cal X}_H,{\cal X}_{P_i}]=+\frac{1}{R^2}{\cal X}_{K_i},\qquad
    [{\cal X}_{K_i},{\cal X}_{K_j}]=0,\qquad
    [{\cal X}_{P_i},{\cal X}_{P_j}]=0,
\end{equation}
\begin{equation}\label{alg02}
    [{\cal X}_{K_i},{\cal X}_{P_j}]=0,\qquad
    [{\cal X}_J,{\cal X}_{P_i}]=\epsilon_{ij}{\cal X}_{P_j},\qquad
    [{\cal X}_J,{\cal X}_{K_i}]=\epsilon_{ij}{\cal X}_{K_j}.
\end{equation}
Se puede ver que no aparecen cargas centrales en esta realización, y coincide (hasta el factor $i$) con la álgebra de la sección \ref{ENH algebra}, pero sin las cargas centrales.

Usando el teorema de Noether obtenemos los generadores de traslaciones y boosts de Newton-Hooke, los cuales aparecen dados en términos de las variables observables (\ref{obserNC}),
\begin{equation}\label{PKNCobs}
    P_i={\cal P}_i\cos\frac{t}{R}+\frac{1}{R}{\cal X}_i\sin\frac{t}{R},\qquad
    K_i={\cal X}_i\cos\frac{t}{R}-R{\cal P}_i\sin\frac{t}{R},
\end{equation}
mientras que el momento angular es
\begin{equation}\label{Jncir}
    J=\epsilon_{ij}(x_ip_j+v_i\pi_j),
\end{equation}
y junto con el Hamiltoniano total (\ref{HtotNC}) generan la algebra,
\begin{equation}\label{alg0}
\{H,J\}=0, \qquad \{J,P_i\}=\epsilon_{ij}P_j,\qquad \{J,K_i\}=\epsilon_{ij}K_j,
\end{equation}
\begin{equation}  \label{alg1}
\{H, K_i\}=-P_i,\qquad \{H,P_i\}=\frac{1}{R^2}K_i,
\end{equation}
\begin{equation}\label{alg4}
    \{K_i,P_j\}=m\delta_{ij},\qquad
    \{K_i,K_j\}=-\kappa\epsilon_{ij}, \qquad \{P_i,P_j\}=-\frac{\kappa}{R^2}
    \epsilon_{ij}.
\end{equation}
Este es el análogo clásico de la álgebra $NH_-$ exótica (\ref{NH3ex})--(\ref{PP}), donde los parámetros $m$ y $\kappa$ juegan el rol de las cargas centrales $Z$ y $\tilde{Z}$, respectivamente.
Haciendo uso de la forma explícita de las integrales $H$, $J$, ${ P}_i$
y ${ K}_i$, se determina que sobre la superficie de vínculos los Casimires (\ref{CasNH1}) y (\ref{CasNH2}) toman valor cero. Como resultado, en la fase no crítica, el Hamiltoniano y el momento angular están representados en términos de las traslaciones y boosts por
\begin{eqnarray}\label{HKPCas}
    H&=&\frac{1}{m(1-\rho^2)}\left(\frac{1}{2}\left(P_i^2+R^{-2}K_i^2\right)-\rho R^{-1}\epsilon_{ij}K_iP_j\right),\\
    J&=&\frac{1}{m(1-\rho^2)}\left(\epsilon_{ij}K_iP_j-\frac{1}{2}\rho R\left(P_i^2+R^{-2}K_i^2\right)\right).\label{JKPCas}
\end{eqnarray}

En espacio de fase reducido con coordenadas $x_i,$ $p_i$, las integrales $P_i$, $K_i$ y $J$ toman la forma
    \begin{equation}  \label{Pi}
    P_i={\cal P}_i\cos\frac{t}{R} +\frac{1}{R}\tilde{\cal X}_i\sin\frac{t}{R},\qquad
    K_i= \tilde{\cal X}_i\cos\frac{t}{R}
    -R{\cal P}_i\sin\frac{t}{R},
\end{equation}
\begin{equation}\label{Jred}
    J=\frac{m^2}{2\kappa}\left(\left(x_i+\frac{\kappa}{m^2}\epsilon_{ij}p_j\right)^2
    -(1-\rho^2)x_i^2\right),
\end{equation}
donde
\begin{equation}\label{tilKi}
     \tilde{\cal X}_i=mx_i\left(1-\frac{1}{2}\rho^2\right)+
    \frac{\kappa}{m}\epsilon_{ij}p_j.
\end{equation}
Junto con el Hamiltoniano (\ref{Hredxp}), ellos forman la misma álgebra clásica (\ref{alg0})--(\ref{alg4}) con respecto de la estructura simpléctica reducida (\ref{xxpx}).

En los casos críticos $\rho=1$ ó $\rho=-1$, la simetría del sistema se reduce a $NH_-$ extendido en una dimensión menor. Aquí, por un lado el Hamiltoniano y el momento angular, y por otro las traslaciones y boosts son linealmente dependientes,
\begin{eqnarray}\label{CritHJKP}
    \rho=1&:&\quad H=R^{-1}J = \frac{1}{2m}P_i^2,\qquad P_i=-R^{-1}\epsilon_{ij}K_j,\\
    \rho=-1&:&\quad H=-R^{-1}J = \frac{1}{2m}P_i^2,\qquad
    P_i=R^{-1}\epsilon_{ij}K_j.\label{CritHJKP1}
\end{eqnarray}
Notar que el signo de $H$ queda ligado al signo del parámetro de masa $m$.

\section{Lagrangiano: una formulación chiral}\label{classicalLag chiral}

Asociadas a los generadores chirales (\ref{NHchiral}) existen también unas coordenadas chirales. Estas son convenientes para el análisis de algunos aspectos clásicos y cuánticos del sistema, ya que son análogas de los modos normales del sistema.

Es posible obtener el Lagrangiano chiral mediante un simple cambio de variables y parámetros
\begin{eqnarray}\label{Xpm}
    X^\pm_i&=&x_i\pm
    R\epsilon_{ij}v_j,\label{Xpm*}\\
    \mu_\pm&=&\frac{1}{2R^2}(mR\pm
    \kappa),\label{Mpm}
\end{eqnarray}
con lo cual (\ref{nhaction}) toma la forma
\begin{equation}\label{Lchir}
    {\cal L}_{ch}={\cal L}_++{\cal L}_-=-\frac{1}{2}
    \mu_+\left(\epsilon_{ij}\dot{X}^+_iX^+_j+\frac{1}{R}{X^+_i}^2\right)
    -\frac{1}{2}
    \mu_-\left(-\epsilon_{ij}\dot{X}^-_iX^-_j+\frac{1}{R}{X^-_i}^2\right).
\end{equation}
que coincide con el Lagrangiano en formulación covariante, ${\cal L}_{nc}$, hasta una derivada total,
\begin{equation}\label{Nc-Ch}
    {\cal L}_{nc}={\cal L}_{ch}+\frac{d}{dt}\Delta {\cal L},\qquad
    \Delta{\cal
    L}=\frac{m}{2}x_iv_i=\frac{m}{4R}\epsilon_{ij}X^+_iX^-_j,
\end{equation}
el cual se ha omitido cuando hemos pasado de (\ref{nhaction}) a
(\ref{Lchir}), pero es importante en la teoría cuántica.

EL Lagrangiano (\ref{Lchir}) también puede ser obtenido por el método de las realizaciones no lineales a partir de la forma chiral de la álgebra exótica de $NH_-$ (\ref{NHchiral}). En este caso parametrizamos el coset por
\begin{equation}\label{cosetchir}
    g=g^+g^-,
\end{equation}
con
\begin{eqnarray}\label{g+-}
    g^+ &=& e^{-ix^0R^{-1}{\cal J}^+}e^{-iR^{-1}X^-_j{\cal J}^+_j}e^{ic^+Z^+}, \nn\\
    g^- &=& e^{ix^0R^{-1}{\cal
    J}^-}e^{iR^{-1}X^+_j{\cal J}^-_j}e^{ic^-Z^-}.
\end{eqnarray}
Notar que sólo hay una variable de evolución en el tiempo $x^0$ para los dos sectores chirales. La diferencia de signo en los exponentes con $x^0$ se origina de la relación $H=({\cal J}^+-{\cal J}^-)/R$.

\subsection{Dinámica clásica: formulación chiral}\label{classicaldyn chiral}

Para $\rho^2\neq 1$ las ecuaciones de movimiento generadas por el Lagrangiano (\ref{Lchir}) son
\begin{equation}\label{X+-Eq}
    \dot{X}^\pm_i\pm \frac{1}{R}\epsilon_{ij}X^\pm_j=0.
\end{equation}
Entonces vemos que las coordenadas $X^+_i$ y $X^-_i$ tienen sentido de modos normales que realizan rotaciones de 'derecha' e 'izquierda' de la misma frecuencia $R^{-1}$,
\begin{equation}\label{X+-motion}
    X^\pm_i(t)=X^\pm_i(0)\cos R^{-1}t\mp \epsilon_{ij}X^\pm_j(0)\sin
    R^{-1}t.
\end{equation}
donde $X^\pm_i(0)$ son las condiciones iniciales.
La evolución de las $x_i$ y $v_i$ es una superposición lineal de esas dos rotaciones,
$$
x_i=\frac{1}{2}(X^+_i+X^-_i),\qquad
v_i=\frac{1}{2R}\epsilon_{ij}(X^-_j-X^+_j).
$$

De (\ref{Mpm}) podemos ver que los casos críticos $\rho=1$ ($\rho=-1$) corresponden a $\mu_-=0$ ($\mu_+=0$), con lo cual las variables $X^-_i$ ($X^+_i$) desaparecen completamente de (\ref{Lchir}), transformándose en variables de gauge puras, mientras tanto $X^+_i$ ($X^-_i$) siguen la misma dinámica dada por (\ref{eqrhocrit}).

Las ecuaciones (\ref{X+-Eq}) obtenidas de la variación de (\ref{Lchir}) in
$X^\pm_i$, pueden ser reescritas de la forma $X^\pm_i\mp R\epsilon_{ij}\dot
X^\pm_j=0$, luego la coordenada $X^\pm_i$ puede ser eliminada en términos de la velocidad de $X^\pm_j, j\neq i$. Entonces cualquier componente (fija) de los vectores $X^+_i$ y $X^-_i$ puede ser tratada como variable auxiliar.
Si seleccionamos $i=2$ para ambos modos chirales, y substituimos
\begin{equation}
X^\pm_2=\mp R\dot X^\pm_1 \label{X2X1dot}
\end{equation}
en el Lagrangiano, obtenemos la forma equivalente (hasta una derivada total)
\begin{equation}\label{Lshift}
    {\cal L}=
    \frac{1}{2}\mu_+R\left((\dot{X}^+_1)^2 -\frac{1}{R^2}X^{+2}_1\right)
    +\frac{1}{2}\mu_-R
    \left((\dot{X}^-_1)^2-\frac{1}{R^2}X^{-2}_1\right).
\end{equation}
Acá podemos ver explícitamente que el Lagrangiano es la suma de dos osciladores armónicos de la misma frecuencia, pero esconde la simetría de rotaciones 2D del sistema.

En el caso no crítico la transformación (\ref{rhosym}), la cual asume el cambio $mR\leftrightarrow \kappa$, produce
\begin{equation}\label{M+M-}
    \mu_\pm\rightarrow \pm \mu_\pm,
\end{equation}
luego el Lagrangiano (\ref{Lchir}) es invariante bajo la transformación (\ref{M+M-}) si es acompañada por la transformación compleja
\begin{equation}\label{X+X-trans}
    X^+_i\rightarrow X^+_i,\qquad X^-_i\rightarrow iX^-_i.
\end{equation}
La aparición del factor imaginario $i$ en la transformación (\ref{X+X-trans}) está relacionado con algunas propiedades no usuales de la fase supercrítica $\rho^2>1$ que luego comentaremos. Sin embargo, es útil mencionar que la transformación discreta (\ref{M+M-},\ref{X+X-trans}) no es una simetría usual del sistema, dado que involucra también una transformación de parámetros. De hecho la transformación (\ref{X+X-trans}) corresponde a una familia de transformaciones de similitud $X^+_i\rightarrow \gamma^+ X^+_i$,
$X^-_i\rightarrow \gamma^- X^-_i$, donde $\gamma^\pm$ son parámetros numéricos, las cuales dejan invariante las ecuaciones de movimiento pero cambian el Lagrangiano. Tales transformaciones no pueden ser promovidas a simetrías a nivel cuántico \cite{Gozzi}.

\subsection{Análisis de vínculos, formulación chiral}

En la base chiral, del Lagrangiano (\ref{Lchir}), aparecen los vínculos
\begin{eqnarray}\label{chi+}
    &\chi^+_i = P^+_i+\frac{1}{2}\mu_+\epsilon_{ij}X^+_j\approx
    0,&\\
    &\chi^-_i = P^-_i-\frac{1}{2}\mu_-\epsilon_{ij}X^-_j\approx
    0.&\label{chi-}
\end{eqnarray}
Aquí $P^\pm_i$ son los momentos canónicos conjugados a $X^\pm_i$,
\begin{equation}\label{Pppi}
    P^\pm_i=\frac{1}{2}\left(p^{ch}_i\pm
    \frac{1}{R}\epsilon_{ij}\pi^{ch}_j\right).
\end{equation}

Los momentos $p^{ch}_i$ y $\pi^{ch}_i$ canónicamente conjugados a $x_i$ y $v_i$ están relacionados a los momentos correspondientes que aparecen del Lagrangiano en formulación convariante (\ref{nhaction}), mediante la transformación canónica
\begin{equation}\label{chi-nochi}
    p_i^{ch}=p_i-\frac{m}{2}v_i,\qquad
    \pi^{ch}_i=\pi_i-\frac{m}{2}x_i,
\end{equation}
la cual está asociada a la derivada total (\ref{Nc-Ch}).

Los paréntesis no triviales entre vínculos son
\begin{eqnarray}\label{mu+}
    \{\chi^+_i,\chi^+_j\}&=&\mu_+\epsilon_{ij},\\
    \{\chi^-_i,\chi^-_j\}&=&-\mu_-\epsilon_{ij}.\label{mu-}
\end{eqnarray}
En el caso no crítico $\rho^2\neq 1$, forman dos conjuntos de vínculos de segunda clase. Cuando $\mu_+=0$ ($\mu_-=0$), los vínculos $\chi^+_i$ ($\chi^-_i$) cambian su naturaleza de segunda a primera clase, tomando la forma $\chi^+_i=P^+_i\approx 0$ ($\chi^-_i=P^-_i\approx 0$). En este caso los grados de libertad correspondientes al modo `$+$' (`$-$') son puro gauge, que es consecuencia de la desaparición de la parte correspondiente ${\cal L}_+$
(${\cal L}_-$) del Lagrangiano (\ref{Lchir}).

El Hamiltoniano canónico correspondiente al Lagrangiano chiral (\ref{Lchir}) es
\begin{equation}\label{H+-}
    H_{can}=\frac{1}{2R}\left(\mu_+{X^+_i}^2+\mu_-{X^-_i}^2\right).
\end{equation}
Definimos el Hamiltoniano total $H=H_{can}+u^+_i\chi^+_i+u^-_i\chi^-_i$. Como consecuencia de la conservación de los vínculos se fijan los multiplicadores, $u^\pm_i=\mp \frac{1}{R}\epsilon_{ij}X^\pm_j$, y queda
\begin{equation}\label{Htotchir}
    H=\frac{1}{R}\epsilon_{ij}(X^+_iP^+_j-X^-_iP^-_j).
\end{equation}
Este Hamiltoniano reproduce las ecuaciones de movimiento Lagrangianas para las coordenadas chirales (\ref{X+-Eq}).

Las variables independientes que conmutan con los vínculos (que serán observables) son
\begin{equation}\label{lambdai}
    \lambda^+_i= P^+_i-\frac{1}{2}\mu_+\epsilon_{ij}X^+_j,\qquad
    \lambda^-_i = P^-_i+\frac{1}{2}\mu_-\epsilon_{ij}X^-_j,
\end{equation}
\begin{equation}\label{lamchi}
    \{\lambda^+_i,\chi^\pm_j\}=\{\lambda^-_i,\chi^\pm_j\}=0.
\end{equation}
Entre ellas poseen paréntesis similares a los de los vínculos,
\begin{equation}\label{lamlam}
    \{\lambda^+_i,\lambda^+_j\}=-\mu_+\epsilon_{ij},\qquad
    \{\lambda^-_i,\lambda^-_j\}=\mu_-\epsilon_{ij},\qquad
    \{\lambda^+_i,\lambda^-_j\}=0.
\end{equation}

Podemos ver que los generadores de traslaciones y boosts de Newton-Hooke están dados en términos de los observables (\ref{lambdai})
\begin{equation}\label{PKchiral}
    P_i=\lambda^+_j\Delta^+_{ji}(t)+\lambda^-_j\Delta^-_{ji}(t),\qquad
    K_i=R\left(\lambda^-_j\Delta^-_{jk}(t)-\lambda^+_j\Delta^+_{jk}(t)\right)
    \epsilon_{ki},
\end{equation}
con lo cual el conjunto de vínculos chirales es invariante bajo aquellas transformaciones
\begin{equation}\label{PKchi}
    \{P_i,\chi^\pm_j\}=\{K_i,\chi^\pm_j\}=0.
\end{equation}

La reducción a la superficie dada por (\ref{chi+}) (cuando $\mu_+\neq 0$) y/o (\ref{chi-}) (cuando $\mu_-\neq 0$) resulta en una exclusión de los momentos $P^+_i$ y/o $P^-_i$.  Los paréntesis de Dirac no triviales son
\begin{eqnarray}\label{XX++}
    \{X^+_i,X^+_j\} & =& -\frac{1}{\mu_+}\epsilon_{ij},\\
    \{X^-_i,X^-_j\} & =& \frac{1}{\mu_-}\epsilon_{ij},\label{XX--}
\end{eqnarray}
y el Hamiltoniano en espacio de fase reducido
\begin{equation}\label{H+-*}
    H^*=\frac{1}{2R}\left(\mu_+{X^+_i}^2+\mu_-{X^-_i}^2\right)
\end{equation}
coincide con la forma del Hamiltoniano canónico (\ref{H+-}). En correspondencia con la formulación covariante, para el caso subcrítico $\rho^2<1$, $\mu_\pm>0$ para $m>0$ y $\mu_\pm<0$ para $m<0$, y la energía toma valores del signo del parámetro $m$. En el caso supercrítico $\rho^2>1$ los vínculos $\mu_+$ y $\mu_-$ tienen signos opuestos, y la energía puede tomar cualquier signo.

\subsection{Simetría de Newton-Hooke exótico clásica: formulación chiral}\label{classicalNH chiral}

En correspondencia con (\ref{trbo1}), (\ref{trbo2}),  en la base chiral las traslaciones y boosts de Newton-Hooke tienen la forma
\begin{equation}\label{NHchiral*}
    X^\pm{}'_i=X^\pm_i+\alpha^\pm_i(t),
\end{equation}
donde
\begin{equation}\label{albet+-}
    \alpha^\pm_i(t)=\Delta^\pm_{ij}(t)\alpha^\pm_j(0)=\Delta^\pm_{ij}(t)(\alpha_j \pm R\epsilon_{jk}\beta_k),
\end{equation}
\begin{equation}\label{Del+-}
    \Delta^\pm_{ij}(t)=\delta_{ij}\cos
    \frac{t}{R}\mp\epsilon_{ij}\sin \frac{t}{R}.
\end{equation}
Las $\Delta^\pm_{ij}(t)$ son matrices de rotación, $\Delta_{ij}(t)\in
SO(2)$.
Los generadores de las transformaciones (\ref{NHchiral*}) corresponden a los generadores chirales\footnote{Hasta un coeficiente multiplicativo que es irrelevante, puesto que va en correlación con los parámetros de transformación $\alpha^\pm_i(0)$.} (\ref{NHchNch}), podemos escribirlos en términos de las variables $\lambda^\pm_i$,
 \begin{equation}\label{calJlam}
    {\cal J}^\pm_i=\mp R\lambda^\pm_j\Delta^\pm_{ji}(t).
\end{equation}
Las ecuaciones
\begin{equation}\label{partJ}
    \frac{\partial {\cal J}^\pm_i}{\partial
    t}=\pm\frac{1}{R}\epsilon_{ij}{\cal J}^\pm_j
\end{equation}
garantizan que ellos son integrales de movimiento, $\frac{d}{dt}{\cal
J}^\pm_i=\frac{\partial {\cal J}^\pm_i}{\partial t}+\{{\cal
J}^\pm_i,H\}=0$. En espacio de fase reducido toman la forma
\begin{equation}\label{JXpm}
    {\cal J}^\pm_i=R\mu_\pm\epsilon_{ij}X_j^\pm (0).
\end{equation}

El momento angular tiene la forma
\begin{equation}\label{Jchir}
    J=\epsilon_{ij}(X^+_iP^+_j+X^-_iP^-_j).
\end{equation}
Notar la estructura similar que tiene con el Hamiltoniano (\ref{Htotchir}). Las relaciones (\ref{PKchi}) y $\{J,\chi^\pm_i\}=\epsilon_{ij}\chi^\pm_j$,
$\{H,\chi^\pm_i\}=\pm\frac{1}{R}\epsilon_{ij}\chi^\pm_j$ muestran explícitamente la invariancia del subespacio físico dado por los vínculos (\ref{chi+}), (\ref{chi-}) bajo las transformaciones de Newton-Hooke.

En espacio de fase reducido, con las variables $X^+_i$ y $X^-_i$ (para $\rho^2\neq 1$), y estructura simpléctica (\ref{XX++},\ref{XX--}), el momento angular toma la forma
\begin{equation}\label{Jchiral}
    J=\frac{1}{2}\left(\mu_+{X^+_i}^2-\mu_-{X^-_i}^2\right),
\end{equation}
comparar con (\ref{H+-}).

Junto a ${\cal J}^\pm=\frac{1}{2}(J\pm RH)$, donde $H$ corresponde al Hamiltoniano total, las integrales (\ref{JXpm}) generan el análogo clásico de la álgebra exótica de Newton-Hooke en su forma chiral (\ref{NHchiral})
\begin{equation}\label{NHchiral}
    \{{\cal J}^\pm_i,{\cal J}^\pm_j\}=\epsilon_{ij}Z^\pm,\qquad
    \{{\cal J}^\pm,{\cal J}^\pm_i\}=\epsilon_{ij}{\cal J}^\pm_j,\qquad,
\end{equation}
donde
\begin{equation}\label{ZZmu}
    Z^+=-R^2\mu_+,\qquad
    Z^-=R^2\mu_-.
\end{equation}

En el caso crítico $\mu_-=0$ o $\mu_+=0$, la simetría se reduce a $NH_-$ centralmente extendido, generado por ${\cal J}^+$, ${\cal
J}^+_i$ y $Z^+$, o por ${\cal J}^-$, ${\cal J}^-_i$ y $Z^-$. En correspondencia con (\ref{CritHJKP},\ref{CritHJKP1}) tenemos
\begin{eqnarray}\label{Crit chiral}
    \mu_-=0&:&\quad {\cal J}^-=0, \qquad {\cal J}^-_i=0,\\
    \mu_+=0&:&\quad {\cal J}^+=0, \qquad {\cal J}^+_i=0.\label{Crit chiral 1}
\end{eqnarray}

\subsection{Simetría SO(3) y SO(2,1) adicional}\label{addsymduality0}

Mostraremos ahora, que el sistema posee una simetría adicional, la cual es explícita en la formulación chiral y, como veremos, su naturaleza depende de la fase.

Por comodidad definiremos las coordenadas adimensionales
\begin{equation}\label{YYY}
\sqrt{|\mu^+|}X^+_1=Y_1,\quad \sqrt{|\mu^+|}X^+_2=Y_2,\quad
\sqrt{|\mu^-|}X^-_1=Y_3,\quad \sqrt{|\mu^-|}X^-_2=Y_4.
\end{equation}
En términos de estas coordenadas el Lagrangiano chiral toma la forma
\begin{equation}\label{Ly}
    {\cal L}_{ch}=-\frac{1}{2}
\left(\dot{Y}^T\W Y+\frac{1}{R}{Y}^T\h Y\right)=-\frac{1}{2}
\left(\dot{Y}^A\W_{AB} Y^B+\frac{1}{R}{Y}^A\h_{AB} Y^B\right),
\end{equation}
donde
\begin{equation}\label{Ly*}
    \eta_{AB}=
\left(
\begin{array}{cccc}
  \varepsilon_+& . & . & . \\
  . & \varepsilon_+ & . & . \\
  . & . & \varepsilon_- & . \\
  . & . & . & \varepsilon_- \\
\end{array}%
\right),\qquad \W_{AB}=
\left(%
\begin{array}{cccc}
  .& \varepsilon_+ & . & . \\
  -\varepsilon_+ & . & . & . \\
  . & . & . & -\varepsilon_- \\
  . & . & \varepsilon_- & . \\
\end{array}%
\right),
\end{equation}
$\varepsilon_\pm$ son los signos de $\mu_\pm$, y las matrices $\eta$ y $\W$ satisfacen las relaciones $\h^2=1,$ $\W^T=-\W,$ y $\W\W=-1$.
El término de potencial es invariante bajo las transformaciones $Y\rightarrow OY$, $O^T\eta O=\eta$, que son rotaciones SO(4) en el caso subcrítico caracterizado por $\varepsilon_+\varepsilon_-=+1$, y pseudo-rotaciones SO(2,2)$\sim$AdS$_3$ en el sector supercrítico $\varepsilon_+\varepsilon_-=-1$. Por otro lado, el término cinético es invariante bajo transformaciones Sp(4), con $\W$ jugando el rol de matriz simpléctica, $Y\rightarrow CY$, $C^T\W C=\W$. Entonces la simetría del Lagrangiano corresponde a la intersección de SO(4) (o SO(2,2)) y Sp(4). Considerando una transformación infinitesimal $\delta Y=\omega Y$, encontramos que la matriz constante $\omega$ tiene que satisfacer las ecuaciones $\omega^T\eta+\eta\omega=0$, $\omega^T\Omega+\Omega\omega=0$. Un análisis simple de esas ecuaciones, con una subsecuente aplicación del teorema de Noether, resulta finalmente en un conjunto de integrales dadas por $H$, $J$ y
\begin{equation}
I_1=\frac 12\varepsilon _{+}\sqrt{\left\vert \frac{\mu _{-}}{\mu
_{+}}\right\vert
}\left( P_{2}^{+}X_{2}^{-}-P_{1}^{+}X_{1}^{-}\right) +\frac 12\varepsilon _{-}\sqrt{%
\left\vert \frac{\mu _{+}}{\mu _{-}}\right\vert }\left(
X_{1}^{+}P_{1}^{-}-X_{2}^{+}P_{2}^{-}\right) ,  \label{a1}
\end{equation}
\begin{equation}
I_2=-\frac 12\varepsilon _{+}\sqrt{\left\vert \frac{\mu _{-}}{\mu _{+}}%
\right\vert }\left( P_{2}^{+}X_{1}^{-}+P_{1}^{+}X_{2}^{-}\right)
+\frac 12\varepsilon _{-}\sqrt{\left\vert \frac{\mu _{+}}{\mu _{-}}\right\vert }%
\left( X_{1}^{+}P_{2}^{-}+X_{2}^{+}P_{1}^{-}\right).  \label{a2}
\end{equation}
En espacio de fase reducido toman la forma
\begin{equation}\label{II12}
    I_1=\frac 12\sqrt{|\mu_+\mu_-|}\,(X^+_1X^-_2+X^+_2X^-_1),\qquad
    I_2=\frac 12\sqrt{|\mu_+\mu_-|}\,(X^+_2X^-_2-X^+_1X^-_1).
\end{equation}
En el caso subcrítico la integral $I_1$ genera rotaciones 2D en los planos $(Y_1,Y_3)$ y $(Y_2,Y_4)$, mientras $I_2$ en los planos $(Y_1,Y_4)$ y $(Y_2,Y_3)$. En el caso supercrítico las rotaciones son cambiadas por transformaciones de Lorentz en los mismos planos. Siendo independientes del tiempo, estas integrales conmutan con el Hamiltoniano,  $\{H,I_1\}=\{H,I_2\}=0$, y junto con el momento angular reescalado
\begin{equation}\label{I3J}
    I_3=\frac{1}{2}J
\end{equation}
generan la siguiente algebra de Lie
\begin{equation}\label{so3so21}
    \{I_1,I_2\}=\varepsilon_+\varepsilon_- I_3,\qquad
    \{I_3,I_1\}=I_2,\qquad
    \{I_2,I_3\}=I_1.
\end{equation}
En el caso subcrítico (\ref{so3so21}) es la álgebra de rotaciones $so(3)$, mientras que en el caso supercrítico es una álgebra de Lorentz $so(2,1)$.
Se puede ver que los paréntesis entre $I_{1,2}$ y las traslaciones y boosts de Newton-Hooke producen ciertas combinaciones de los últimos.


\subsection{Dualidad}\label{addsymduality}

\begin{figure}[h]
\begin{center}
  \includegraphics[width=.6\textwidth]{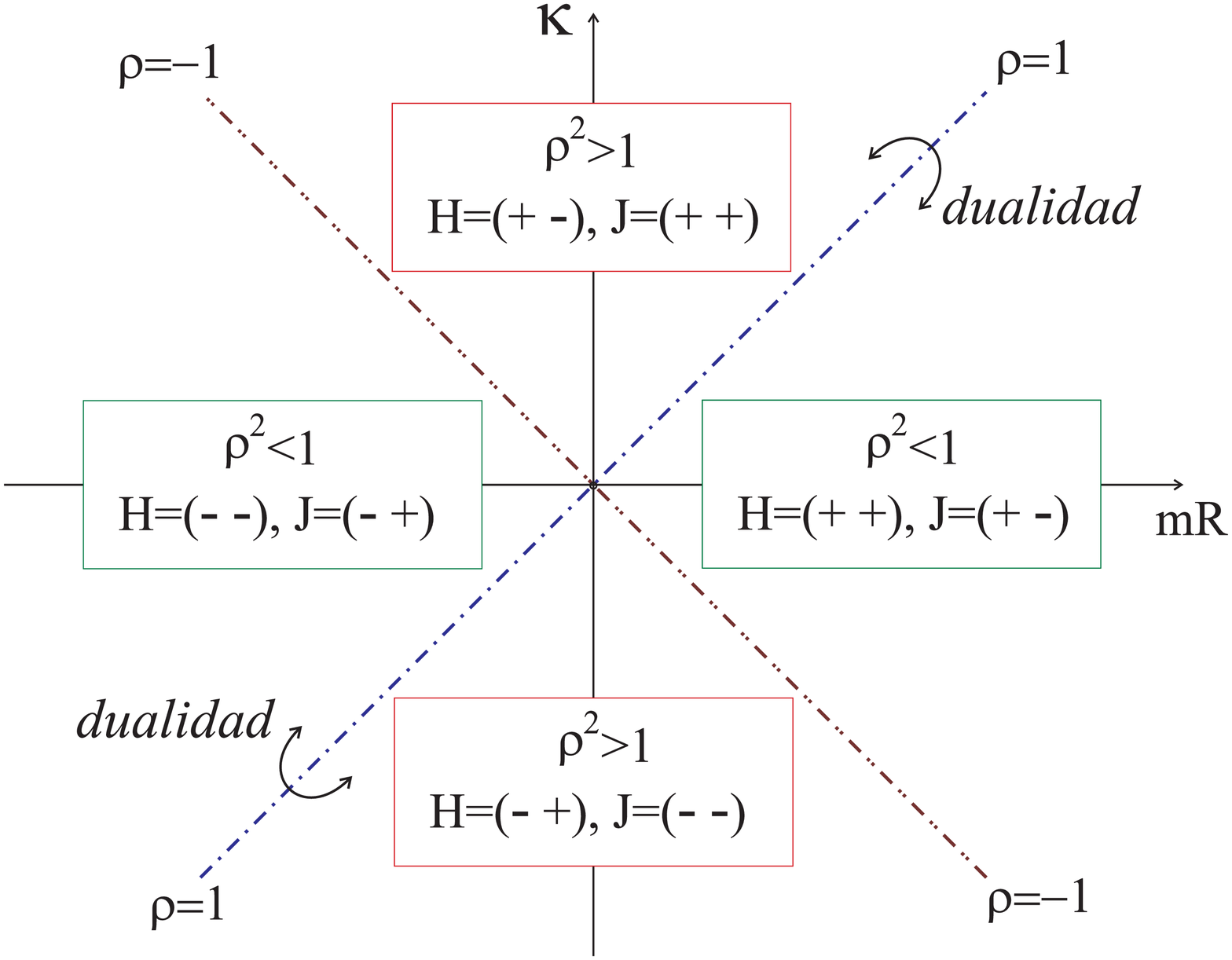}\\
  \caption{Fases y dualidad.}\label{duality_fig}
     \end{center}
\end{figure}

Volvamos a analizar la relación entre la fase sub y supercrítica, a la luz de la transformación de dualidad, la cual puede ser presentada de tres formas equivalentes
\begin{equation}\label{duality3}
    \rho\rightarrow \rho^{-1};\qquad  mR\leftrightarrow
    \kappa;\qquad \mu_\pm\rightarrow \pm\mu_\pm.
\end{equation}
Esta dualidad induce una transformación mutua entre las fases sub y supercrítica, y entre las dos fases críticas dadas por $\mu_+=0$, $\mu_-<0$ y $\mu_+=0$, $\mu_->0$, mientras que deja invariantes las fases críticas con $\mu_-=0$, ver figura \ref{duality_fig}


De acuerdo con (\ref{NHchNch},\ref{Mpm}), la dualidad (\ref{duality3}) induce las transformaciones
\begin{equation}\label{JHdual}
    J\leftrightarrow RH,
\end{equation}
\begin{eqnarray}\label{PBo1}
    P_i&\rightarrow& P_i'=\frac{1}{2}\left(
    (P_1+P_2)+(-1)^iR^{-1}(K_1+K_2)\right),\\
    K_i&\rightarrow& K_i'=\frac{1}{2}\left((-1)^{i+1}
    (K_1-K_2)+R(P_2-P_1)\right).\label{PBo4}
\end{eqnarray}
En términos de las integrales chirales (\ref{NHchNch}), estas transformaciones son equivalentes a
\begin{equation}\label{chirJJJ}
    {\cal J}^\pm\rightarrow \pm{\cal J}^\pm,\qquad
    {\cal J}^+_i\rightarrow {\cal J}^+_i,\qquad
    {\cal J}^-_1\rightarrow {\cal J}_2^-,\qquad
     {\cal J}^-_2\rightarrow {\cal J}_1^-.
\end{equation}
Recordando también (\ref{ZZmu}), tenemos
\begin{equation}\label{ZZZZ}
    Z^\pm\rightarrow \pm Z^\pm.
\end{equation}
Usando las relaciones (\ref{chirJJJ}), (\ref{ZZZZ}), concluimos que la transformación de dualidad no cambia la álgebra (\ref{NHchiral}) de las integrales de movimiento, entonces es un automorfismo de la álgebra de Newton-Hooke exótica.

Pero donde aparece una transformación no trivial, es en el sector de la simetría adicional, donde produce la transformación mutua $so(3) \leftrightarrow so(2,1)$, en acuerdo con (\ref{I+I+I3}). Como veremos en la siguiente sección, esto es acompañado por un cambio radical de la degeneración de los niveles de energía, de finita a infinita o viceversa.

\section{Cuantización en espacio de fase reducido}\label{reducedquant}

De acuerdo con las relaciones (\ref{XX++},\ref{XX--}), en el caso subcrítico $\rho^2<1$ con $\mu_+>0$, $\mu_->0$, uno define dos conjuntos independientes de operadores de creación y aniquilación
\begin{equation}
    a_{+}=\sqrt{\frac{|\mu_+|}{2}}\, (X^+_2+ i X^+_1),\qquad
    a_-=\sqrt{\frac{|\mu_-|}{2}}\, (X^-_1+ i X^-_2),\qquad
    a_\pm^\dagger=\left(a_\pm\right)^\dagger,
 \label{a+a-}
\end{equation}
\begin{equation}\label{aaaa}
[a_+,a^\dagger_+]=1,\qquad [a_-,a^\dagger_-]=1,\qquad
[a_+,a_-]=[a_+,a^\dagger_-]=0.
\end{equation}
Hemos puesto $\mu_\pm$ bajo signos de modulo teniendo en mente una futura generalización. Construiremos el Hamiltoniano y momento angular eligiendo el orden normal en (\ref{H+-*}) y (\ref{Jchiral}), obtenemos
\begin{equation}\label{HX+X-redsub}
    H=R^{-1}\left(a^\dagger_+a_++
    a^\dagger_-a_-+1\right),\qquad
    J=a^\dagger_+a_+ -
    a^\dagger_-a_-\,.
\end{equation}
El sistema cuántico es un oscilador armónico isótropo de frecuencia $R^{-1}$.

La energía toma valores positivos, $E_{n_+,n_-}=R^{-1}(n_++n_-+1)$, mientras que el momento angular puede tomar valores de ambos signos, $j_{n_+,n_-}=n_+-n_-$, donde $n_\pm=0,1,\ldots$ son los autovalores de los operadores de número $N_\pm=a_\pm^\dagger a_\pm$, $N_\pm|n_+,n_-\rangle=n_\pm |n_+,n_-\rangle$.

En la fase subcrítica $\rho^2<1$ con $\mu_+<0$, $\mu_-<0$, los operadores $a_+$ y $a_-$ son realizados como en (\ref{a+a-}) con la substitución $X^+_1\leftrightarrow X^+_2$, $X^-_1\leftrightarrow X^-_2$. Esto provoca cambios globales de signo en el Hamiltoniano y momento angular,
\begin{equation}\label{HX+X-redsub*}
    H=-R^{-1}\left(a^\dagger_+a_+ +
    a^\dagger_-a_-+1\right),\qquad
    J=-a^\dagger_+a_+ + a^\dagger_-a_-\,.
\end{equation}

En el caso supercrítico $\rho>1$ ($\mu_+>0$, $\mu_-<0$), el operador $a_+$ está definido como en (\ref{a+a-}) mientras que $ a_-$ es obtenido mediante la substitución $X^-_1\leftrightarrow X^-_2$. Entonces
\begin{equation}\label{Hrho>1}
    H=R^{-1}\left(a^\dagger_+a_+ -
    a^\dagger_-a_-\right),\qquad
    J=a^\dagger_+a_+ + a^\dagger_- a_- +1.
\end{equation}
Este es un oscilador exótico con el Hamiltoniano y momento angular intercambiados. La energía puede tomar valores positivos y negativos $E_{n_+,n_-}=R^{-1}(n_+-n_-)$, $n_\pm=0,1,\ldots$, y no es restringido de abajo, en cambio el momento angular puede tomar sólo valores positivos, $j_{n_+,n_-}=n_++n_-+1$, es decir, ambos modos son de `derecha'.

En el caso supercrítico $\rho<-1$ ($\mu_+<0$, $\mu_->0$), el rol de los modos es cambiado, la energía del modo $X^+$ es negativa, mientras que para $X^-$ es positiva. En este caso $a_-$ está definido como en (\ref{a+a-}), mientras que $ a_+$ está realizado mediante el cambio $X^+_1\leftrightarrow X^+_2$. Como resultado tenemos,
\begin{equation}\label{Hrho<-1}
    H=R^{-1}\left(-a^\dagger_+ a_+ +
    a^\dagger_-a_-\right),\qquad
    J=-\left(a^\dagger_+a_+ +
    a^\dagger_-a_-+1\right).
\end{equation}
Nuevamente tenemos un oscilador exótico, pero en este caso ambos modos resultan ser de `izquierda'.

Podemos escribir en forma condensada las expresiones para el Hamiltoniano y el momento angular para los cuatro casos,
\begin{equation}\label{HJuni}
    H=R^{-1}\left(\varepsilon_+a^\dagger_+a_+ +
    \varepsilon_-a^\dagger_- a_-+\frac{1}{2}(\varepsilon_++\varepsilon_-)\right),\qquad
    J=\varepsilon_+a^\dagger_+ a_+ -
    \varepsilon_-a^\dagger_-a_-+\frac{1}{2}(\varepsilon_+-\varepsilon_-),
\end{equation}
donde $\varepsilon _{\pm }=\mathrm{sign}\,\mu _{\pm }$.
Las fases sub- y supercríticas son esencialmente diferentes desde el punto de vista de la estructura de los niveles de la energía y el momento angular. De acuerdo con (\ref{HJuni}), los niveles de energía y momento angular son
\begin{equation}\label{Esub+-}
    E_{n_+,n_-}=\frac{1}{R}\left(\varepsilon_+n_+
    +\varepsilon_-n_-+\frac{1}{2}(\varepsilon_++\varepsilon_-)\right),\qquad
    j_{n_+,n_-}=\varepsilon_+n_+-\varepsilon_-n_-+
    \frac{1}{2}(\varepsilon_+-\varepsilon_-),
\end{equation}
donde $n_\pm=0,1,\ldots$.

En la fase crítica con $\mu_-=0$ tenemos un oscilador de un modo con
   $\quad E_{n_+}=\varepsilon_+
    R^{-1}\left(n_++\frac{1}{2}\right),$
    $j_{n_+}=\varepsilon_+\left(n_++\frac{1}{2}\right)$, mientras que para
    $\mu_+=0$ tenemos
    $ E_{n_-}=\varepsilon_-R^{-1}\left(n_-+\frac{1}{2}\right),$
    $j_{n_-}=-\varepsilon_-\left(n_-+\frac{1}{2}\right)$.

\subsection{Degeneración de los niveles y simetría adicional}

Los operadores de subida y bajada, $I_+=I_1+i I_2$ y  $I_-=I_+^\dagger$, de la simetría adicional SO(3) (para $\varepsilon_+\varepsilon_-=+1$), o SO(2,1) (para $\varepsilon_+\varepsilon_-=-1$),
\begin{equation}\label{I+I+I3}
    [I_+,I_-]=\varepsilon_+\varepsilon_-i I_3,\qquad
    [I_3,I_\pm]=\pm iI_\pm,
\end{equation}
corresponden a combinaciones complejas de las integrales clásicas (\ref{II12}) dadas por
\begin{equation}\label{HJunia1}
\varepsilon_+\varepsilon_-=+1:\qquad
I_1+i\varepsilon_+I_2=a^\dagger_+a_-,
\end{equation}
\begin{equation}\label{HJunia2}
\varepsilon_+\varepsilon_-=-1:\qquad
I_1+i\varepsilon_+I_2=ia^\dagger_+a_-^\dagger.
\end{equation}

\begin{figure}[h]
\begin{center}
  \includegraphics[width=.55\textwidth]{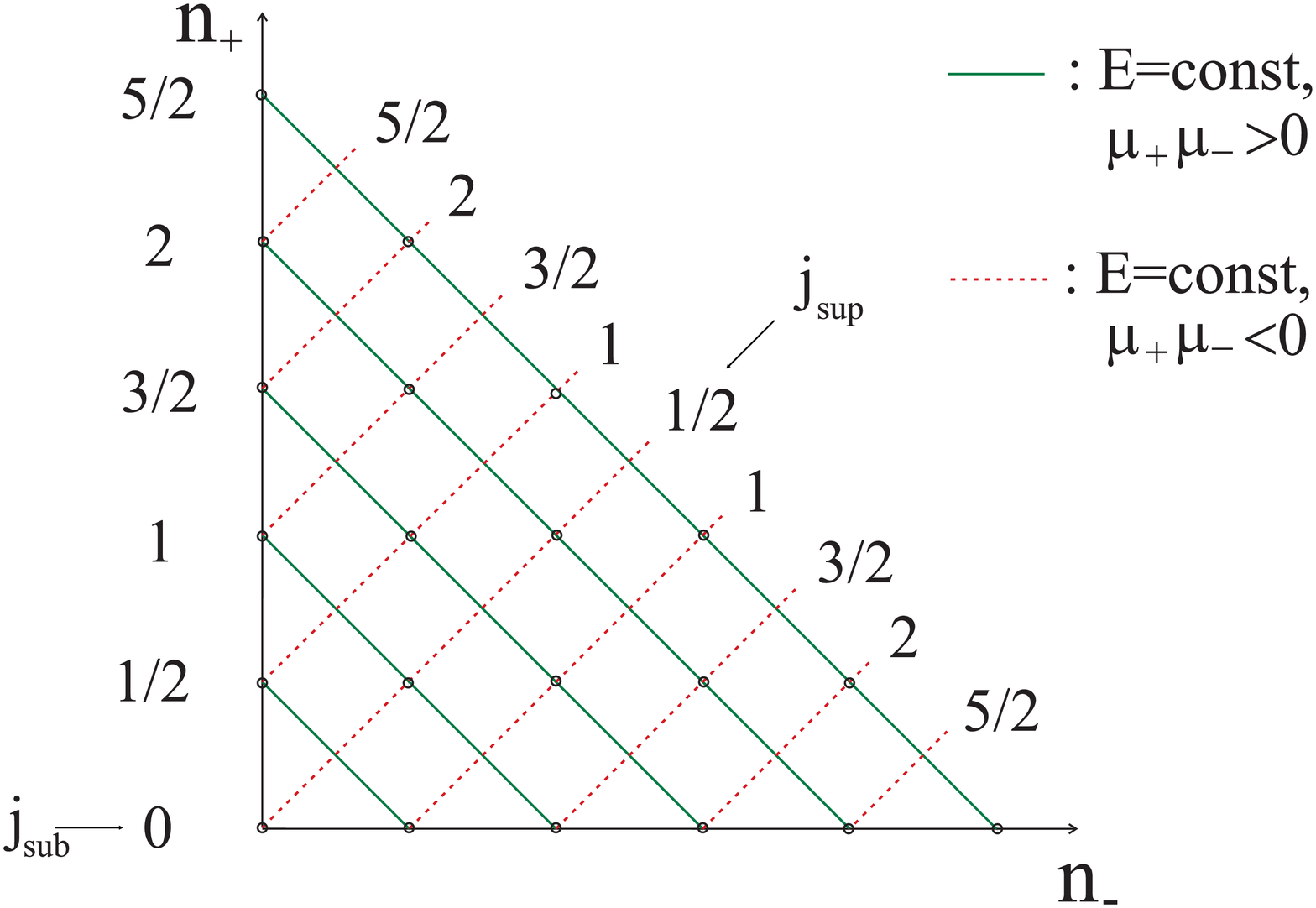}\\
  \caption{Niveles de energía constante.}\label{cons energy fig}
   \end{center}
\end{figure}

 En el caso subcrítico $\varepsilon_+\varepsilon_-=+1$ cada nivel tiene degeneración igual a $2j_{sub}+1$, donde
\begin{equation}\label{j<def}
    j_{sub}=\frac{1}{2}\left(n_++n_-\right),
\end{equation}
y los correspondientes autoestados de energía están caracterizados por los números cuánticos $n_+$ y $n_-$ situados en líneas rectas restringidas por ejes verticales y horizontales, ver figura \ref{cons energy fig}. Los operadores de subida y bajada de $so(3)$ dados por (\ref{HJunia1}) actúan a lo largo de esas líneas, caracterizadas por $n_++n_-=2j_{sub}$, donde el Casimir de $so(3)$, $C_{so(3)}=I_1^2+I_2^2+I_3^2$ toma el valor $j_{sub}(j_{sub}+1)$.

En la fase supercrítica $\varepsilon_+\varepsilon_-=-1$, y los niveles de energía constante corresponden a líneas rectas punteadas $n_+-n_-=\mathrm{const}$ en la figura \ref{cons energy fig}. Los operadores de escalera de $so(2,1)$ (\ref{HJunia2}) actúan a lo largo de aquellas líneas. Cada nivel es infinitamente degenerado, y el Casimir de $so(2,1)$, dado por $C_{so(2,1)}=I_1^2+I_2^2-I_3^2$ toma los valores $-j_{sup}(j_{sup}-1)$, donde \begin{equation}\label{j>so21}
    j_{sup}=\frac{1}{2}\left(|n_+-n_-|+1\right).
\end{equation}
Cada una de esas líneas tienen asociadas representaciones unitarias infinito-dimensionales semi-acotadas de la álgebra $so(2,1)$, en las cuales el generador compacto de $so(2,1)$, correspondiente a $I_3$ toma los valores $i_3=\varepsilon_+\frac{1}{2}(n_++n_-+1)$. Esto puede ser presentado equivalentemente por $i_3=\varepsilon_+ (j_{sup}+k),$ donde $k=n_-=0,1,\ldots$ para $n_+\geq n_-$, y $k=n_+=0,1,\ldots$ para $n_-\geq n_+$. Entonces, tenemos las llamadas series de representaciones discretas $D^+_{j_{sup}}$ (para $\varepsilon_+=+1$) y $D^-_{j_{sup}}$ (para $\varepsilon_+=-1$) de $SL(2,R)$, en la terminología de Bargmann \cite{BarSL2R}.


\section{Ecuaciones de onda, cuadro de Schrödinger}\label{waveEquat}

En esta sección cuantizaremos el sistema mediante el método de Gupta-Bleuler, lo cual nos permitirá construir un conjunto de ecuaciones de onda que realizan la simetría exótica de Newton-Hooke. Usando estas identificaremos las fases proyectivas asociadas a las traslaciones y boosts de Newton-Hooke exótico.

Por simplicidad, esta vez comenzaremos con la formulación chiral, y después la covariante, la cual es necesaria para mirar el límite plano (Galileo exótico).

\subsection{Base chiral}

Debemos considerar combinaciones complejas de los vínculos (\ref{chi+}), (\ref{chi-}) tales que:
\begin{itemize}
\item conmuten entre sí,
\item a nivel cuántico sean de naturaleza de operadores de aniquilación, teniendo así, núcleos no triviales,
\item el conjunto sea consistente con la dinámica, es decir, que los conmutadores con el Hamiltoniano sean proporcionales a las combinaciones elegidas de los vínculos.
\end{itemize}

\noindent
Entonces, a nivel cuántico, estas combinaciones van a separar el subespacio físico.

En correspondencia con la estructura de los paréntesis de los vínculos chirales (\ref{chi+}), (\ref{chi-}), para $\rho^2\neq 1$ es necesario distinguir cuatro casos en dependencia de los signos de $\mu_\pm$, los cuales son los cuadros en la figura \ref{duality_fig}. Para $\mu_+>0$, $\mu_->0$, la elección apropiada para la combinación lineal de vínculos es
\begin{equation}\label{chir+-sub}
\chi^+_1+i\chi^+_2\approx 0,\qquad \chi^-_1-i\chi^-_2\approx 0.
\end{equation}
En representación de coordenadas toman la forma de ecuaciones de onda
\begin{equation}\label{++case}
    \left(\partial_{\bar{\cal Z}^+}+\frac{\mu_+}{4}{\cal Z}^+\right)\Psi
    =0,\qquad
    \left(\partial_{{\cal Z}^-}
    +\frac{\mu_-}{4}\bar{\cal Z}^-\right)\Psi
    =0,
\end{equation}
donde $\partial_{\bar{\cal Z}^+}=\partial/\partial \bar{\cal Z}^+$,
$\partial_{{\cal Z}^-}=\partial/\partial {\cal Z}^-$, y hemos introducido dos variables complejas independientes
\begin{equation}\label{Z+Z-}
    {\cal Z}^+=X^+_1+iX^+_2,\qquad
    {\cal Z}^-=X^-_1+iX^-_2.
\end{equation}
La dinámica es dada por la ecuación de Schr\"odinger
\begin{equation}\label{Schchir}
    \left(i\partial_t -\frac{1}{R}\left(
    {\cal Z}^+\partial_{{\cal Z}^+}
    -\bar{\cal Z}^+\partial_{\bar{\cal Z}^+}
    -{\cal Z}^-\partial_{{\cal Z}^-}
    +\bar{\cal Z}^-\partial_{\bar{\cal Z}^-}
    \right)\right)\Psi=0,
\end{equation}
donde el operador diferencial lineal en variables chirales complejas es el análogo cuántico del Hamiltoniano total (\ref{Htotchir}).  Las ecuaciones (\ref{++case}) y (\ref{Schchir}) corresponden al conjunto de ecuaciones de onda para la partícula con simetría exótica de Newton-Hooke.

La solución de las ecuaciones de onda (\ref{++case}) es
\begin{equation}\label{psiphys}
    \Psi_{phys}=\exp\left(
    -\frac{\mu_+}{4}|{\cal Z}^+|^2
    -\frac{\mu_-}{4}|{\cal Z}^-|^2
    \right)\psi({\cal Z}^+,\bar{\cal Z}^-).
\end{equation}
La acción de los operadores de Hamiltoniano y momento angular sobre los estados físicos (\ref{psiphys})  se reduce a
\begin{equation}\label{HJPsi}
    H\Psi_{phys}=\exp(.)\frac{1}{R}\left(
    {\cal Z}^+\partial_{{\cal Z}^+}
    +\bar{\cal Z}^-\partial_{\bar{\cal Z}^-}\right)\psi({\cal Z}^+,\bar{\cal
    Z}^-),
\end{equation}
\begin{equation}\label{HJPsi*}
    J\Psi_{phys}=\exp(.)\frac{1}{R}\left(
    {\cal Z}^+\partial_{{\cal Z}^+}
    -\bar{\cal Z}^-\partial_{\bar{\cal Z}^-}\right)\psi({\cal Z}^+,\bar{\cal
    Z}^-),
\end{equation}
donde $\exp(.)$ se refiere a la exponencial de (\ref{psiphys}). Las autofunciones de la energía y el momento angular son dadas por los estados (\ref{psiphys}) con funciones de onda monomiales $\psi({\cal Z}^+,\bar{\cal Z}^-)$,
\begin{equation}\label{psinn}
    \psi_{n_+,n_-}({\cal
    Z}^+,\bar{\cal Z}^-)=
    \frac{1}{\sqrt{n_+!n_-!}}\left(\sqrt{\frac{\mu_+}{2}}{\cal Z}^+\right)^{n_+}
    \left(\sqrt{\frac{\mu_-}{2}}\bar{\cal
    Z}^-\right)^{n_-},
    \qquad n_+,n_-=0,1,\ldots,
\end{equation}
que corresponden a los autovalores de energía y momento angular $E_{n_+,n_-}=\frac{1}{R}(n_++n_-)$, $j_{n_+,n_-}=n_+-n_-$. Notar que un simple desplazamiento de la energía (igual a $-\frac{1}{R}\hbar$, $\hbar=1$) aparece en comparación con la cuantización  en espacio de fase reducido, el cual es consecuencia de una elección particular del ordenamiento cuántico, aquí hemos tomado un Hamiltoniano correspondiente a la expresión (\ref{Htotchir}).

Estos resultados están en completa correspondencia con la cuantización en espacio de fase reducido, (hasta el corrimiento de la energía), y lo que hemos obtenido acá es la representación holomorfa para el sistema del oscilador 2D.

El esquema de cuantización se generaliza directamente para los otros tres casos asociados a los signos $\varepsilon_+$ y $\varepsilon_-$ de los parámetros $\mu_+$ y $\mu_-$. Las ecuaciones de onda que separan los estados físicos están dadas por las combinaciones lineales de vínculos $\chi^+_i$ y $\chi^-_i$,
\begin{equation}\label{quantconst}
    \left(\chi^+_1+i\varepsilon_+\chi^+_2\right)\Psi=0,\qquad
    \left(\chi^-_1-i\varepsilon_-\chi^-_2\right)\Psi=0.
\end{equation}
En la tabla \ref{polari tabla} resumimos todos los casos

\begin{table}[h]
\begin{center}
\begin{tabular}{|c|c|c|c|c|}
  \hline
  Fase & $\varepsilon_+,$ $\varepsilon_- $ & $m,$ $\kappa$ & Vínculos & $\psi(.,.)$  \\
  \hline
  $\rho^2<1$& $+1,+1$ & $m>0,$ $|\kappa|<mR$ & $\chi^+_1+i\chi^+_2,$ $\chi^-_1 -i\chi^-_2$& $\overset{}{\psi({\cal Z}^+,\bar{\cal Z}^-)}$ \\
  $\rho^2<1$ & $-1,-1$ & $m<0$, $|\kappa|<mR$ &$\chi^+_1-i\chi^+_2,$ $\chi^-_1 +i\chi^-_2$&  $\psi(\bar{\cal Z}^+,{\cal Z}^-)$ \\
  $\rho^2>1$ & $+1,-1$ & $-\infty<m<\infty$, $\kappa>|m|R$ & $\chi^+_1+i\chi^+_2,$ $\chi^-_1 +i\chi^-_2$& $\psi({\cal Z}^+,{\cal Z}^-)$ \\
  $\rho^2>1$ & $-1,+1$  & $-\infty<m<\infty$, $\kappa<|m|R$ & $\chi^+_1-i\chi^+_2,$
  $\chi^-_1 - i\chi^-_2$& $\psi(\bar{\cal Z}^+,\bar{\cal Z}^-)$ \\
  \hline
\end{tabular}
\caption{Diferentes condiciones de polarización.}\label{polari tabla}
\end{center}
\end{table}

\noindent Las soluciones correspondientes que describen estados físicos tienen forma similar a (\ref{psiphys}),
\begin{equation}\label{psiphys*}
    \Psi_{phys}=\exp\left(
    -\frac{|\mu_+|}{4}|{\cal Z}^+|^2
    -\frac{|\mu_-|}{4}|{\cal Z}^-|^2
    \right)\psi(.,.),
\end{equation}
con los argumentos de la función $\psi$ especificados en la tabla.
La acción de $H$ y $J$ sobre esos estados tiene forma similar a (\ref{HJPsi}), (\ref{HJPsi*}): ellos son reducidos a la suma de dos operadores diferenciales de primer orden, en variables correspondientes a los argumentos de la función $\psi(.,.)$. Los signos que aparecen antes de los operadores indicados en la figura \ref{duality_fig}, y ellos corresponden a los signos de (\ref{HJuni}). Los análogos de las funciones (\ref{psinn}), los parámetros $\mu_\pm$ son cambiados por sus valores absolutos. El producto escalar es definido en todos los casos con la medida
$$
{\cal D}=\frac{|\mu_+\mu_-|}{(2\pi)^2}dX^+_1dX^+_2dX^-_1dX^-_2.
$$
Con respecto a ese producto escalar las funciones de onda de la forma (\ref{psiphys}), (\ref{psinn}) representan una base ortonormal en el subespacio físico del sistema.

En la fase crítica tenemos $\mu_+=0$, ó $\mu_-$=0, y de acuerdo con el esquema clásico de las ecuaciones cuánticas complejas que especifican los estados físicos son cambiadas por
\begin{equation}\label{dX+-}
    \frac{\partial}{\partial X^+_i}\Psi=0,\quad {\rm or}\quad
    \frac{\partial}{\partial X^-_i}\Psi=0.
\end{equation}
El modo correspondiente $X^+$ ó $X^-$ desaparece completamente de la teoría, y entonces obtenemos un oscilador de un modo en representación holomorfa.

Notar que las ecuaciones cuánticas (\ref{quantconst}), que aquí especifican los estados del subespacio físico, tienen el sentido del polarizaciones complejas para el oscilador 2D tratado con el marco de la cuantización geométrica \cite{GQ}, mientras las ecuaciones (\ref{dX+-}) tienen naturaleza de polarizaciones reales.

\subsection{Ecuaciones de onda en la base covariante y límite plano}

Las ecuaciones de onda y los estados físicos en la base covariante pueden ser obtenidos directamente de los de la base chiral, tomando en consideración una transformación de fase (unitaria) asociada con la derivada total en la cual difieren los Lagrangianos (\ref{Nc-Ch}). Por ejemplo, en la fase supercrítica caracterizada por $\varepsilon_+\varepsilon_-=-1$, las funciones de onda físicas tienen la forma
\begin{equation}\label{physnonCh}
    \Psi_{phys}=\exp\left(-\frac{|\kappa|}{4}\left(v_i^2+\frac{x_i^2}{R^2}\right)
    -\frac{m}{2}\epsilon_{ij}x_iv_j+i\frac{m}{2}x_iv_i\right)\psi(x_1+i\varepsilon_+x_2,v_1+i\varepsilon_+v_2,t).
\end{equation}
las cuales satisfacen las ecuaciones de onda
\begin{equation}\label{V+-Pi+-}
    (V_1+i\varepsilon_+V_2)\Psi(x,v,t)=0,\qquad
    (\Pi_1+i\varepsilon_+\Pi_2)\Psi(x,v,t)=0,
\end{equation}
donde $V_i$ y $\Pi_i$ están dados por (\ref{constNon1}), (\ref{constNon2}), y hemos asumido que  $p_i=-i\partial/\partial x_i$ y $\pi_i=-i\partial/\partial v_i$.
La dinámica está dada por la ecuación de Schr\"odinger
\begin{equation}\label{HtotNC*}
    \left(i\frac{\partial}{\partial t}
    -iv_j\frac{\partial}{\partial
    x_j}+i\frac{1}{R^2}x_j\frac{\partial}{\partial v_j}
    -\frac{m}{2}\left(
    \frac{x_j^2}{R^2}-v_j^2\right)\right)\Psi=0,
\end{equation}
en la cual, de acuerdo con las relaciones de consistencia (\ref{NCconsistency}), el Hamiltoniano total (\ref{HtotNC}) juega el rol de Hamiltoniano cuántico.

Recordando las relaciones (\ref{chi-nochi}), vemos que hay la siguiente correspondencia entre los vínculos en la formulación covariante, (\ref{constNon1},\ref{constNon2}), y chiral, (\ref{chi+},\ref{chi-}),
\begin{equation}\label{nonChRel}
    V_i\leftrightarrow (\chi^+_i+\chi^-_i),\qquad
    \Pi_i\leftrightarrow R\epsilon_{ij}(\chi_j^--\chi^+_j).
\end{equation}
Esto muestra que las ecuaciones (\ref{V+-Pi+-}) son combinaciones lineales de los vínculos chirales cuánticos (\ref{quantconst}) con $\varepsilon_+\varepsilon_-=-1$.

De manera similar, uno puede derivar ecuaciones de onda y estados físicos en la base covariante para la fase subcrítica, donde $\varepsilon_+\varepsilon_-=1$. Sin embargo, desde el punto de vista de el límite plano $R\rightarrow \infty$ (que puede ser tomado sólo en la fase subcrítica), las ecuaciones de vínculos cuánticas (\ref{quantconst}) no son un punto de partida apropiado. La razón es que los vínculos del caso subcrítico (\ref{quantconst}) tienen signos diferentes para los modos izquierdo y derecho, mirar (\ref{chir+-sub}) para $\varepsilon_+=\varepsilon_-=1$. Como consecuencia, no existe una combinación lineal de esos vínculos tal que en el límite plano nos entregue dos ecuaciones de vínculos independientes. Para buscar los vínculos complejos apropiados, es más conveniente proceder directamente desde los vínculos en formulación covariante (\ref{constNon1}) y (\ref{constNon2}). Para fijar ideas, tomemos $\kappa>0$, y entonces, $\varepsilon_+=\varepsilon_-=1$. La combinación lineal $\Pi_1+i\Pi_2$ tiene núcleo no trivial a nivel cuántico y puede ser elegida para una de las ecuaciones que andamos buscando,
\begin{equation}\label{Pi+Phys}
    (\Pi_1+i\Pi_2)\Psi=0.
\end{equation}
Para fijar el otro vínculo, consideremos una combinación lineal de $V_i$ y $\Pi_i$,
\begin{equation}
    \tilde{V}_{i}=V_{i}-\frac{m}{\kappa }\epsilon _{ij}\Pi _{j}=
    p_{i}+\frac{%
\kappa }{2R^{2}}\epsilon _{ij}x_{j}-\frac{m}{\kappa }\epsilon
_{ij}\left( \pi _{j}-\frac{\kappa }{2}\epsilon _{jk}v_{k}\right),
\end{equation}
que es desacoplada de los vínculos $\Pi_i$ en el sentido de los paréntesis,
\begin{eqnarray}
    &\{ \tilde{V}_{i},\Pi _{j}\} =0,\label{tilViPi}&\\
    &\{ \tilde{V}_{i},\tilde{V}_{j}\} =-\frac{m^2}{\kappa}(1-\rho^2)\epsilon_{ij}.\label{tilVitilVi}&
\end{eqnarray}
Las relaciones (\ref{tilViPi}), (\ref{tilVitilVi}) indican que lacombinación $\tilde{V}_i$ es una extension de Dirac de $V_i$ respecto a los vínculos de segunda clase $\Pi_i\approx 0$ \footnote{Comparar (\ref{tilVitilVi}) con los paréntesis de Dirac (\ref{VV}).}. De acuerdo con (\ref{tilVitilVi}), a nivel cuántico la combinación lineal $\tilde{V}_1-i\tilde{V}_2$ tiene núcleo no trivial (es de naturaleza de operador de aniquilación), y el vínculo cuántico (\ref{Pi+Phys}) puede ser suplantado por la ecuación
\begin{equation}\label{Til-Phys}
    (\tilde{V}_1-i\tilde{V}_2)\Psi=0.
\end{equation}

Las ecuaciones (\ref{NCconsistency}) nos dan los siguientes paréntesis de Poisson para el Hamiltoniano total (\ref{HtotNC}) con las combinaciones de vínculos seleccionadas
\begin{eqnarray}
    \{H,\Pi_1+i\Pi_2\}&=&-i\frac{m}{\kappa}(\Pi_1+i\Pi_2)+(\tilde{V}_1+i\tilde{V}_2),\label{HtotViPi1}\\
    \{H,\tilde{V}_1-i\tilde{V}_2\}&=&-i\frac{m}{\kappa}(\tilde{V}_1-i\tilde{V}_2)+\frac{m^2}{\kappa^2}
    (1-\rho^2)(\Pi_1-i\Pi_2).\label{HtotViPi2}
\end{eqnarray}
Al lado derecho aparecen las combinaciones complejas conjugadas de los vínculos elegidos. Esto significa que la dinámica generada por la ecuación de Schr\"odinger, con el Hamiltoniano total como operador de Hamiltoniano, no será consistente con los vínculos (\ref{Pi+Phys},\ref{Til-Phys}). Uno puede sobrepasar este obstáculo si pasamos del Hamiltoniano total a uno corregido $H\rightarrow \tilde H$, mediante la suma de términos cuadráticos en los vínculos,
\begin{equation}\label{H*}
    \tilde H=H-\frac{1}{\kappa}\epsilon_{ij}\Pi_i\tilde{V}_j+
    \frac{m}{2\kappa^2}(\Pi_1-i\Pi_2)(\Pi_1+i\Pi_2)+\frac{1}{2m(1-\rho^2)}(\tilde{V}_1+i\tilde{V}_2)
    (\tilde{V}_1-i\tilde{V}_2).
\end{equation}
El Hamiltoniano corregido conmuta fuertemente con los vínculos $\Pi_i$ y $\tilde{V}_i$, y, en particular,   con las combinaciones de los vínculos elegidas. Entonces la ecuación de Schr\"odinger
\begin{equation}\label{SchH*}
    \left(i\partial_t-\tilde H\right)\Psi=0
\end{equation}
será consistente con los vínculos cuánticos (\ref{Pi+Phys}), (\ref{Til-Phys}): un estado físico que satisface las ecuaciones de vínculos a $t=0$ también lo hará para $t>0$. Notar que debido a la conmutatividad del Hamiltoniano corregido con los vínculos, este es una extension de Dirac del Hamiltoniano canónico con respecto a todos los vínculos de segunda clase.

No mostraremos una forma explícita de los estados físicos que satisfacen las ecuaciones cuánticas (\ref{Pi+Phys},\ref{Til-Phys}), en cambio discutiremos el límite plano del sistema. Para $R\rightarrow \infty$, el Lagrangiano (\ref{nhaction}) se reduce al de la partícula libre en el plano no-conmutativo,
\begin{equation}\label{Lflat}
{\cal L}_{nc}=
    mv_i\dot x_i-m\frac{v_i^2}{2}
    +\frac 12\kappa\epsilon_{ij}v_i\dot v_j.
\end{equation}
El sistema (\ref{Lflat}) tiene simetría de Galileo exótico dado por las ecuaciones (\ref{NH3ex}-\ref{PP}) con $R=\infty$. En espacio de fase reducido el sistema es descrito por una estructura simpléctica con coordenadas no conmutativas,
\begin{equation}\label{xxpfree}
    \{x_i,x_j\}=\frac{\kappa}{m^2}\epsilon_{ij},\qquad
    \{x_i,p_j\}=\delta_{ij},\qquad \{p_i,p_j\}=0,
\end{equation}
y por un Hamiltoniano libre
\begin{equation}\label{H*free}
    H^*=\frac{1}{2m}p_i^2,
\end{equation}
los cuales son los límites planos de (\ref{xxpx}) y  (\ref{Hredxp}).

Por otro lado, los vínculos (\ref{Pi+Phys}), (\ref{Til-Phys}) y la ecuación de Schrödinger (\ref{SchH*}) se reducen en el límite plano, respectivamente, a \begin{equation}\label{ef1}
\left( \frac{\partial }{\partial v_{-}}+\frac{\kappa
}{4}v_{+}\right) \Psi \left( x,v,t\right) =0,
\end{equation}
\begin{equation}\label{ef2}
\left( \frac{\partial }{\partial x_{+}}-i\frac{m}{4}v_{-}-i\frac{m}{\kappa }%
\frac{\partial }{\partial v_{+}}\right) \Psi \left( x,v,t\right) =0,
\end{equation}
\begin{equation}\label{ef3}
\left( i\frac{\partial }{\partial t}+\frac{2}{m}\frac{\partial ^{2}}{%
\partial x_{+}\partial x_{-}}\right) \Psi \left( x,v,t\right) =0,
\end{equation}
donde usamos las notaciones $x_\pm=x_1\pm ix_2$, $v_\pm=v_1\pm iv_2$.
La solución general a las ecuaciones de vínculos es
\begin{equation}\label{Psiphys2}
    \Psi_{phys}=\exp\left(-\frac{\kappa}{4}v_+v_-\right)
    \psi\left(x_+-i\frac{\kappa}{m}v_+,x_-,t\right),
\end{equation}
y una solución general a la ecuación de Schrödinger siendo un autoestado del operador de momento con autovalor $\vec{p}$ puede ser escrita de la forma
\begin{eqnarray}
    \Psi_{\vec{p}}\,(\vec{x},\vec{v},t)&=&\exp\left(-\frac{\kappa}{4}\left(v_+v_-+\frac{\vec{p}\,{}^2}{m^2}
    \right)
    -i\frac{\vec{p}\,^2}{2m}t+i\vec{p}\vec{x}
    +\frac{1}{2}\frac{\kappa}{m}p_- v_+\right),\label{Psiop1}\\
    &=&\exp\left(-\frac{\kappa}{4}\left(\vec{v}-\frac{1}{m}\vec{p}\right)^2
    +\frac{i}{2}\frac{\kappa}{m}\epsilon_{ij}p_iv_j
    -i\frac{\vec{p}\,^2}{2m}t+i\vec{p}\vec{x}\right),\label{Psiop2}
\end{eqnarray}
donde $p_-=p_1-ip_2$. Una dependencia específica en $v_i$ garantiza la normalizabilidad del estado (\ref{Psiop1}) en las variables de velocidad. Notar que (\ref{Pi+Phys}) puede ser presentada en la forma $(a_1-a_2)\Psi=0$, donde $a_{1,2}$ son operadores de aniquilación construidos en términos de $v_{1,2}$ y sus derivadas. Entonces, esta ecuación significa que los estados físicos corresponden a $n_1=n_2$, donde $n_1$, $n_2$ son autovalores de los operadores de número $N_1=a^\dagger_1a_1$ y $N_2=a^\dagger_2a_2$, es decir, los estados físicos tienen una ``polarización circular'' definida en las variables de velocidad. Para $\kappa=0$ la función de onda (\ref{Psiop2}) se vuelve una onda plana para la partícula libre planar usual.

En conclusión de esta sección encontramos una relación del estado (\ref{Psiop2}) con función de onda de la partícula libre exótica en representación de Fock \cite{PHMP}. En el apronte de la representación de Fock, primero eliminamos a nivel clásico los momentos $\pi_i$ y luego cuantizamos a partir de la estructura simpléctica dada por (\ref{xpvv}). El subespacio físico is dado en este caso por la ecuación
\begin{equation}\label{physFock}
\left(-2i\frac{\partial}{\partial x_+}-m\hat v_-\right)\Psi=0,
\end{equation}
donde $[\hat v_-,\hat v_+]=\frac{2}{\kappa}$. Nosotros construimos $\hat v_\pm$ mediante operadores oscilatorios, $\hat v_-=\sqrt{2\kappa^{-1}}\,a$, $\hat v_+=(v_-)^\dagger$, $[a,a^\dagger]=1$, descomponiendo el estado en términos de los autoestados del operador de número de la velocidad,
\begin{equation}\label{sum0inf}
    \Psi=\sum_{n=0}^{\infty}\phi_n(x)\vert n\rangle,
\end{equation}
y encontramos que para los estados físicos todas las componentes $\phi_n$ con $n>1$ pueden ser representadas en términos de $\phi_0$,
\begin{equation}\label{phin0}
    \phi_n=(-1)^n\left(\frac{\kappa}{2}\right)^{\frac{n}{2}}
    \frac{1}{\sqrt{n!}}\left(\frac{\hat p_-}{m}\right)^n\phi_0,
\end{equation}
donde $\hat p_-=-2i\partial/\partial x_+$, ver ecuación (3.6) en \cite{PHMP}. Teniendo en mente la correspondencia entre el espacio de Fock y las representaciones holomorfas,
\begin{equation}\label{Fockhol1}
    a^\dagger\leftrightarrow z,\quad  a\leftrightarrow
    \frac{d}{dz},\quad
    \vert n\rangle \leftrightarrow \exp\left(-\frac{1}{2}\vert z\vert^2\right)
    \frac{z^n}{\sqrt{n!}},
\end{equation}
donde $z$ es una variable compleja, identificamos $v_+=\sqrt{\frac{2}{\kappa}}\,z$, y descomponemos
\begin{equation}\label{coherent}
    \exp\left(\frac{1}{2}\frac{\kappa}{m}p_- v_+\right)
\end{equation}
en serie de Taylor. Como resultado encontramos que aquel estado físico (\ref{Psiop1}) es equivalente al estado (\ref{sum0inf}) con $\phi_n$ dado por (\ref{phin0}). Notar que, de acuerdo con (\ref{physFock}), el estado (\ref{sum0inf}), (\ref{phin0}) es un estado coherente del operador de aniquilación en las velocidades $\hat v_-$ con autovalor operador-valuado $\frac{1}{m}\hat p_-$, y en la función de onda (\ref{Psiop1}), es el factor (\ref{coherent}) el que refleja la naturaleza de un estado físico.

\section{Fase proyectiva y dos-cociclo}\label{Phaseproj}

Ahora calcularemos la fase proyectiva correspondiente a la simetría exótica de Newton-Hooke. Esta fase está relacionada al dos-cociclo no trivial del grupo exótico de Newton-Hooke. Este cociclo puede ser obtenido por cálculo directo, o de la cuasi-invariancia del Lagrangiano bajo traslaciones y boosts \cite{bargmann,azcarragabook,levyleblond69}. Su presencia garantiza la invariancia de las ecuaciones de onda.

Consideremos el operador unitario
\begin{equation}\label{Utrbo}
    U(\alpha,\beta)=\exp i(\alpha_iP_i-\beta_iK_i)
\end{equation}
en la base covariante, donde $P_i$ y $K_i$ se supone que son los análogos cuánticos de las integrales clásicas (\ref{PKNCobs}). Usando las relaciones de conmutación (\ref{KPK}) y (\ref{PP}) y $ e^{A}e^B=e^{A+B}e^{\frac{1}{2}[A,B]}$ \footnote{Válida para cualquiera operadores  $A$ y $B$ tales que satisfacen $[A,[A,B]]=[B,[A,B]]=0.$}, obtenemos la ley de composición siguiente
\begin{equation}\label{UUaabb}
    U(\alpha,\beta)U(\alpha',\beta')=e^{-i\omega_2(\alpha,\beta;\alpha',\beta')}
    U(\alpha+\alpha',\beta+\beta'),
\end{equation}
donde un factor de fase no trivial es igual a (modulo $2\pi n$, $ n\in
\Z$) a
\begin{equation}\label{omega2}
    -\omega_2(\alpha,\beta;\alpha',\beta')=-\frac{1}{2}\left(\kappa\epsilon_{ij}
    \left(\frac{1}{R^2}\alpha_i\alpha'_j+\beta_i\beta'_j\right)
    +m(\alpha_i\beta'_i-\beta_i\alpha'_i)\right).
\end{equation}
Un calculo directo muestra que (\ref{omega2}) satisface una condición de cero coborde \begin{equation}\label{}
    \Delta \omega_2\equiv
    \omega_2(g_2,g_3)-\omega_2(g_1g_2,g_3)+\omega_2(g_1,g_2g_3)-\omega_2(g_1,g_2)=0,
\end{equation}
la cual garantiza la asociatividad del producto (\ref{UUaabb}). Aquí $g_{1}$, $g_2$ son elementos del grupo, los cuales en nuestro caso están caracterizados por el conjunto de parámetros $(\alpha_i,\beta_i)$, $(\alpha'_i,\beta'_i)$, con una ley de composición $g_1g_2\rightarrow (\alpha_i+\alpha'_i,\beta_i+\beta'_i)$. Entonces $\omega_2(\alpha,\beta;\alpha',\beta')$ es el dos-cociclo del grupo de $ENH_-$ asociado a traslaciones y boosts de Newton-Hooke.

Ahora consideremos la acción del operador unitario (\ref{Utrbo}) en las coordenadas $x_i, v_i$. En correspondencia con (\ref{trbo1},\ref{trbo2}), este genera las traslaciones y boosts de Newton-Hooke,
\begin{equation}\label{xxtrbo}
    x'_i=Ux_iU^{-1}=x_i+ R {\cal A}_i(t),\qquad
    v'_i=Uv_iU^{-1}=v_i+{\cal B}_i(t),
\end{equation}
donde hemos introducido una notación compacta para la rotación en el `plano' $(\frac{1}{R}\alpha_i,\beta_i)$ de los parámetros adimensionales,
\begin{equation}\label{ABalbeta}
    {\cal A}_i(t)=\frac{1}{R}\alpha_i\cos R^{-1}t + \beta_i\sin
    R^{-1}t,\qquad
    {\cal B}_i(t)=\beta_i\cos R^{-1}t - \frac{1}{R}\alpha_i\sin
    R^{-1}t,
\end{equation}
 ${\cal A}_i(0)=\frac{1}{R}\alpha_i$,  ${\cal B}_i(0)=\beta_i$.
En términos de (\ref{ABalbeta}), tenemos
\begin{equation}\label{PKAB}
    \alpha_iP_i-\beta_iK_i= A(x,v;\alpha,\beta) +B(p,\pi;\alpha,\beta),
\end{equation}
donde hemos separado las partes de coordenada y momento,
\begin{eqnarray}
    A(x,v;\alpha,\beta)&=&-m{\cal
    B}_i(t)x_i-\frac{\kappa}{2}\epsilon_{ij}\left(R^{-1}{\cal
    A}_i(t)x_j+{\cal B}_i(t)v_j\right),\nn\\
    B(p,\pi;\alpha,\beta)&=&R{\cal A}_i(t)p_i+{\cal B}_i(t)\pi_i.\label{ABxvt}
\end{eqnarray}
Como resultado, la acción de (\ref{Utrbo}) en una función de onda puede ser presentada en la forma
\begin{equation}\label{Upsixv}
    \tilde{\Psi}(x,v,t)\equiv
    U(\alpha,\beta)\Psi(x,v,t)=e^{-i\omega_1(x,v,t;\alpha,\beta)}\Psi(x',v',t),
\end{equation}
donde $x'$ y $v'$ son dados por (\ref{xxtrbo}), y la fase $\omega_1$ es dada (modulo $2\pi n$, $n\in \Z$)  por
\begin{equation}\label{omega_1F}
    \omega_1(x,v,t;\alpha,\beta)=m{\cal
    B}_i(t)\left(x_i+\frac{1}{2}R{\cal A}_i(t)\right)
    +\frac{\kappa}{2R}\epsilon_{ij}\left({\cal A}_i(t)x_j+R{\cal
    B}_i(t)v_j\right).
\end{equation}
$\omega_1(x,v,t;\alpha,\beta)$ es un uno-cociclo real valuado, fase proyectiva de ${\cal NH}_-$.

De (\ref{UUaabb}) y (\ref{Upsixv}) podemos ver que el dos-cociclo (\ref{omega2}) puede ser escrito en términos de la fase proyectiva (\ref{omega_1F}),
\begin{equation}\label{ome1om2}
    \omega_2(g_1,g_2)=\Delta\omega_1\equiv \omega_1(q^{g_1};g_2)+\omega_1(q;g_1)-\omega_1(q;g_1g_2),
\end{equation}
donde $q$ significa el conjunto $(x_i,v_i,t)$, y $q^g=gq$ una aplicación de $g$ a las coordenadas $q$, que en nuestro caso corresponde a $(x'_i,v'_i,t')$ con $t'=t$ y las coordenadas transformadas (\ref{xxtrbo}).

La fase proyectiva también esta asociada a la cuasi-invariancia del Lagrangiano respecto de las correspondientes transformaciones clásicas de simetría \cite{azcarragabook}.
Un chequeo directo muestra que en nuestro caso bajo las transformaciones (\ref{xxtrbo}), el Lagrangiano en formulación covariante (\ref{nhaction}) transforma como
\begin{equation}\label{omFL}
    {\cal L}'_{nc}={\cal L}_{nc}+\delta {\cal L}_{nc},\qquad
    \delta {\cal L}_{nc}=\frac{d}{dt}\left(\omega_1(x,v,t;\alpha,\beta)\right).
\end{equation}
\vskip0.3cm

En el límite plano $R\rightarrow \infty$  (\ref{omega_1F}) se reduce a
\begin{equation}\label{Fflat}
    \omega_1^0(x,v,t;\alpha,\beta)=m\beta_ix_i+\frac{1}{2}m\beta_i^2t+\frac{\kappa}{2}\epsilon_{ij}\beta_iv_j.
\end{equation}
El dos-cociclo en este caso es
\begin{equation}\label{omega20}
    -\omega_2^0(\alpha,\beta;\alpha',\beta')=-\frac{1}{2}\left(\kappa\epsilon_{ij}
    \beta_i\beta'_j
    +m(\alpha_i\beta'_i-\beta_i\alpha'_i)\right).
\end{equation}

\vskip0.3cm

Del mismo modo uno puede calcular la fase proyectiva en la base chiral. Aparece bajo la acción del análogo chiral del operador unitario (\ref{Utrbo}) en la función de onda chiral,
\begin{equation}\label{Upsixv+-}
    \tilde{\Psi}(X^+,X^-,t)\equiv
    U_{ch}(\alpha,\beta)\Psi(X^+,X^-,t)=e^{-i\tilde{\omega}_1(X^+,X^-,t;\alpha,\beta)}\Psi(X^+{}',{X^-}{}',t),
\end{equation}
\begin{equation}\label{FaseChir2}
    \tilde{\omega}_1(X^+,X^-,t;\alpha,\beta)=-\frac{1}{2}\mu_+\epsilon_{ij}X^+_i\alpha^+_j(t)+
    \frac{1}{2}\mu_-\epsilon_{ij}X^-_i\alpha^-_j(t).
\end{equation}
Aquí $\alpha^\pm_i(t)$ está definido por \bref{albet+-} y los generadores de traslaciones y boosts están construidos de acuerdo a (\ref{PKchiral}). La transformación \bref{xxtrbo} es cambiada en variables chirales como en \bref{NHchiral*}. Al igual que formulación covariante, para la base chiral el corrimiento dependiente de $t$ que es simetría bajo traslaciones y boosts \bref{NHchiral*} produce un cambio en el Lagrangiano chiral\bref{Lchir}, que es dado por la fase proyectiva
\begin{equation}\label{Lchir3}
    {\cal L}'_{ch}={\cal L}_{ch} +
    \frac{d}{dt}\tilde{\omega}_1(X^+,X^-,t;\alpha,\beta).
\end{equation}

Un cálculo del dos-cociclo usando la fórmula de Baker-Campbell-Hausdorff da el mismo resultado que en la formulación covariante \bref{omega2}, dado que este está basado en la misma álgebra exótica de Newton-Hooke, y en particular, en las mismas relaciones de conmutación \bref{KPK} y \bref{PP}. La igualdad de los dos-cociclos en ambas formulaciones también puede ser entendida desde el punto de vista de transformaciones canónicas asociadas a la diferencia de derivada total (\ref{Nc-Ch}) entre las dos versiones de Lagrangianos. De hecho, en correspondencia con la relación clásica (\ref{Nc-Ch}), las fases proyectivas en ambas formulaciones están relacionadas por
\begin{equation}\label{phasechirnonchir}
    \tilde{\omega}_1-\omega_1\equiv
     \rho(x,v,t;\alpha,\beta)
    =-\frac{1}{2}m(x'_iv'_i-x_iv_i)=
     \frac{mR}{2}\left({\cal
    A}_i(t)v_i+{\cal B}_i(t)x_i+{\cal A}_i(t){\cal
    B}_i(t)\right),
\end{equation}
donde $x'_i$ y $v'_i$ son dados en \bref{xxtrbo}. La diferencia es un uno-cociclo real-valuado de la forma
$f(q^g)-f(q)$.

Finalmente, comentaremos que la transformación canónica \bref{chi-nochi} está detrás de la relación siguiente entre las versiones covariante y chiral del operador unitario \bref{Utrbo}, que representa traslaciones y boosts de Newton-Hooke,
\begin{equation}\label{UchirUnc}
 {U}_{ch}=e^{-i\rho(x,v,t;\alpha,\beta)}U_{nc},
\end{equation}
donde $\rho(x,v,t;\alpha,\beta)$ es definido en
\bref{phasechirnonchir}.

\vskip 0.3cm
Veamos ahora la covarianza de las ecuaciones que hemos introducido. Aquellas pueden ser escritas como
\begin{equation}\label{covEq1}
    {\cal D}_a\Psi(s,t)=0,
\end{equation}
\begin{equation}
    \left(i\partial_t -H\right)\Psi(s,t)=0,\label{covEq2}
\end{equation}
donde $s$ denota las coordenadas $x_i,\, v_i$ o $X^+_i,\, X^-_i$ para las formulaciones covariante o chiral, \bref{covEq1} es un conjunto de dos vínculos cuánticos, cuya forma depende del caso del sistema (sub o supercrítico), y el Hamiltoniano cuántico $H$ es ajustado con ellos (para tener consistencia). La invariancia de la teoría bajo traslaciones en tiempo es obvia, y dado que para cualquier vínculo cuántico $[J,{\cal D}_a]\propto {\cal D}_a, $ la invariancia de rotaciones también es obvia. Una conclusión análoga es válida para la simetría adicional asociada a las integrales $I_1,$ $I_2$. Además, dado que los operadores $P_i$ y $K_i$ conmutan con los vínculos cuánticos, las ecuaciones de onda \bref{covEq1} son invariantes bajo las traslaciones y boosts de Newton-Hooke. Finalmente , en correspondencia con la relación clásica $\partial \Gamma_i/\partial t+\{\Gamma_i,H\}=0$, $\Gamma_i=P_i,K_i$, a nivel cuántico el operador \bref{Utrbo} conmuta con el operador $i\partial_t-H$.
Entonces tenemos
\begin{equation}
    (i\partial_t-H)\Psi(s,t)=0,\quad\to\quad
    (i\partial_t-H)U\Psi(s,t)=(i\partial_t-H)e^{-i\omega_1(s,t;g)}\Psi(s^g,t)=0,
\end{equation}
es decir la función de onda transformada \bref{Upsixv} o \bref{Upsixv+-} se vuelve una solución de la ecuación de Schrödinger.

Resumiendo, la fase proyectiva garantiza la covarianza de nuestras ecuaciones de onda.

\subsection{Cohomología de Eilenberg-Chevalley de $NH_-$}\label{E-C coho}
En esta sección calcularemos la cohomología de Eilenberg-Chevalley para las 2-formas del grupo de Newton-Hooke en 2+1 dimensiones. El análogo 2+1 dimensional de la álgebra de Newton-Hooke no extendida (\ref{NH 3+1_1}-\ref{NH 3+1_4}) está dado por
\begin{equation}\label{NH3ex*}
    [H,J]=0,
\end{equation}
\begin{equation}\label{HPK*}
    [H,K_i]=-iP_i,\qquad [H,P_i]=i\frac{1}{R^2}K_i,
\end{equation}
\begin{equation}\label{JKP*}
    [J,P_i]=i\epsilon_{ij}P_j,\qquad
    [J,K_i]=i\epsilon_{ij}K_j,
\end{equation}
\begin{equation}\label{KPK*}
    [K_i,P_j]=
    [K_i,K_j]=0,
\end{equation}
\begin{equation}\label{PP*}
    [P_i,P_j]=0.
\end{equation}
Consideremos un elemento del grupo
\begin{equation}
\label{coset1*} g=e^{-iH
x^0}\;e^{iP_ix^i}\;e^{iK_jv^j}e^{-iJ\theta},
\end{equation}
donde $\theta$ es una coordenada local en $S^1$.  La 1-forma de Maurer-Cartan está dada por
 \begin{equation}\label{MCar*}
    \Omega=-ig^{-1}dg= -L_H H +L^i_P P_i+L^i_K K_i+L_J J,
\end{equation}
donde
\begin{eqnarray}
    L_H &=& dx^0,
    \qquad
    L^i_P ={ R^i}_j(\theta)(dx^j-v^j dx^0),
    \qquad
    L^i_K = {R^i}_j(\theta) \left(dv^j+\frac{x^j}{R^2}\,dx^0\right),
    \qquad
    L_J = -d\theta.
\nn\\
\end{eqnarray}
con ${R^i}_j(\theta)=\left(\begin{array}{cc} \cos\T&\sin\T\\-\sin\T&\cos\T
\end{array}\right)$
siendo una rotación $SO(2)$.

Existen dos 2-formas cerradas invariantes de rotaciones,
\begin{eqnarray}
\Omega_2=L^i_K \wedge L^i_P,\qquad d\Omega_2=0,
\end{eqnarray}
\begin{eqnarray}
\tilde{\Omega}_2=\frac 12\epsilon_{ij}\left[L^i_K\wedge L^j_K+
\frac{1}{R^2}L^i_P \wedge L^j_P\right],\qquad d\tilde{\Omega_2}=0.
\end{eqnarray}
La cuales pueden ser expresadas localmente por
\begin{equation}
\Omega_2=d\Omega_1,\qquad
\Omega_1=v^idx^i-\frac{v_i^2}{2}dx^0-\frac{x_i^2}{2R^2}\,dx^0,
\end{equation}
\begin{equation}
\tilde{\Omega}_2=d\tilde\Omega_1,\qquad
    \tilde\Omega_1=\frac{1}{2}\epsilon_{ij}
    \left(v^idv^j+\frac{1}{R^2}x^i dx^j-
    \frac{2}{R^2}x^iv^jdx^0\right).
\end{equation}
Las 1-formas $\Omega_1$ y $\tilde\Omega_1$ no son invariantes de izquierda (``left invariant'').

De hecho también hay una tercera forma cerrada invariante de rotaciones,
\begin{equation}\label{Om3}
\check{\Omega }_{2}=L_{H}\wedge L_{J},\qquad d\check{\Omega }_{2}=0,
\end{equation}
que localmente se puede expresar
\begin{equation}\label{Om3*}
    \check{\Omega }_{2}=d\check{\Omega }_{1},\qquad
    \check{\Omega }_{1}=\theta dx^{0}.
\end{equation}
La 1-forma $ \check{\Omega }_{1}$ tampoco es invariante de izquierda.
Entonces la cohomología de Eilenberg-Chevalley de grado dos es no trivial. Esto implica que la álgebra de Newton-Hooke en 2+1 dimensiones admite una extensión central triple
\begin{equation}\label{HJg}
    \left[ H,J\right] =\check{Z}.
\end{equation}
\begin{equation}
    \lbrack K_{i},P_{j}]=i\delta _{ij}Z,\qquad \lbrack
    K_{i},K_{j}]=-i\epsilon _{ij}\tilde{Z},  \label{KPK**}
\end{equation}%
\begin{equation}
    \lbrack P_{i},P_{j}]=-i\frac{1}{{R}^{2}}\epsilon _{ij}\tilde{Z},
\label{PP**}
\end{equation}
Sin embargo, nosotros hemos descartado la extensión central asociada al generador $\check{Z}$. Tal extensión triple de la álgebra NH$^3$ no puede ser obtenida mediante una contracción de AdS$_3$, comparar (\ref{P0M12}) con (\ref{HJg}). En la presencia del tercer generador central $\check Z$, los únicos Casimires de la álgebra son los tres elementos centrales, y entonces, $H$ y $J$ no pueden ser presentados en términos de los generadores de traslaciones y boosts, ver \cite{Brihaye:1995nv,Gao}. La tercera extensión central no nos da ninguna contribución no trivial para nuestro Lagrangiano de la partícula exótica. Por otro lado, si agregamos una exponencial extra con una nueva extensión central en el elemento del grupo (\ref{coset1}), la forma de Maurer-Cartan adquiere un nuevo término $dy$, donde $y$ es el campo de Goldstone correspondiente, asociado a la tercera carga central. Sin embargo la contribución de este término al Lagrangiano es una derivada total.

Otra implicación de la cohomología no trivial es que tenemos dos términos de Wess-Zumino,
\begin{equation}
S_1=\int \Omega_1^*,\qquad \tilde S_1=\int \tilde\Omega_1^*,
\end{equation}
cuya combinación lineal describe la dinámica de la partícula exótica; aquí $*$ significa el pullback sobre la línea de mundo de la partícula.

\section{Simetrías en espacio de fase extendido}\label{ext space sec}
Analizaremos las transformaciones de simetría globales del Lagrangiano
extendido (\ref{nhactionZZ}). La idea es recuperar la invariancia del Lagrangiano bajo las traslaciones y boosts de Newton-Hooke, para esto consideraremos inicialmente $\delta \tau =0$, pero $\delta c,\ \delta \tilde{c},\ \delta
x_{i},\ \delta v_{i},\ \delta x^{0}$ arbitrarias, y fijaremos $\delta c$ y $\delta \tilde{c}$ de tal modo de cancelar los términos de derivada total. La transformación
del Lagrangiano se puede escribir en la forma
\begin{eqnarray}
\delta L &=&x_{i}\left[ -m\left\{ \delta \dot{v}_{i}+\frac{1}{R^{2}}\dot{x}^{0}\delta x_{i}\right\}+\frac{\kappa }{R^{2}}\epsilon _{ij}\left\{ \delta\dot{x}_{j}-\dot{x}^{0}\delta v_{j}\right\} \right]\notag\\
&& +v_{i}\left[ m\left\{
\delta \dot{x}_{i}-\dot{x}^{0}\delta v_{i}\right\} +\kappa \epsilon
_{ij}\left\{ \delta \dot{v}_{j}+\frac{1}{R^{2}}\dot{x}^{0}\delta
x_{j}\right\} \right]   \notag \\
&&+\delta \dot{x}^{0}H+\frac{d}{d\tau }\left[ m\delta c+\frac{\kappa }{2}
\delta \tilde{c}+x_{i}\left( m\delta v_{i}-\frac{\kappa }{2R^{2}}\epsilon
_{ij}\delta x_{j}\right) -\frac{\kappa }{2}\epsilon _{ij}v_{i}\delta v_{j}
\right] ,  \label{dl}
\end{eqnarray}
donde $H$ es el Hamiltoniano canónico dado por (\ref{HcanNC}).

Las transformaciones de simetría más obvias están dadas por $x^{0}$%
-traslaciones, $c$-traslaciones, $\tilde{c}$-traslaciones, y rotaciones en
índice $i$, es decir $\delta x_{i}=-\epsilon _{ij}x_{j}$ y $\delta
v_{i}=-\epsilon _{ij}v_{j}$. Independiente de estas aparecen las
traslaciones y boosts de Newton-Hooke, si exigimos%
\begin{equation}
\left\{ \delta \dot{v}_{i}+\frac{1}{R^{2}}\dot{x}^{0}\delta x_{i}\right\}
=0,\ \ \ \ \ \left\{ \delta \dot{x}_{i}-\dot{x}^{0}\delta v_{i}\right\} =0,\
\ \ \ \ \ \ \delta \dot{x}^{0}=0,  \label{d1}
\end{equation}
\begin{equation}
\delta c=-x_{i}\delta v_{i},\ \ \ \ \ \ \ \delta \tilde{c}=\epsilon
_{ij}\left( \frac{1}{R^{2}}x_{i}\delta x_{j}+v_{i}\delta v_{j}\right) ,
\label{dc}
\end{equation}
el Lagrangiano es invariante. Las relaciones (\ref{dc}) se obtienen de cancelar independientemente los términos proporcionales a $m$ y $\kappa $ en la derivada total en (\ref{dl}). Las transformaciones linealmente independientes que satisfacen (\ref{d1}) están dadas por
\begin{equation}
\text{\textit{Traslaciones}\textbf{:}}\qquad \delta
x_{i}=\alpha _{i}\cos \frac{x^{0}}{R}, \qquad \delta v_{i}=-\frac{1}{R}
\alpha _{i}\sin \frac{x^{0}}{R},
\end{equation}
\begin{equation}
\text{\textit{Boosts}\textbf{:}}\qquad \delta x_{i}=R\beta
_{i}\sin \frac{x^{0}}{R},\qquad \delta v_{i}=\beta _{i}\cos \frac{x^{0}
}{R}.
\end{equation}

Vía teorema de Noether es posible obtener los generadores a todas estas
simetrías

\begin{equation}
P_{0}=p_{0}, \qquad P_{c}=p_{c}, \qquad P_{\tilde{c}}=p_{\tilde{c}},
\label{p0}
\end{equation}
\begin{equation}
J=\epsilon _{ij}x_{i}p_{j}+\epsilon _{ij}v_{i}\pi _{j},  \label{J}
\end{equation}
\begin{equation}
P_{i}=\left[ p_{i}-\frac{1}{R^{2}}p_{\tilde{c}}\epsilon _{ij}x_{j}
\right] \cos \frac{x^{0}}{R}-\frac{1}{R}\left[ \pi _{i}-p_{c}x_{i}-p_{\tilde{
c}}\epsilon _{ij}v_{j}\right] \sin \frac{x^{0}}{R},
\end{equation}
\begin{equation}
K_{i}=-R\left[ p_{i}-\frac{1}{R^{2}}p_{\tilde{c}}\epsilon
_{ij}x_{j}\right] \sin \frac{x^{0}}{R}-\left[ \pi _{i}-p_{c}x_{i}-p_{\tilde{c%
}}\epsilon _{ij}v_{j}\right] \cos \frac{x^{0}}{R}.  \label{ki}
\end{equation}
Estos generadores satisfacen la algebra de Lie de Newton-Hooke exótico con $P_{c}$, $P_{\tilde{c}}$ y $P_{0}$ jugando el rol de cargas centrales
\begin{equation}\label{K,K extend}
\left\{K_i,K_j\right\} =-2P_{\tilde{c}}\epsilon
_{ij},\qquad \left\{K_i,P_j\right\}
=P_{c}\delta _{ij},\qquad \left\{P_i,P_j\right\} =-\frac{2}{R^{2}}P_{\tilde{c}}\epsilon _{ij},
\end{equation}
\begin{equation}
\left\{ P_{0},K_i\right\} =P_i,\qquad \left\{
P_{0},P_i\right\} =-\frac{1}{R^2}K_i,\qquad
\left\{ P_{0},P_{c}\right\} =0=\left\{ P_{0},P_{\tilde{c}}\right\},
\end{equation}
\begin{equation}
\left\{ P_{c},K_i\right\} =0=\left\{ P_{c},P_i\right\},
\end{equation}
\begin{equation}
\left\{ P_{\tilde{c}},K_i\right\} =0=\left\{ P_{\tilde{c}},P_i\right\},
\end{equation}
\begin{equation}
\left\{ P_{c},P_{\tilde{c}}\right\} =0,
\end{equation}
\begin{equation}
\left\{ P_{0},J\right\} =0=\left\{ P_{c},J\right\} =\left\{ P_{\tilde{c}},J\right\},\qquad \left\{P_i,J\right\} =-\epsilon _{ij}P_j,\qquad \left\{K_i,J\right\} =-\epsilon_{ij}K_{j},
\end{equation}
el factor 2 en (\ref{K,K extend}) se debe sólo a un asunto convencional en la definición (\ref{p0}).
\section{Mapeo al caso hiperbólico}\label{map hyp case}
Hemos analizado el caso ``trigonométrico'', o oscilante, de la simetría exótica de Newton-Hooke. Los resultados pueden ser traducidos de un modo simple para el caso ``hiperbólico'', o expansivo, de la simetría exótica de Newton-Hooke, el cual aparece como la contracción de dS$_3$ \cite{Mariano,Gao}. La álgebra de dS$_3$ se puede obtener de la de AdS$_3$ (\ref{P0M12}-\ref{PiPj}) mediante la substitución $R^2\rightarrow -R^2$. El Lagrangiano correspondiente puede ser obtenido a partir del Lagrangiano en formulación covariante (\ref{nhaction}), mediante la misma substitución. Como resultado, en el análogo de la relación (\ref{detA}) que caracteriza la álgebra de los vínculos, la cantidad $m^4(1-\rho^2)^2$ se transforma en $m^4(1+\rho^2)^2$. Entonces en el caso hiperbólico los vínculos son de segunda clase para cualquier elección de los parámetros $m$ y $\kappa$, y el sistema exótico de Newton-Hooke tiene una sola fase, la subcrítica.
\section{Conclusiones del capítulo}\label{conclus ENH}
Antes de dar algunas conclusiones incluimos una tabla que resume algunas de las notaciones usadas en el capítulo.\\
\begin{table}[h]
\begin{center}
\scalebox{0.8}[0.8]{
\begin{tabular}{l|c|c}
Notaciones usadas en: & Base covariante & Base Chiral \\ \hline\hline
Generadores de simetría y cam-& $H,\ J,\ K_{i},\ P_{i},\ Z,\ \tilde{Z
}$ & \multirow{3}{*}[3 mm]{$\mathcal{J}^{\pm },\ \mathcal{J}_{i}^{\pm },\ Z^{\pm }$} \\
pos vectoriales correspondientes & $\mathcal{X}_{H},\ \mathcal{X}_{J},\ \mathcal{
X}_{K_{i}},\ \mathcal{X}_{P_{i}},\ \mathcal{X}_{Z},\ \mathcal{X}_{\tilde{Z}}$
&  \\ \hline
Variables de espacio de fase (y & \multirow{3}{*}[3 mm]{ $x_{i},\ (p_{i}),\ v_{i},\ (\pi _{i})$} & \multirow{3}{*}[3 mm]{$X_{i}^{\pm },\ (P_{i}^{\pm })$} \\
momentos canónicos conjugados) &  &  \\ \hline
Vínculos & $\Pi _{i},\ V_{i}$ & $\chi _{i}^{\pm }$ \\ \hline
Variables observables & $\mathcal{P}_{i},\ \mathcal{X}_{i}$ & $\lambda
_{i}^{\pm }$ \\ \hline
Generadores de simetría adicional & \multicolumn{2}{|c}{$I_{1},\
I_{2},\ I_{3}$} \\ \hline
Operadores de creación y aniqui- & \multicolumn{2}{|c}{\multirow{3}{*}[3 mm]{\small$a_{+},\ a_{+}^{\dag },\ a_{-},\ a_{-}^{\dag }$}} \\
lación en espacio de fase reducido & \multicolumn{2}{|c}{} \\ \hline
Coordenadas complejas de las& $x_{\pm }=x_{1}\pm ix_{2}$ & \multirow{3}{*}[3 mm]{$\mathcal{Z}^{\pm
}=X_{1}^{\pm }+iX_{2}^{\pm }$} \\
ecuaciones de onda & $v_{\pm }=v_{1}\pm iv_{2}$ &
\end{tabular}}
\caption{Resumen con algunas de las notaciones usadas en este capítulo.}\label{notaciones tbl}
\end{center}
\end{table}

En este capítulo hemos estudiado el sistema de Newton-Hooke exótico, el cual aparece como límite no relativista de los modelos cosmológicos más simples, los espacios de de Sitter.

La álgebra de simetría posee dos extensiones centrales. Una aparece en el conmutador entre traslaciones y boosts, y en ese sentido es usual, ya que aparece para cualquier dimensión de espacio tiempo. Su valor está asociado a la masa de la partícula no relativista, $m$.

La segunda sólo existe en 2+1 dimensiones, por esto es algo más peculiar, y se le llama exótica. Aparece en el conmutador de los boosts, y también hasta un coeficiente , en el de momentos
\be
[K_i,K_j]=-i \kappa \epsilon_{ij}, \qquad [P_i,P_j]=-i\frac{\kappa}{R^{2}}\epsilon _{ij}.
\ee
El análogo plano, $R \rightarrow \infty$, del grupo de Newton-Hooke exótico es Galileo exótico. Este ha sido estudiado y, por ejemplo, se ha visto que $\kappa$ puede ser asociada a coordenadas no conmutativas. Igualmente para el caso de Newton-Hooke exótico aparece no conmutatividad de coordenadas, pero adicionalmente aparece una estructura de fases ligada a la razón $\rho=\frac{\kappa}{mR}$, en las cuales el sistema exhibe propiedades físicas distintas. Subcrítica para $\rho^2>1$, supercrítica $\rho^2<1$ y crítica $\rho^2=1$. Evidentemente el caso plano $\rho^2=0$, Galileo exótico, sólo vive en la fase subcrítica.

Los parámetros $m$ y $\kappa$ pueden ser promovidos a variables
dinámicas si los tratamos en el Lagrangiano (\ref{nhactionZZ}) como
el momento canónico conjugado de las variables $c$ y $\tilde c$.
Como resultado, diferentes fases del modelo se realizarán en
diferentes sectores del espacio de fase extendido. Notemos que este
fenómeno es similar al observado anteriormente en la gravedad de
Lovelock \cite{TeiZan} y teorías de Chern-Simons puras en altas
dimensiones \cite{BanHen}.

Hemos analizado el caso ``trigonométrico" (periódico) de la simetría
de Newton-Hooke exótica en 2+1 dimensiones. El resultado puede interpretarse
en una forma sencilla para el caso ``hiperbólico" de NH$_3$ exótico,
la que aparece bajo contracción de dS$_3$ \cite{Mariano,Gao}. La álgebra dS$_3$ aparece a partir de la álgebra AdS$_3$
(\ref{P0M12})--((\ref{PiPj}) mediante la sustitución simple de
$R^2\rightarrow -R^2$. Es posible encontrar el correspondiente
Lagrangiano a partir de nuestro Lagrangiano no-chiral
(\ref{nhaction}) mediante la misma sustitución. Como resultado, en
analogía con la relación (\ref{detA}) que caracteriza la álgebra de
las ligaduras, la cantidad $m^4(1-\rho^2)^2$ se cambiará por
$m^4(1+\rho^2)^2$. Por tanto, en el caso hiperbólico, las ligaduras
forman un conjunto de ligaduras de segunda clase para cualquier
elección de los parámetros $m$ y $\kappa$, y el correspondiente
sistema de Newton-Hooke exótico tiene sólo una fase.

\chapter{Oscilador anisótropo exótico}\label{aho chapter}

En el capítulo anterior hemos mencionado algunas similitudes entre los sistemas de Newton-Hooke exótico y el problema de Landau no conmutativo. En el presente capítulo construiremos explícitamente un sistema de oscilador armónico anisótropo que incluye como casos particulares Newton-Hooke exótico (caso isótropo) y el problema de Landau no conmutativo (una frecuencia no nula y la otra igual a cero).

En la sección \ref{eho lag chap} derivaremos el Lagrangiano del oscilador planar anisótropo. En la sección \ref{sim chir ani} discutiremos la forma chiral de la simetría exótica de Newton-Hooke del sistema. En la sección \ref{eho cov chap} analizaremos la simetría del oscilador anisótropo exótico en la formulación covariante. Luego, en la sección \ref{Chern-Simons AHO} mostraremos que es posible obtener el sistema mecánico del oscilador anisótropo a partir de un sistema de campos de Chern-Simons masivos. En la sección \ref{eho hyp ani chap} analizamos la posibilidad de tener el análogo hiperbólico exótico del sistema. En la sección \ref{hidden ads chap} discutimos una simetría de tipo AdS$_4$ oculta. Finalmente, en la sección \ref{conclu eho chap} cerramos el capítulo con algunas conclusiones.

\section{Oscilador armónico planar y el problema de Landau no conmutativo}\label{eho lag chap}

El Lagrangiano canónico en el ``espacio de fases'' para el oscilador armónico uno-dimensional de masa $m$ y frecuencia $\omega$ está dado por
\begin{equation}\label{Lharm}
    L_{can}=\frac{\mu}{2}\left(\epsilon_{ij}\dot{X}_iX_j-
    \frac{\alpha}{R}X_i^2\right),
\end{equation}
donde $m=\alpha^{-1}\mu R$, $\omega=\alpha R^{-1}$, y $\alpha$
es un parámetro adimensional. La variable $X_1$ puede ser identificada como la coordenada de una partícula uno-dimensional, y entonces $X_2$ proporcional a su momento lineal. La estructura simpléctica, $\{X_i,X_j\}=\frac{1}{\mu}\epsilon_{ij},$ y el Lagrangiano (\ref{Lharm}) poseen una simetría de rotaciones dos-dimensional en el \emph{espacio de fase}. Tomando la suma de $n$ copias de (\ref{Lharm}) con parámetros independientes $\mu$'s y $\alpha$'s, obtenemos un sistema generalizado de $n$ osciladores armónicos no interactuantes con diferentes frecuencias.

Consideremos el caso $n=2$, y tomemos el Lagrangiano en la forma
\begin{equation}
    L_{+-}=-\frac{\mu _{+}}{2}\left( \epsilon_{ij}
    \dot{X}_{i}^{+}X_{j}^{+}+ \frac{\alpha_+}{R}{X_{i}^{+}}^{2}\right)
    -\frac{\mu _{-}}{2}\left( -\epsilon_{ij}
    \dot{X}_{i}^{-}X_{j}^{-}+\frac{\alpha_-}{R}{X_{i}^{-}}
    ^{2}\right),
    \label{L_gen}
\end{equation}
donde suponemos que $R>0$ y que $\mu_\pm$ puede tomar valores de cualquier signo. Por el momento no asumiremos restricciones en los parámetros $\alpha_\pm$. La dinámica de (\ref{L_gen}) es generada por
\begin{equation}
    \dot{X}_{i}^{\pm }\pm \omega_\pm\epsilon _{ij}X_{j}^{\pm }=0.\qquad
    \omega_\pm=\alpha_\pm R^{-1},
    \label{eqs}
\end{equation}
mientras la estructura simpléctica es
\begin{equation}
    \{X_i^+,X_j^+\} =- \frac{1}{\mu _+}\epsilon _{ij},\qquad
    \{X_i^-,X_j^-\} = \frac{1}{\mu _-}\epsilon _{ij}, \qquad
    \{X_{i}^+,X_{j}^-\}=0. \label{Xi+-,Xj+-}
\end{equation}
En el caso especial elegido, $n=2$, el índice \emph{de espacio de fase} $i$ puede ser reinterpretado como un índice de la parte \emph{espacial} de un espacio-tiempo de tres dimensiones. Con tal reinterpretación, el Lagrangiano (\ref{L_gen}), como también las ecuaciones de movimiento (\ref{eqs}) y la estructura simpléctica (\ref{Xi+-,Xj+-}), poseen simetría de rotaciones \emph{espaciales} $SO(2)$ explícita. Esto corresponde a la parte diagonal de la la evidente simetría rotacional chiral $SO(2)\times SO(2)$ de (\ref{L_gen}). La parte no diagonal es identificada con la simetría de traslaciones en el tiempo, ver (\ref{trans2ch}) y (\ref{hj}) más abajo.

Como resultado, el sistema (\ref{L_gen}) nos provee con una descripción \emph{invariante de rotaciones} para el oscilador armónico planar \emph{anisótropo}.

En este punto, queremos aclarar bajo que condiciones el oscilador anisótropo puede ser interpretado como un sistema de \emph{partícula} planar con coordenadas $x_i$ y velocidad $v_i$, relacionadas por la ecuación dinámica usual
\begin{equation}\label{xdotv}
    \dot{x}_i=v_i.
\end{equation}
De acuerdo con (\ref{eqs}), para tal sistema de partícula, las variables $X^+_i$ y $X^-_i$ tienen sentido de modos normales, o chirales.

Para este objetivo, primero notemos que la transformación $\alpha_\pm\rightarrow -\alpha_\mp$, $\mu_\pm\rightarrow -\mu_\mp$, $X^+_i\leftrightarrow X^-_i$ no cambia las ecuaciones de movimiento y el Lagrangiano, y es suficiente asumir que $(\alpha_++\alpha_-)\geq 0$. Si $(\alpha_++\alpha_-)=0$, los dos modos chirales tienen exactamente la misma evolución. Este caso puede ser excluido dado que no nos permite introducir $x_i$ y $v_i$ relacionadas por (\ref{xdotv}). Teniendo también en consideración que $\alpha_\pm$ aparece sólo en la combinación $\alpha_\pm/R$, sin pérdida de generalidad podemos poner
\begin{equation}\label{alpha}
    \alpha_\pm(\chi)=\cos\chi(\cos\chi\pm\sin\chi),\qquad
    -\frac{\pi}{2}<\chi<\frac{\pi}{2}.
\end{equation}
Con la normalización elegida en (\ref{alpha}), tenemos\footnote{Sin embargo, en lo que sigue será más conveniente trabajar en términos de $\alpha_\pm$, implicando la representación uno-paramétrica (\ref{alpha}).} $0<(\alpha_++\alpha_-)\leq 1$ y $-1\leq (\alpha_+-\alpha_-)\leq 1$.

Ahora podemos introducir las coordenadas y velocidades para un sistema de partícula en un espacio de dos dimensiones,
\begin{equation}\label{XXxv}
    X^\pm_i=\alpha_{{}_\mp}  x_i\pm R\epsilon_{ij}v_j,
\end{equation}
\begin{equation}\label{xXX}
    x_i=\frac{1}{\alpha_++\alpha_-}\left(X^+_i+X^-_i\right),
    \qquad
    v_i=\frac{1}{R (\alpha_++\alpha_-)}\,
    \epsilon_{ij}\left(\alpha_-X^-_j-\alpha_+X^+_j\right),
\end{equation}
las cuales satisfacen (\ref{xdotv}).
El el caso genérico, en correspondencia con (\ref{Xi+-,Xj+-}), las coordenadas $x_i$ describen un plano \emph{no conmutativo},
\begin{equation}\label{xxnon}
    \left\{x_i,x_j\right\}=\frac{1}{(\alpha_++\alpha_-)^2}\,
    \frac{\mu_+-\mu_-}{\mu_+\mu_-}\epsilon_{ij}.
\end{equation}
Las componentes del vector de coordenadas $x_i$ son conmutativas solo cuando $\mu_+=\mu_-$. Como veremos, sólo en este caso los boosts de Galileo conmutan mutuamente. Los otros paréntesis son
\begin{equation}\label{xxvv}
    \{x_i,v_j\}=\frac{1}{R^2 (\alpha_++\alpha_-)^2}\,
    \frac{\mathcal{M}}{\mu_+\mu_-}\delta_{ij},\qquad
    \{v_i,v_j\}=\frac{1}{R^2 (\alpha_++\alpha_-)^2 }\,
   \frac{\mathcal{B}}{\mu_+\mu_-} \epsilon_{ij},
\end{equation}
donde
\begin{equation}\label{MNma}
    \mathcal{M}=R\left(\mu_+\alpha_- +
    \mu_-\alpha_+\right),\qquad
     \mathcal{B}=\mu_+\alpha_-^2 -\mu_-\alpha_+^2\,.
\end{equation}
La dos-forma simpléctica correspondiente a (\ref{xxnon}) y (\ref{xxvv}) tiene una estructura simple,
\begin{equation}\label{2form}
    \sigma={\cal M}dv_i\wedge dx_i+
    \frac{1}{2}R^2(\mu_+-\mu_-)\epsilon_{ij}dv_i\wedge dv_j
    +\frac{1}{2}{\cal B}\epsilon_{ij}dx_i\wedge dx_j.
\end{equation}

En términos de las coordenadas $x_i$, la anisotropía se revela en la dinámica acoplada de las componentes $x_1$ y $x_2$,
\begin{equation}\label{couposc}
    \ddot{x_i}+(\omega_+-\omega_-)
    \epsilon_{ij}\dot{x}_j+\omega_+\omega_-x_i=0,
\end{equation}
donde $ \omega_\pm$ están definidos en (\ref{eqs}). Esto puede ser comparado con las ecuaciones de segundo orden para los modos chirales, $\ddot{X}^\pm_i+\omega_\pm^2 X^\pm_i=0$. La dinámica de $x_1$ y $x_2$ es desacoplada sólo en el caso isótropo ($\chi=0$)
\begin{equation}\label{aa=1}
    \alpha_+=\alpha_-=1.
\end{equation}

En términos de las variables $x_i$ y $v_i$, el Lagrangiano (\ref{L_gen}) toma, hasta un término de derivada total, una forma covariante
\begin{equation}
    L =\mathcal{M}\left(\dot{x}_{i}
    v_{i}-\frac{\alpha_+\alpha_-}{2R^2}x_{i}^{2}-
    \frac{\mathcal{N}}{2\mathcal{M}}v_{i}^{2}\right)+\left(
    \mu_{-}-\mu_{+}\right)\left(\alpha_+\alpha_-\,\epsilon_{ij}
    x_{i}v_{j}-\frac{R^{2}}{2}\epsilon _{ij}v_i\dot{v}_{j}\right)+
    \frac{\mathcal{B}}{2}\epsilon _{ij}x_{i}\dot{x}_{j}\, ,
    \label{Lncom}
\end{equation}
donde $\mathcal{N}=R\left(\mu_+\alpha_+ +  \mu_-\alpha_-\right)$.
Notar que $\mathcal{M}$ y $\mathcal{N}$ tienen unidades de masa, mientras $\mathcal{B}$ de campo magnético. El término $\epsilon_{ij}{v}_{i}\dot{v}_{j}$, con coeficiente proporcional a $(\mu_+-\mu_-)$ en (\ref{Lncom}) es responsable de la no conmutatividad de coordenadas (\ref{xxnon}). Físicamente este describe un acoplamiento tipo magnético para las velocidades. Los términos $\dot{x}_{i}v_{i}$ y $\epsilon _{ij}x_{i}\dot{x}_{j}$, con coeficientes $\mathcal{M}$ y $\frac{1}{2}\mathcal{B}$, correspondientes a los paréntesis no triviales entre $x_i$ y $v_i$, y a la no conmutatividad de velocidades ver la primera y segunda relación en (\ref{xxvv}).

Las ecuaciones de movimiento obtenidas por la variación de (\ref{Lncom}) en  $x_i$ y $v_i$, son, respectivamente,
\begin{eqnarray}
    \mathcal{M}\left(\dot{v}_i+\frac{\alpha_+\alpha_-}{R^2}x_i\right)=
    \epsilon_{ij}\large(\mathcal{B}\dot{x}_j+(\mu_--\mu_+)\alpha_+\alpha_-
    \, v_j\large),\label{ec x_i}\\
    R^2(\mu_+-\mu_-)\left(\dot{v}_i+\frac{\alpha_+\alpha_-}{R^2}x_i\right)=
    \epsilon_{ij}\left(\mathcal{M}\dot{x}_j
    -\mathcal{N}v_j\right).
\label{ecvi}
\end{eqnarray}

Si $\mu_+=\mu_-$, entonces, $\left\{x_i,x_j\right\}=0$, (\ref{ecvi}) toma la forma (\ref{xdotv}), es decir una ecuación algebraica para $v_i$. En este caso $v_1$ y $v_2$ se vuelven variables auxiliares y consecuentemente pueden ser eliminadas usando sus ecuaciones de movimiento a partir de (\ref{Lncom}). Como resultado, (\ref{Lncom}) pasa a ser un Lagrangiano regular, de segundo orden $L(x,\dot{x})$, que describe el problema de Landau usual en la presencia de un término de potencial armónico adicional.

Es necesario notar que, aunque para $\mu_+\neq \mu_-$ la ecuación (\ref{xdotv}) también aparece como consecuencia del sistema de ecuaciones (\ref{ec x_i}) y (\ref{ecvi}), esta no es producida por la sola variación del Lagrangiano en $v_i$. En caso de substituir $v_i$, usando (\ref{xdotv}), en el Lagrangiano (\ref{Lncom}), podemos obtener un Lagrangiano con altas derivadas, no equivalente, que genera ecuaciones diferentes de (\ref{couposc}).

En el caso isótropo (\ref{aa=1}), el sistema (\ref{Lncom}) se reduce a la partícula exótica de Newton-Hooke, que hemos discutido en el capítulo anterior \cite{AGKP}.

Ahora mostraremos que el caso de anisotropía máxima,
\begin{equation}\label{aa0}
    \chi=\varepsilon\frac{\pi}{4}: \qquad \alpha_\varepsilon=1,\quad
    \alpha_{-\varepsilon}=0,\qquad
    \varepsilon=+,-,
\end{equation}
corresponde al problema de Landau no conmutativo, el cual está descrito por el Hamiltoniano
\begin{equation}
    H=\frac{1}{2m}{\mathcal{P}}_{i}^{2},
    \label{Hnclp}
\end{equation}
la estructura simpléctica
\begin{equation}
    \left\{ {\mathcal{X}}_{i},{\mathcal{X}}_{j}\right\} =\frac{\theta}
    {1-\beta } \epsilon _{ij},\qquad \left\{
    {\mathcal{X}}_{i},{\mathcal{P}}_{i}\right\} = \frac{1}{1-\beta}
    \delta _{ij},\qquad \left\{ {\mathcal{P}}_{i},{\mathcal{P}}_{j}
    \right\} =\frac{B}{1-\beta }\epsilon _{ij},
    \label{Poissonnclp}
\end{equation}
y las ecuaciones de movimiento
\begin{equation}
    \dot{\mathcal{X}}_{i}=\frac{1}{m^*}\mathcal{P}_{i},\qquad
    \dot{\mathcal{P}}_{i}=\frac{B}{m^*}\epsilon_{ij}\mathcal{P}_{j}.
    \label{ecmovnclp}
\end{equation}
Aquí $B$ es el campo magnético, $\beta=\theta B$, $m^*=m(1-\beta)$ juega el rol de masa efectiva, mientras que $\theta$ es un parámetro, que para $B=0$ caracteriza la no conmutatividad de coordenadas y también la de boosts para la partícula libre exótica \cite{Duval:2000xr,Duval:2001hu,Horvathy:2004fw}.
El Lagrangiano correspondiente a (\ref{Hnclp}) y (\ref{Poissonnclp}) está dado por
\begin{equation}\label{NLPlag}
    L_{{NLP}}={\cal P}_i\dot{\cal X}_i -\frac{1}{2m}{\cal P}_i^2
    +\frac{1}{2}\theta\epsilon_{ij}{\cal P}_i\dot{\cal P}_j+
    \frac{1}{2}B\epsilon_{ij}{\cal X}_i\dot{\cal X}_j\,.
\end{equation}

Comparando (\ref{Lncom}) y (\ref{NLPlag}), encontramos que en el caso maximalmente anisótropo (\ref{aa0}) el primer sistema se reduce al segundo, haciendo la siguiente correspondencia entre las variables y parámetros:
\begin{equation}\label{xXvP}
    x_i={\cal X}_i,\qquad v_i=\frac{1}{m^*}{\cal P}_i,
\end{equation}
\begin{equation}
    \mu_{\varepsilon}=\vert B(1-\theta B)\vert,\qquad
      \mu_{-\varepsilon}=\vert B\vert {\rm sgn}(1-\theta B),\qquad
    R=\left\vert\omega^{-1}\right\vert,\label{ahoNLP}
\end{equation}
\begin{equation}
    \varepsilon={\rm sgn} \left(B(\beta-1)\right),\qquad
       \omega=\frac{B}{m^*}.
  \label{omega}
\end{equation}

Teniendo esta correspondencia, y usando la transformación (\ref{xXX}), nosotros obtenemos la forma chiral del Lagrangiano (\ref{NLPlag}) para el caso $\alpha_+=0$, $\alpha_-=1$ ($\varepsilon=-1$),
\begin{equation}
    L_{NLP}^{+-}=-\frac{B}{2} \epsilon_{ij}
    \dot{X}_{i}^{+}X_{j}^{+} -\frac{B(1-\beta)}{2}\left(
    -\epsilon _{ij}\dot{X}_{i}^{-}X_{j}^{-}+
    \omega{X_{i}^{-}}^{2}\right),
    \label{NLPchir}
\end{equation}
el cual genera las ecuaciones de movimiento
\begin{equation}
    \dot{X}_{i}^{+}=0, \qquad \dot{X}_{i}^{-}-\omega
    \epsilon_{ij}X_{j}^{-}=0.
    \label{NLPeqX}
\end{equation}
En términos de las variables ${\cal X}_i$ y ${\cal P}_i$, los modos chirales están dados por
\begin{equation}
    X_{i}^{+}={\cal X}_i+\frac{1}{m^*|\omega|}\epsilon_{ij}{\cal P}_j, \qquad
    X_{i}^{-}=-\frac{1}{m^*|\omega|}\epsilon_{ij}{\cal P}_j.
\end{equation}
El modo chiral $X^+_i$ es una integral de movimiento que no depende explícitamente del tiempo (comparar con las integrales de movimiento chirales (\ref{calJi+-}) en el caso genérico). Este puede ser identificado como coordenada del centro de giro.

El caso $\alpha_+=1$, $\alpha_-=0$ ($\varepsilon=+1$) puede ser obtenido vía cambios obvios en correspondencia con las relaciones (\ref{ahoNLP}), (\ref{omega}). En este caso el modo chiral $X^-_i$ juega el role de coordenada de centro de giro, mientras $X^+_i$ tiene la misma ley de evolución que el modo chiral $X^-_i$ en el caso previo $\varepsilon=-1$.

El límite plano
\begin{equation}
    R\rightarrow\infty,\qquad (\mu_+-\mu_-)\rightarrow 0,
    \qquad
     R\mu_+(\alpha_++\alpha_-)\rightarrow m,\qquad
     R^2(\mu_+-\mu_-)\rightarrow \theta m^2,\label{freeEx}
\end{equation}
aplicado a (\ref{Lncom}), produce la partícula libre exótica,
\begin{equation}\label{Lfree}
    L_\theta=m\left(\dot{x}_iv_i-\frac{1}{2}v_i^2\right)+\frac{1}{2}\theta
    m^2\epsilon_{ij}v_i\dot{v}_j,
\end{equation}
que está descrita por las ecuaciones de movimiento $\dot{x}_i=v_i$,
$\dot{v}_i=0$, y lleva la álgebra de Galileo doblemente extendida \cite{Duval:2000xr,Duval:2001hu}. Si, como en el caso genérico (\ref{Lncom}), tratamos de substituir $v_i$ usando la ecuación $v_i=\dot{x}_i$ producida por la variación de (\ref{Lfree}) en $x_i$, obtenemos un modelo de altas derivadas no equivalente \cite{LSZ} con grados de libertad de spin adicionales \cite{Horvathy:2002vt,PHMP}.

\section{Simetrías: formulación chiral}\label{sim chir ani}

Para identificar las simetrías de nuestro sistema, procederemos desde el Lagrangiano en formulación chiral (\ref{L_gen}). Integramos las ecuaciones de movimiento (\ref{eqs}),
\begin{equation}\label{Xt}
    X_i^{\pm}(t)=\Delta_{ij}^{\pm}(t)X_j^{\pm}(0), \qquad
    \Delta_{ij}^{\pm}(t)=\delta_{ij}\cos(\alpha_{\pm}t/R)\mp\epsilon_{ij}\sin(\alpha_{\pm}t/R),
\end{equation}
y construimos las integrales de movimiento,
\begin{equation}
    \mathcal{J}_{i}^{\pm }\equiv
    R\mu_{\pm}\epsilon_{ij}X_{j}^{\pm}(0)=R\mu_{\pm}\epsilon_{ij}
    \Delta_{jk}^{\pm}(-t)X_{k}^{\pm},
    \label{calJi+-}
\end{equation}
\begin{equation}
    \mathcal{J}^{\pm }\equiv
    \pm\frac{\mu_{\pm}}{2}\left(X_{i}^{\pm}(0)\right)^{2}
    =\pm\frac{\mu_{\pm}}{2}\left(X_{i}^{\pm}\right)^{2},
    \label{calJ+-}
\end{equation}
donde $X^\pm_i=X^\pm_i(t)$. Las cantidades $\mathcal{J}_{i}^{\pm}$ son integrales de movimiento que incluyen dependencia explícita del tiempo y satisfacen las ecuaciones $\frac{d}{dt}\mathcal{J}_{i}^{\pm }= \frac{\partial}{\partial t}\mathcal{J}_{i}^{\pm }+\{\mathcal{J}_{i}^{\pm },H\}=0$, donde
\begin{equation}\label{Hamil}
    H=\frac{1}{2R}\left(\mu_+\alpha_+X_i^+{}^2+\mu_-\alpha_-X_i^-{}^2\right)
\end{equation}
juega el rol de Hamiltoniano. A diferencia de las integrales (\ref{calJi+-}) lineales en $X^\pm_i$, las integrales cuadráticas (\ref{calJ+-}) no incluyen dependencia explícita del tiempo.

Las integrales (\ref{calJi+-}) y (\ref{calJ+-}) generan la álgebra
\begin{equation}
    \{\mathcal{J}^{+},\mathcal{J}_{i}^{+}\}=\epsilon _{ij}\mathcal{J}_{j}^{+},
    \qquad
    \{\mathcal{J}_{i}^{+},\mathcal{J}_{j}^{+}\}=Z^{+}\epsilon _{ij},
    \label{chiral+}
\end{equation}
\begin{equation}
    \{\mathcal{J}^{-},\mathcal{J}_{i}^{-}\}=\epsilon _{ij}\mathcal{J}_{j}^{-},
    \qquad
    \{\mathcal{J}_{i}^{-},\mathcal{J}_{j}^{-}\}=Z^{-}\epsilon _{ij},
    \label{chiral-}
\end{equation}
donde $Z^\pm=\mp R^2\mu_{\pm}$ tienen sentido de cargas centrales, y los paréntesis restantes son iguales a cerro. Esta es la forma chiral de la álgebra exótica de Newton-Hooke (2+1) dimensional, presentada como la suma directa de dos álgebras de Newton-Hooke centralmente extendidas 1+1 dimensiones. Los Casimires cuadráticos de esta álgebra son
\begin{equation}\label{Casimir}
    {\cal C}_\pm= {{\cal J}^\pm_i}^2 + 2Z^\pm {\cal J}^\pm \, .
\end{equation}

Nos gustaría remarcar que la álgebra exótica de Newton-Hooke (\ref{chiral+}), (\ref{chiral-}) en el caso genérico (\ref{alpha}) tiene exactamente la misma forma que en el caso particular isótropo (\ref{aa=1}), la cual recordemos que en el capítulo anterior apareció como límite no relativista de AdS${}_3$.

Las integrales (\ref{calJi+-}) y (\ref{calJ+-}) generan las transformaciones de simetría de los modos chirales\footnote{Donde hemos omitido los parámetros de transformación infinitesimales.},
\begin{eqnarray}
   & \{ X_{i}^{+ },\mathcal{J}_{j}^{+ }\} =- R\Delta_{ij}^{+}(t), \qquad
    \{X_{i}^{- },\mathcal{J}_{j}^{- }\} = R\Delta_{ij}^{-}(t),&
\label{trans1ch}\\
    &\{ X_{i}^{+ },\mathcal{J}^{+}\} = -\epsilon _{ij}X_{j}^{+},\qquad \{
    X_{i}^{- },\mathcal{J}^{-}\} =-\epsilon_{ij}
    X_{j}^{-}.&\label{trans2ch}
\end{eqnarray}
Debido a la presencia de la dependencia explícita del tiempo de las integrales chirales (\ref{calJi+-}), las transformaciones de simetría (\ref{trans1ch}) son también dependientes del tiempo. Bajo ellas, el Lagrangiano (\ref{L_gen}) es cuasi-invariante.

El Hamiltoniano y el momento angular son identificados como ciertas combinaciones lineales de las integrales cuadráticas, que generan traslaciones en el tiempo y rotaciones espaciales,
\begin{equation}
    H=\frac{1}{R}\left( \alpha _{+}\mathcal{J}^{+}-\alpha_{-}
    \mathcal{J} ^{-}\right) ,\qquad
    J=\mathcal{J}^{+}+\mathcal{J}^{-}.
    \label{hj}
\end{equation}
Estas formas de $H$ y $J$ están detrás de la anisotropía en la dinámica y la simetría rotacional del sistema (\ref{L_gen}).

Para el caso (\ref{aa0}), uno de los generadores chirales,
${\cal J}^+$ or ${\cal J}^-$, desaparece del Hamiltoniano. Como resultado, el modo chiral correspondiente tiene dinámica trivial como coordenada de centro de giro, $\dot{X}^+=0$, o $\dot{X}^-_i=0$, ver (\ref{NLPeqX}), y uno de los dos vectores (\ref{calJi+-}) se transforma en integral de movimiento que no depende explícitamente de del tiempo.

Notar que en el caso excluido $\alpha_+=-\alpha_-$, los generadores de traslaciones en el tiempo, $H$, y rotaciones, $J$, se vuelven (hasta una constante multiplicativa) el mismo, lo cual imposibilita la introducción de la coordenada $x_i$ y la velocidad $v_i$ relacionados por (\ref{xdotv}). \vskip0.2cm

Discutamos brevemente las simetrías a nivel cuántico. Definimos los operadores
\begin{equation}
    a^{-}=\sqrt{\frac{\left\vert \mu _{+}\right\vert }{2}}\left(
    X_{2}^{+}+iX_{1}^{+}\right) ,\quad a^{+}=\left( a^{-}\right) ^{\dag},
    \qquad b^{-}=\sqrt{\frac{\left\vert \mu _{-}\right\vert}{2}}
    \left( X_{1}^{-}+iX_{2}^{-}\right) ,\quad b^{+}=\left(b^{-}\right) ^{\dag },
\end{equation}
los cuales obedecen las relaciones de conmutación $\left[ a^{-},a^{+}\right] =\epsilon_+$, $\left[ b^{-},b^{+}\right] =\epsilon_-$ y $[a^\pm,b^\pm]=0$, donde $\epsilon_\pm={\rm sgn} (\mu_\pm)$. Para $\mu_+>0$ ($\mu_+<0$) y $\mu_->0$ ($\mu_-<0$), los operadores de tipo oscilatorio $a^+$ ($a^-$) y $b^+$ ($b^-$) son identificados como creación y aniquilación. Con la prescripción de orden simétrico, (\ref{calJ+-}) y (\ref{hj}) nos da el Hamiltoniano y momento angular en la forma
\begin{equation}
    R \ H= \epsilon_+\alpha_{+}\,a^{+}a^{-}+
    \epsilon_-\alpha_{-}\,b^{+}b^{-}
    +\frac{1}{2}\left(\epsilon_+\alpha_{+}+
    \epsilon_-\alpha_{-}\right),
\end{equation}
\begin{equation}
    J=\epsilon_+\,a^{+}a^{-}-
    \epsilon_-\,b^{+}b^{-}+\frac{1}{2}\left(\epsilon_+-
    \epsilon_-\right).
\end{equation}
{}De aquí se ve que en el caso anisótropo genérico, así como también para $ENH_-$, \cite{AGKP} y NLP \cite{Horvathy:2004fw}, el sistema exhibe tres tipos de comportamiento en dependencia de los valores de los parámetros $\mu_\pm$.

Dado que los autovalores de los operadores de número toman valores enteros no negativos, $n_a$, $n_b=0,1,...$, encontramos que cuando $\epsilon_+=\epsilon_-$, $J$ puede tomar valores de ambos signos, es decir su espectro es no acotado. Este caso lo llamaremos fase \emph{subcrítica}. El espectro de $H$ en esta fase es acotado de abajo (cuando $\mu_+,\mu_->0$), o de arriba cuando (para $\mu_+,\mu_-<0$).

Cuando los signos de $\mu_+$ y $\mu_-$ son opuestos, el espectro de $J$ es acotado de un lado. Esta será la fase \emph{supercrítica}. En esta fase el espectro de $H$ es no acotado, excepto el caso de NLP. El caso de NLP es \emph{muy especial}: su energía está acotada en ambas fases sub y supercrítica, debido a que uno de los modos chirales tiene frecuencia cero, y consecuentemente no contribuye a la energía.

Hay otra fase, donde los parámetros $\mu_+$ o $\mu_-$ toman valor cero. En tal fase uno de los modos desaparece del Lagrangiano (volviéndose un grado de libertad de gauge puro), y el sistema se transforma en un oscilador uno-dimensional, cuya simetría es descrita por la álgebra de Newton-Hooke 1+1 dimensiones centralmente extendida \cite{AGKP}. Esto es una fase \emph{crítica} que separa las otras dos, y es caracterizada por el valor trivial que toma una de las cargas centrales, $Z^+$, or $Z^-$. En el NLP esto corresponde a la fase relacionada al efecto de Hall cuántico. Nosotros notamos aquí que en la fase crítica, en cambio, dos casos deben ser distinguidos. Cuando digamos, $\mu_-=0$ y $\alpha_+=1$, el Hamiltoniano es no trivial y genera una rotación del modo chiral restante, $X^+_i$, que coincide (hasta un desplazamiento de gauge) con $x_i$. Cuando $\mu_-=0$ y $\alpha_+=0$, el Hamiltoniano es igual a cero, y el modo chiral tiene dinámica trivial, $X^+_i(t)=X^+_i(0)$ (para una discusión de la fase crítica en el NLP, mirar \cite{Horvathy:2004fw}). En ambos casos, $\alpha_+=1$ y $\alpha_+=0$, $x_i$ y $v_i$ satisfacen la relación (\ref{xdotv}), pero ellas son variables linealmente dependientes en correspondencia con la disminución del número de grados de libertad físicos del sistema.

{}Desde el punto de vista de la dinámica y las simetrías, como en el caso del problema usual del oscilador anisótropo dado por un Lagrangiano regular, también es necesario distinguir casos especiales. Como se sigue de (\ref{xXX}) y (\ref{Xt}), la trayectoria de partícula $x_i(t)$ es cerrada sólo cuando $\alpha_+/\alpha_-$ es un número racional. Detrás de esta propiedad está la simetría adicional.

En el caso isótropo (\ref{aa=1}), el sistema de la partícula de ENH es caracterizado por una simetría adicional asociada con las integrales $a^+b^-$ y $a^-b^+$, las cuales, como $H$ y $J$, no incluyen dependencia explícita del tiempo. En términos de los modos chirales, ellos son combinaciones lineales de los operadores hermíticos $(X^+_1X^-_2+X^+_2X^-_1)$ y  $(X^+_2X^-_2-X^+_1X^-_1)$. En las fases sub y supercrítica, ellos, junto al momento angular $J$ generan las simetrías $so(3)$ y $so(2,1)$, las cuales son responsables de la degeneración finita o infinita de los niveles de energía. Esta simetría adicional es de la misma naturaleza de la discutida en el capítulo anterior \cite{AGKP}.

Análogamente, en el problema de Landau no conmutativo, tenemos una simetría adicional $so(2,1)$. Si, digamos, $\alpha_+=0$, $\alpha_-=1$, y $\mu_+>0$, los generadores $so(2,1)$ son dados por las integrales cuadráticas $I_0=\frac{1}{4}\{a^+,a^-\}$, $I_+=\frac{1}{2}a^+{}^2$ y $I_-=\frac{1}{2}a^-{}^2$,
\begin{equation}
    [I_0,I_\pm]=\pm I_\pm,\qquad
    [I_+,I_-]=-2I_0.
    \label{III}
\end{equation}
Todos los niveles de energía son infinitamente degenerados.

En el caso anisótropo con $\alpha_+/\alpha_-=p/q$, $p,q=1,2,\ldots$, $p\neq q$, las frecuencias de los modos chirales son conmensurables, y el sistema tiene las integrales de movimiento adicionales $j_+=(a^+)^q(b^-)^p$ y $j_-=(a^-)^q(b^+)^p$. En este caso los niveles tienen degeneración adicional finita, o infinita, en dependencia de si estamos en la fase sub o supercrítica. Estas integrales, junto con el momento angular generan una deformación no lineal (de hecho polinomial) de $so(3)$, o $so(2,1)$ de la álgebra,
\begin{equation}
    [J,j_\pm]=\pm (q+p) j_\pm\,, \label{Jjj}
\end{equation}
\begin{equation}\label{jjJH}
    [j_+,j_-]=\prod_{k=1}^{q}[a^+a^-
    +(1-k)\epsilon_+]
                \prod_{\ell=1}^{p}[b^+b^-+\ell\epsilon_-]-
\prod_{k=1}^{q}[a^+a^-+k\epsilon_+]
                \prod_{\ell=1}^{p}[b^+b^-+(1-\ell)\epsilon_-]\,,
\end{equation}
donde
\begin{equation}
  a^+a^- =\frac{\epsilon_+}{(\alpha_++\alpha_-)}\left(RH+\alpha_-J\right)-
\frac{1}{2},\qquad
  b^+b^- =\frac{\epsilon_-}{(\alpha_++\alpha_-)}\left(RH-\alpha_+J\right)-
\frac{1}{2},
\end{equation}
en las cuales el Hamiltoniano juega el rol de elemento central, $[H,J]=[H,j_\pm]=0$. Eso es completamente análogo a la propiedad bien conocida del oscilador planar anisótropo usual (no exótico), con frecuencias conmensurables \cite{Jauch,Louck,Boer,Kij}.

\section{Simetría exótica de Newton-Hooke: formulación covariante}\label{eho cov chap}

Acá determinaremos las transformaciones de simetría en términos de las variables asociadas a la coordenada y velocidad de la partícula, $x_i$, y $v_i$, y los generadores correspondientes. En particular, identificaremos las integrales que en el límite plano se transforman en los generadores de traslaciones conmutativas y boosts no conmutativos, y encontraremos la álgebra que forman con $H$ yd $J$.

Para identificar los generadores de traslaciones y boosts, notemos que debido a la naturaleza vectorial, ellos deben ser combinaciones lineales de las integrales $\mathcal{J}_i^{\pm}$. De acuerdo con (\ref{trans1ch}) y (\ref{xXX}), las transformaciones producidas por $\mathcal{J}_i^{\pm}$, son
\begin{equation}
    \{
    x_{i},\mathcal{J}_{j}^{\pm}\}=\mp R
    (\alpha_++\alpha_-)^{-1}\Delta_{ij}^{\pm}(t),\qquad
    \{v_{i},\mathcal{J}_{j}^{\pm}\}=\alpha_\pm
     (\alpha_++\alpha_-)^{-1}\epsilon_{ik}\Delta_{kj}^{\pm}(t),
    \label{xvJJ}
\end{equation}
donde $\Delta_{ij}^{\pm}(t)$ fue definido en (\ref{Xt}). Entonces, en orden de recuperar las transformaciones de Galileo en el límite plano
(\ref{freeEx}), $
    \{ x_{i},P_{j}\}=\delta_{ij},
$
$ \{x_{i},K_{j}
    \}=-\delta_{ij}t,
$ $\{ v_{i},P_{j} \}=0,$ $\{v_{i},K_{j} \}=-\delta_{ij},$
obtenemos
\begin{equation}
    P_{i}=\frac{1}{R}\left(\alpha_+\mathcal{J}_{i}^{-}-\alpha_-
    \mathcal{J}_{i}^{+}\right),
    \qquad K_{i}=-\epsilon_{ij}\left(\mathcal{J}_{j}^{+}+
    \mathcal{J}_{j}^{-}\right). \label{PK}
    \end{equation}
Notar que la relación (\ref{PK}) entre $P_i$, $K_i$ y ${\cal J}^\pm_i$ tiene estructura similar a la entre $H$, $J$ y ${\cal J}^\pm$, ver (\ref{hj}). Las transformaciones de simetría generadas por $P_i$ y $K_i$ pueden ser calculados mediante (\ref{xvJJ}) y (\ref{PK}).

En correspondencia con (\ref{xdotv}) y (\ref{couposc}), las traslaciones en el tiempo toman la forma
\begin{equation}\label{xvH}
    \{x_i,H\}=v_i,\qquad
    \{v_i,H\}=-\frac{\alpha_+\alpha_-}{R^2}
    x_i-\frac{\alpha_+-\alpha_-}{R}\epsilon_{ij}v_j,
\end{equation}
donde, en correspondencia con (\ref{Lncom}),
\begin{equation}\label{Hxv}
    H=\alpha_+\alpha_-\left(\frac{\mathcal{M}}{2R^2}x_i^2+
    (\mu_+-\mu_-)\epsilon_{ij}x_iv_j\right)
    +\frac{\cal N}{2}v_i^2.
\end{equation}
El segundo término en la ley de transformación para la velocidad es proporcional a las transformaciones de rotaciones $\{v_i,J\}=-\epsilon_{ij}v_j$, y desaparece en el caso isótropo. Es útil notar que la estructura del momento angular,
\begin{equation}\label{Jxv}
    J=\frac{{\cal B}}{2}x_i^2 +\frac{1}{2}R^2(\mu_+-\mu_-)v_i^2+
    {\cal M}\epsilon_{ij}x_iv_j,
\end{equation}
reproduce la estructura de dos-forma simpléctica (\ref{2form}).

La álgebra de simetría generada por $H$, $J$, $P_i$ y $K_i$ es
\begin{equation}\label{KPK}
    \left\{ K_{i},K_{j}\right\} =-\tilde{Z}\epsilon _{ij},\quad
     \left\{ P_{i},P_{j}\right\} =- \frac{1}{R^{2}}\left( R\left(
    \alpha _{+}-\alpha _{-}\right) Z+\alpha _{+}\alpha_{-}
    \tilde{Z}\right) \epsilon _{ij},
    \quad
    \left\{
    K_{i},P_{j}\right\} =Z\delta _{ij},
\end{equation}
\begin{equation}\label{KPJ}
    \left\{ K_{i},J\right\} =-\epsilon _{ij}K_{j},\qquad \left\{ P_{i},J\right\}
    =-\epsilon_{ij}P_{j,}\qquad \left\{ H,J\right\} =0\, ,
\end{equation}
\begin{equation}\label{HJPK}
    \left\{ K_{i},H\right\} =P_{i}+\frac{\left( \alpha _{+}-\alpha _{-}\right) }{R}
    \epsilon_{ij}K_{j},\qquad \left\{ P_{i},H\right\} =-\frac{\alpha _{+}
    \alpha _{-}}{R^{2}}K_{i}\, ,
\end{equation}
donde
\begin{equation}\label{ZZZ}
    Z=(\alpha_+Z^--\alpha_-Z^+)R^{-1}={\cal M},\qquad
    \tilde{Z}=-(Z^++Z^-)=R^2(\mu_+-\mu_-),
\end{equation}
y ${\cal M}$ es definida en (\ref{MNma}). Los Casimires (\ref{Casimir}) toman acá la forma equivalente
\begin{equation}\label{Cas2}
    \mathcal{C}_{\pm}=\left(P_i \mp
    \frac{\alpha_{\pm}}{R}\epsilon_{ij}K_j\right)^2
    -2\left(Z \pm
    \frac{\alpha_{\pm}}{R}\tilde{Z}\right) \left(H \pm
    \frac{\alpha_{\mp}}{R}J\right).
\end{equation}

Cuando $\mu_+=\mu_-$, la carga central $\tilde{Z}$ toma valor cero, y los boosts de Galileo conmutan. Es exactamente el mismo caso cuando las coordenadas de la partícula son conmutativas, mirar (\ref{xxnon}). Análogamente, la conmutatividad de los boosts y traslaciones aparece cuando las otras cargas centrales desaparecen, $Z={\cal M}=0$. En este caso las coordenadas $x_i$ y velocidades $v_i$ conmutan, ver la primera relación en (\ref{xxvv}).

Para el caso particular (\ref{aa0}) del problema de Landau no conmutativo la forma explícita de la álgebra generada por $H$, $J$, $P_i$ y $K_i$ puede ser obtenida de (\ref{KPK}-\ref{Cas2}) mediante las relaciones de correspondencia (\ref{xXvP}-\ref{omega}). Aquí sólo notamos que los generadores de traslaciones se reducen al modo chiral conservado, el cual es identificado con las coordenadas de centro de giro, mirar (\ref{PK}). Este genera traslaciones rígidas (independientes del tiempo), $\delta x_i=\delta a_i$, $\delta v_i=0$, bajo las cuales el Lagrangiano (\ref{NLPlag}) es cuasi-invariante.

En el caso genérico, los generadores $H$, $J$, $P_i$ y $K_i$, y las cargas centrales $Z$ y $\tilde{Z}$ son combinaciones lineales de las integrales chirales ${\cal J}^\pm$ y ${\cal J}^\pm_i$ y las cargas centrales $Z^+$ y $Z^-$, mirar (\ref{hj}), (\ref{PK}) y (\ref{ZZZ}). Entonces, existe una transformación lineal entre los generadores en base covariante del oscilador exótico anisótropo, caracterizado por los parámetros $\alpha_\pm(\chi)$, y los generadores en base covariante del grupo exótico de Newton-Hooke (caso simétrico) (\ref{aa=1}). Explícitamente tenemos
\begin{equation}
    \binom{RH}{J}_{\alpha_+,\alpha_-}=
    \begin{pmatrix}
    {\cal A}_+  & {\cal A}_-  \\
    0 & 1
    \end{pmatrix}
    \binom{RH}{J}_{\alpha_+=\alpha_-=1},
    \label{A1}
\end{equation}
\begin{equation}
    \binom{RP_{i}}{K_{i}}_{\alpha_+,\alpha_-}=
    \begin{pmatrix}
    {\cal A}_+ \delta _{ij} &
    {\cal A}_- \epsilon _{ij} \\
    0 & \delta _{ij}
    \end{pmatrix}
    \binom{RP_{j}}{K_{j}}_{\alpha_+=\alpha_-=1},
    \label{A2}
\end{equation}
\begin{equation}
    \binom{RZ}{\tilde{Z}}_{\alpha_+,\alpha_-}=
    \begin{pmatrix}
    {\cal A}_+  & -{\cal A}_-  \\
    0 & 1
    \end{pmatrix}
    \binom{RZ}{\tilde{Z}}_{\alpha_+=\alpha_-=1},
    \label{A3}
\end{equation}
donde $${\cal A}_\pm=\frac{1}{2}(\alpha_+\pm\alpha_-).$$

Estas relaciones implican que el caso general del oscilador exótico anisótropo, incluyendo el problema de Landau no conmutativo, son descritos, de hecho, por la misma álgebra de simetría. Por otro lado, haciendo uso de (\ref{A1}-\ref{A3}), los generadores y cargas centrales del caso anisótropo genérico también pueden ser expresados como combinación lineal de los del problema de Landau no conmutativo.

\section{Lagrangiano de Chern-Simons masivo}\label{Chern-Simons AHO}
En esta sección mostraremos como es posible obtener el sistema del oscilador anisótropo exótico a partir del límite de una teoría de campos de Chern-Simons masivo.

Consideremos un Lagrangiano de Chern-Simons U(1) con un termino de masa
\begin{equation}
\mathcal{L}=\mathcal{L}\left( A_{\mu };\mu ,\omega \right)=-\mu \left[ \epsilon ^{\mu \nu \lambda }A_{\mu }\partial _{\nu
}A_{\lambda }+\omega \eta ^{\mu \nu }A_{\mu }A_{\nu }\right],\label{L}
\end{equation}
donde $\mu $ y $\omega $ son parámetros (por suposición distintos de cero) y $\eta ^{\mu \nu }=\mathrm{diag}(-,+,+)$. Para realizar el conteo de grados de libertad de este sistema es necesario hacer el procedimiento de Dirac para sistemas degenerados.

Reescribamos (\ref{L})
\begin{equation}
\mathcal{L} =-\mu \left[ \epsilon ^{0ij}\left( \dot{A}_{i}A_{j}+2A_{0}\partial
_{i}A_{j}\right) +\omega \eta ^{\mu \nu }A_{\mu }A_{\nu }\right] -\mu
\epsilon ^{0ij}\partial _{j}\left( A_{0}A_{i}\right),
\end{equation}
hasta un termino de divergencia total podemos considerar la densidad
Lagrangiana dada por
\begin{equation}
\mathcal{L}=-\mu \left[ \epsilon ^{ij}\left( \dot{A}_{i}A_{j}+2A_{0}\partial
_{i}A_{j}\right) +\omega \eta ^{\mu \nu }A_{\mu }A_{\nu }\right],   \label{Ld}
\end{equation}
donde $\epsilon ^{ij}=\epsilon ^{0ij}$. Definimos los momentos canónicos
conjugados
\begin{equation}\label{simpl CS}
\pi ^{\mu }:=\frac{\partial \mathcal{L}}{\partial \dot{A}_{\mu }},\qquad
\left\{ A_{\mu }\left( \mathbf{x}\right) ,\pi ^{\nu }\left( \mathbf{x}^{\prime }\right)
\right\} _{t=t^{\prime }}=\delta _{\mu }^{\nu }\delta \left( \mathbf{x}-\mathbf{x}^{\prime}\right),
\end{equation}
usando (\ref{Ld}) obtenemos
\begin{equation}
\pi ^{0}=0,\qquad \pi ^{i}=-\mu \epsilon ^{ij}A_{j},
\end{equation}
de donde aparecen tres vínculos primarios
\begin{equation}\label{vinc prim}
\phi ^{0}=\pi ^{0}\approx 0, \qquad \phi ^{i}=\pi ^{i}+\mu \epsilon ^{ij}A_{j}\approx 0,
\end{equation}
la densidad Hamiltoniana canónica esta dada por
\begin{equation}
\mathcal{H}_{c}=\mu \left[ 2A_{0}\epsilon ^{ij}\partial _{i}A_{j}+\omega
\eta ^{\mu \nu }A_{\mu }A_{\nu }\right] =\mu \left[ 2A_{0}\epsilon
^{ij}\partial _{i}A_{j}-\omega A_{0}^{2}+\omega A_{i}^{2}\right],
\end{equation}
de la condición de consistencia para $\pi ^{0}\approx 0$ obtenemos un nuevo
vínculo que junto a (\ref{vinc prim}) forman un conjunto de segunda clase
\begin{equation}
\phi ^{3}=\epsilon ^{ij}\partial _{i}A_{j}-\omega A_{0}\approx 0,
\label{vin sec}
\end{equation}
este vínculo corresponde a la ecuación de movimiento que se obtiene de (\ref{Ld}) para $\mu =0$.

En resumen tenemos el sistema definido la estructura simpléctica (\ref{simpl CS}), cuatro de vínculos de segunda clase dados por (\ref{vinc prim},\ref{vin sec}) y por el Hamiltoniano total dado por
\begin{equation}
\mathcal{H}=\mathcal{H}_{c}+u_{a}\phi ^{a},
\end{equation}
donde $u_{a}$ son multiplicadores de Lagrange cuyos valores quedan fijos
mediante la exigencia de la conservación de los vínculos. Entonces el número de grados de libertad físicos por punto del espacio son dos. De este conteo podemos esperar que en cierto límite podamos recuperar el Lagrangiano (\ref{L_gen}) a partir de la suma de dos Lagrangianos de Chern-Simons masivos. En efecto, podemos ver que cuando $A_{\mu }$ no depende de las componentes espaciales de $x$ podemos reproducir el Lagrangiano del oscilador armónico anisótropo exótico mediante la combinación
\begin{equation}
L_{+-}=\lim_{\partial _{i}=0}\left\{ \mathcal{L}\left( A_{\mu }^{+};-\mu
_{+},\omega _{+}\right) +\mathcal{L}\left( A_{\mu }^{-};\mu _{-},-\omega
_{-}\right) \right\}.
\end{equation}

\section{¿Existe el caso anisótropo hiperbólico?}\label{eho hyp ani chap}
Una pregunta natural se relaciona con la posibilidad de obtener el análogo anisótropo exótico con soluciones hiperbólicas. Al igual que en la sección \ref{map hyp case}, para pasar al caso hiperbólico es necesario hacer la substitución
\be
R\rightarrow iR.\label{cambio}
\ee
Si realizamos este cambio en las ecuaciones de movimiento (\ref{couposc}) obtenemos
\begin{equation}
\ddot{x}_{i}+i\left( \lambda _{+}-\lambda _{-}\right) \epsilon _{ij}\dot{x}_{j}-\lambda _{+}\lambda _{-}x_{i}=0  \label{eomh}
\end{equation}%
donde $\lambda _{\pm }=\alpha _{\pm }/R$, notar que (\ref{eomh}) es un
sistema de ecuaciones complejas. La solución se obtiene de forma análoga que
para (\ref{couposc}). Podemos reemplazar el ansatz $x_{i}=c_{i}\ e^{\lambda t}$,
donde $c_{i}$ son constantes (reales si consideramos valores reales para las
coordenadas), como resultado obtenemos dos ecuaciones algebraicas
(homogéneas) para $\lambda $. Para tener soluciones no triviales para los $c_{i}$ el sistema debe ser degenerado ($\det =0$), de esto obtenemos una
ecuación de orden cuatro para $\lambda $ cuyas soluciones están dadas por%
\begin{equation}
\lambda =\pm \lambda _{+}\text{, }\pm \lambda _{-}.
\end{equation}%
De acá vemos que (\ref{eomh}) tiene las soluciones de tipo hiperbólicas
buscadas. Pero la anisotropía, $\left( \lambda _{+}-\lambda _{-}\right)
\neq 0$ tuvo como consecuencia una ecuación compleja.

Por otra parte es claro que podemos tener un sistema hiperbólico anisótropo correspondiente, por ejemplo, a ecuaciones del tipo
\begin{equation}
\ddot{x}_{1}-\lambda _{+}^{2}x_{1}=0,
\end{equation}
\begin{equation}
\ddot{x}_{2}-\lambda _{-}^{2}x_{2}=0,
\end{equation}
pero claramente en este caso perdemos simetría de rotaciones espaciales (interpretación de partícula).

De hecho es posible ver que ocurre al considerar una ecuación del tipo (\ref{eomh}) no es posible obtener soluciones hiperbólicas anisótropas si se consideran coeficientes reales. Pongamos
\begin{equation}
\ddot{x}_{i}+A\epsilon _{ij}\dot{x}_{j}+Bx_{i}=0  \label{general}
\end{equation}
donde $A$ y $B$ son coeficientes arbitrarios. Reemplazando el ansatz $x_{i}=c_{i}\ e^{\lambda t}$ y resolviendo la ecuación para $\lambda $ obtenemos
\begin{equation}
\lambda =\pm \frac{\sqrt{-A^{2}}+\sqrt{-\left( A^{2}+4B\right) }}{2},\ \pm
\frac{\sqrt{-A^{2}}-\sqrt{-\left( A^{2}+4B\right) }}{2},
\end{equation}
si $A=0$ entonces tenemos un caso isótropo (ya sea trigonométrico o hiperbólico). Entonces $A$ no puede ser cero, luego el factor $\sqrt{-A^{2}}$ nos dice que la única posibilidad de tener soluciones de tipo hiperbólico es que $A$ sea imaginario puro como en (\ref{eomh}).

Es importante considerar (\ref{general}) porque es la ecuación más general posible que se puede obtener de hacer modificaciones en los parámetros del Lagrangiano (\ref{Lncom}).

Otra posibilidad es analizar la álgebra de simetría, a nivel clásico tenemos
\begin{equation}
\left\{ K_{i},K_{j}\right\} =-\tilde{Z}\epsilon _{ij},\quad \left\{
P_{i},P_{j}\right\} =-\frac{1}{R^{2}}\left( R\left( \alpha _{+}-\alpha
_{-}\right) Z+\alpha _{+}\alpha _{-}\tilde{Z}\right) \epsilon _{ij},\quad
\left\{ K_{i},P_{j}\right\} =Z\delta _{ij},  \label{KPK}
\end{equation}%
\begin{equation}
\left\{ K_{i},J\right\} =-\epsilon _{ij}K_{j},\qquad \left\{ P_{i},J\right\}
=-\epsilon _{ij}P_{j,}\qquad \left\{ H,J\right\} =0\,,  \label{KPJ}
\end{equation}%
\begin{equation}
\left\{ K_{i},H\right\} =P_{i}+\frac{\left( \alpha _{+}-\alpha _{-}\right) }{%
R}\epsilon _{ij}K_{j},\qquad \left\{ P_{i},H\right\} =-\frac{\alpha
_{+}\alpha _{-}}{R^{2}}K_{i}\,,  \label{HJPK}
\end{equation}%
de acá vemos algo muy importante: si consideramos el caso anisótropo, $\left(
\alpha _{+}-\alpha _{-}\right) \neq 0$, entonces aparecen dos términos
lineales en $R$. Entonces, desde este punto de vista, considerar el cambio $R\rightarrow iR$ produce problemas con la hermiticidad de algunos generadores de simetría.

\section{Simetría AdS${}_4$ oculta}\label{hidden ads chap}
Los generadores de traslaciones, $P_i$, y boosts, $K_i$, de la álgebra exótica de Newton-Hooke son combinaciones lineales de las integrales chirales ${\cal J}_i^\pm$ (\ref{calJi+-}), que incluyen dependencia explícita del tiempo. Por el contrario, $H$ y $J$ son combinaciones lineales de las integrales cuadráticas (\ref{calJi+-}), (\ref{calJ+-}), que no dependen explícitamente del tiempo. Las simetrías adicionales discutidas al final de la sección \ref{sim chir ani} son también generadas por las integrales que no dependen explícitamente del tiempo, que son cuadráticas en ${\cal J}^\pm_i$.
Si nosotros complementamos esas cuatro integrales cuadráticas que no dependen explícitamente del tiempo, con otras seis, que siempre se pueden construir por combinaciones cuadráticas de ${\cal J}^\pm_i$, y que en el caso genérico si dependen explícitamente del tiempo, obtenemos una simetría mas amplia $AdS_4\sim so(3,2)\sim sp(4)$. Con respecto a los diez generadores de $so(3,2)$, ciertas combinaciones lineales de las cantidades cuadráticas $L_aL_b$, $a=1,\ldots, 4$, $L_a=({\cal J}^+_i,{\cal J}^-_j)$, las integrales ${\cal J}^\pm_i$ forman un spinor de Majorana \cite{Sudar,Horvathy:2007pm}. {}Del punto de vista de la álgebra $so(3,2)$, el Hamiltoniano $H$ y el momento angular $J$ del oscilador exótico anisótropo son simplemente una combinación lineal de los generadores ``espaciales'' del espacio-tiempo abstracto 3+2 dimensional, y del generador de rotaciones en el plano de las dos coordenadas tipo tiempo \cite{Horvathy:2007pm}. Notamos que el sistema (\ref{L_gen}) puede estar relacionado a la versión con gauge fijado del modelo de la partícula invariante de gauge $Sp(4)$ \cite{GomisKamimura}.
\section{Conclusiones del capítulo}\label{conclu eho chap}
Hemos mostrado que el oscilador armónico exótico es caracterizado por exactamente la misma forma chiral de la álgebra exótica de Newton-Hooke (\ref{chiral+},\ref{chiral-}), como en el caso isótropo. En la formulación chiral la anisotropía se revela en las transformaciones de simetría, ver (\ref{trans1ch}) y (\ref{Xt}), que están asociadas con la anisotropía de la estructura del Hamiltoniano (\ref{hj}). En la formulación covariante la anisotropía se manifiesta en las transformaciones de simetría, y en la estructura de la álgebra de simetría. Lo que es importante es que la anisotropía y la estructura exótica se conjugan de tal forma de que la invariancia rotacional no se pierde, a diferencia del oscilador anisótropo usual. La presencia de dos parámetros independientes $\mu_+$ y $\mu_-$ esta detrás de la naturaleza no conmutativa del modelo (\ref{xxnon}).

\chapter{Problema de Landau no conmutativo}\label{senclp chapter}

El problema de Landau no conmutativo (NCLP) ha sido estudiado en el contexto del efecto de Hall cuántico \cite{Duval:2000xr,Duval:2001hu,DH2}, y física de ayones \cite{Horvathy:2004fw}. En el caso sin spin este posee tres fases diferentes \cite{Horvathy:2004fw} \footnote{Las fases sub y supercríticas han sido estudiadas antes en \cite{BNS}. Sin embargo, el marco de trabajo utilizado allí es sólo valido para campo magnético homogéneo, ver \cite{Horvathy:2002wc}. Ni las extensiones conformes ni supersimétricas fueron consideradas en esos artículos.}, una subcrítica ($\beta<1$) y una supercrítica ($\beta>1$) separadas por una fase crítica cuando $\beta=1$. Estas fases son análogas a las de Newton-Hooke exótico. Aquí $\beta=\theta B$ es el producto de un campo magnético homogéneo $B=const$ y el parámetro de no conmutatividad $\theta$. Dimensionalmente tenemos $[B]=\ell^{-2}$ y $[\theta]=\ell^{2}$ con lo cual $\beta$ es adimensional. En el caso genérico de campo magnético inhomogéneo $B(x)$, todas las tres fases pueden estar presentes en el espacio de fase total.

La simetría conforme no relativista \cite{NRConf,HMS,Duval:1993hs} está atrayendo mucho interés, particularmente en el contexto de la correspondencia AdS/CFT \cite{Son:2008ye,Balasubramanian:2008dm,Duval:2008jg,LeivaPl}. El problema de Landau usual ($\theta=0$) tiene simetría conforme no relativista. Lo que nos lleva a la pregunta ¿qué sucede con esta simetría en el caso no conmutativo?. A priori la respuesta no es obvia. La razón es que la simetría conforme involucra invariancia de escalas, mientras que el parámetro de no conmutatividad $\theta$ introduce una escala de longitud independiente a las que introducen los parámetros de masa y de campo magnético. Sin embargo la invariancia de escala es esperable, al menos, en la fase sub crítica, cuyas propiedades son, en muchos aspectos, similares a las del problema de Landau usual con $\theta=0$. Pero no es claro que ocurre con esta simetría en la fase subcrítica, y en particular para una partícula de spin-1/2.

El problema de Landau conmutativo para un electrón no relativista de spin $\half $ tiene supersimetría $N=2$ \cite{LandauSUSY,CIMT}. La simetría conforme $so(2,1)$ es extendida a una simetría superconforme $osp(2|2)$, con el momento angular como carga central \cite{Duval:1993hs,AnPl}. Sin embargo, en el problema de Landau no conmutativo, el momento angular se comporta en una manera esencialmente distinta para las fases sub y supercríticas: toma valores de ambos signos para $\beta<1$, pero de un solo signo cuando $\beta>1$ \cite{Horvathy:2004fw}.

Algunas referencias donde discuten supersimetría y no conmutatividad son \cite{Harikumar:2004qc}-\cite{Geloun}, en particular se discuten la superextensión de Galileo exótico y las simetrías de Schrödinger, y también algunos aspectos del problema de Landau no conmutativo. Cuestiones relacionadas a la superextensión de las simetrías de conformes de Newton-Hooke y el problema de Landau no conmutativo no son consideradas allí.

Este capítulo está dedicado al estudio de las simetrías del problema de Landau no conmutativo superextendido para una particula de spin-1/2 con razón giromagnética $g=2$.

Este capítulo contiene lo siguiente. En la sección \ref{super chap} construimos el sistema supersimétrico para el caso de campo magnético arbitrario. Luego, para el estudio de las fases y las simetrías nos restringiremos al caso homogéneo. En la sección \ref{fases chap} discutimos las fases, mostramos sus diferencias y como la supersimetría ``regulariza'' la fase crítica. En la sección \ref{sim NCLP chap} discutimos la extensión de las simetrías por las dilataciones y las expansiones. En la sección \ref{base alt nclp chap} cambiamos a otra base de generadores de la álgebra de simetría que nos permite observar diferencias esenciales en la subálgebra conforme respecto de las diferentes fases. En la sección \ref{conclu nclp chap} damos las conclusiones del capítulo.

\section{Extensión supersimétrica}\label{super chap}

Construiremos una extension $N=2$ supersimétrica para el caso de campo magnético constante arbitrario, es decir que pueda depender de $x$.

El Lagrangiano de primer orden de una partícula ``exótica'' sin spin, en un campo magnético $B=B(x)$ \cite{NH-NCL,Duval:2000xr,Duval:2001hu}, esta dado por
\begin{equation}
    L=P_i \dot{x}_i - \frac{1}{2m}
    P_i^2-\frac{\theta}{2}\epsilon_{ij}\dot{P}_iP_j-
    \frac{B}{2}\epsilon_{ij}\dot{x}_ix_j,
\end{equation}
del cual salen, a nivel cuántico, las relaciones de conmutación,
\begin{equation}
    \left[ x_{i},x_{j}\right] =i\frac{\theta }{1-\beta }\epsilon _{ij},
    \qquad \left[ x_{i},P_{j}\right]
    =i\frac{1}{1-\beta }\delta _{ij},\qquad
    \left[ P_{i},P_{j}\right] =i\frac{B}{1-\beta}\epsilon
    _{ij},
    \label{conmutadores1}
\end{equation}
donde $\epsilon _{ij}$ es el tensor antisimétrico con $\epsilon_{12}=1$ \footnote{Hemos elegido unidades $\hbar=1=c$ y carga eléctrica igual a uno.}. El parámetro $\theta $ está relacionado con la no conmutatividad de coordenadas y tiene unidades de longitud al cuadrado $\ell^2$. La dimensión del campo magnético, $B(x)$, es $\ell^{-2}$. $\beta=\beta(x)=\theta B(x)$ es entonces adimensional. Los grados de libertad de spin se introducen complementando (\ref{conmutadores1}) con
\begin{equation}
    \left\{ S_{i},S_{j}\right\} =\frac{1}{2}\delta _{ij},
    \qquad \left[ S_{i},x_{j}\right] =0,\qquad
    \left[ S_{i},P_{j}\right] =0. \label{conmutadores2}
\end{equation}
Las coordenadas $x_{i}$ y el momento $P_{i}$ son bosónicos, mientras que los operadores de spin-1/2
$S_{i}$ son variables fermiónicas. Las relaciones (\ref{conmutadores1})-(\ref{conmutadores2})
especifican una estructura cuántica consistente (las identidades de Jacobi son válidas para un
campo magnético inhomogéneo arbitrario \cite{Horvathy:2002wc}).

En el caso crítico $\beta=1$ la dos-forma diferencial bosónica asociada con  (\ref{conmutadores1})
se vuelve degenerada y, en el caso sin spin, requiere una consideración especial \cite{Duval:2000xr,Duval:2001hu,Horvathy:2004fw}.
En el NCLP extendido que consideraremos a continuación, la fase crítica $\beta=1$ se obtendrá al
reducir el sistema al subespacio (infinitamente degenerado) de energía cero.

Los operadores fermiónicos
\begin{equation}
    Q_{1} =\sqrt{\frac{2}{m}}\,P_{i}S_{i},
    \qquad Q_{2} =\sqrt{\frac{2}{m}}\,\epsilon _{ij}P_{i}S_{j}
    \label{Q's}
\end{equation}
generan una $sl(1|1)$ super-álgebra \cite{KhareSukhatme},
\begin{equation}\label{Hgen}
    \left\{ Q_{a},Q_{b}\right\}=2H\delta_{ab} ,
    \qquad
    \left[ Q_{a},H\right]=0,
\end{equation}
donde
\begin{equation}
    H=\frac{1}{2m}P_{i}^{2}-\omega S_{3},
    \label{H}
\qquad
    \omega=\frac{B}{m^*},\qquad
     m^*=m(1-\beta),
\end{equation}
y $S_3$ está definida por
\begin{equation}\label{S3}
    S_{3}=-i\epsilon _{ij}S_{i}S_{j}.
\end{equation}
Escogiendo $H$ como el Hamiltoniano, obtenemos un sistema que generaliza al caso no conmutativo la supersimetría $N=2$
usual de una partícula de spin-$1/2$, con una razón giromagnética $g=2$, en un campo magnético
arbitrario \cite{KhareSukhatme}. El operador $S_3$, como las supercargas
$Q_a$, es una integral de movimiento, la que actúa como un operador $\Gamma $ de $\Z_2$-graduación
para la supersimetría de $N=2$, $\Gamma=2S_3$, $\Gamma^2=1$. Se puede escoger, en particular, una
representación donde  $S_3$ es proporcional a la matriz de Pauli diagonal,
$S_3=\frac{1}{2}\sigma_3$.

\section{Tres fases del NCLP superextendido}\label{fases chap}

{}De ahora en adelante consideraremos campo homogéneo $B=const\neq0$. Es conveniente definir una combinación lineal de los operadores bosónicos $x_i$ y $P_i$,
\begin{equation}\label{calP}
    {\cal P}_i=P_i-B\epsilon_{ij}x_j.
\end{equation}
Para campo magnético distinto de cero, el conjunto formado por los ${\cal P}_i$ y $P_i$ es una alternativa al conjunto inicial de variables bosónicas. La ventaja es que ${\cal P}_i$ conmutan con los $P_j$. {}De la forma del Hamiltoniano (\ref{H}) podemos inferir que ${\cal P}_i$ son integrales de movimiento. Dado que $[x_i,{\cal P}_j]=i\delta_{ij}$ y
\begin{equation}
\label{calPP}
    \left[ \mathcal{P}_{i},\mathcal{P}_{j}\right] =
    -iB\epsilon_{ij},
\end{equation}
los ${\cal P}_i$, $i=1,2$, son identificados como generadores de traslaciones no conmutantes. Otro operador bosónico
\begin{eqnarray}
    J=\frac{1}{2B}\left({\cal P}_i^2-(1-\beta)P_i^2\right)+S_3
    =
    \frac{B}{2}\left(x_i+\frac{1}{B}
    \epsilon_{ij}P_j\right)^2-\frac{1-\beta}{2B}P_i^2+S_3,
    \label{J(x,P)}
\end{eqnarray}
es identificado como el momento angular, dado que este genera las rotaciones,
$[J,R_i]=i\epsilon_{ij}R_j$ for $R_i=x_i,\, P_i,\, S_i$.

Poniendo $\varepsilon_z=\mathrm{sgn}(z)$, definimos los operadores de creación y aniquilación bosónicos
\begin{eqnarray}
    &a^{+}=(a^-)^\dagger=\sqrt{\frac{|1-\beta|}{2\left\vert B \right\vert
    }} \left( P_{1}- i \varepsilon_{B}\varepsilon_{\left( 1-\beta
    \right)} P_{2}\right) ,\qquad
    b^{+}=(b^-)^\dagger=\frac{1}{\sqrt{2\left\vert B\right\vert }}\left(
    \mathcal{P}_{1}+i \varepsilon_B \mathcal{P}_{2}\right),&\label{ab+-}\\
    &f^{+}=(f^-)^\dagger=S_{1}- i \varepsilon_{B}\varepsilon_{\left( 1-\beta \right)}
    S_{2}.&
    \label{f+-}
\end{eqnarray}
Sus relaciones de (anti)-conmutación no triviales son
\begin{equation}
    \left[ a^{-},a^{+}\right] =1,\qquad \left[ b^{-},b^{+}\right] =1,
    \qquad \left\{ f^{-},f^{+}\right\}
    =1.
    \label{crea-anhi}
\end{equation}
La definición de los operadores de creación y aniquilación depende, en vista de (\ref{conmutadores1}) y (\ref{calPP}), en los signos del campo magnético y de la cantidad $1-\beta$ que define la fase del sistema.
La dependencia es incluida en la definición de los operadores fermiónicos. Esto no permite representar el Hamiltoniano en ambas fases sub- ($\beta<1$) y super- ($\beta>1$) críticas en forma universal,
\begin{equation}\label{HNN}
    H=\left\vert \omega \right\vert
    \left( N_{a}+N_{f}\right).
\end{equation}
Aquí hemos introducido los operadores de número bosónicos $N_a=a^+a^-$ y $N_b=b^+b^-$, y fermiónico $N_f=f^+f^-$, con autovalores $n_a,n_b=0,1,\ldots$, y $n_f=0,1$, respectivamente.
El momento angular tiene la forma,
\begin{equation}\label{JNN}
    J=\varepsilon_B \left( N_{b}+\frac{1}{2}\right)
    -\varepsilon_{ B}\varepsilon_{\left( 1-\beta \right)}\left(
    N_{a}+N_{f}\right).
\end{equation}

De acuerdo a (\ref{HNN}), en ambas fases no críticas el sistema tiene un espectro supersimétrico $N=2$ típico, con estado básico correspondiente a $n_a=n_f=0$, y dobletes supersimétricos de energía con números cuánticos $n_a>1$, $n_f=0$, y $n_a-1$, $n_f=1$, respectivamente. Cada nivel del energía tiene una degeneración de energía infinita adicional ($n_b=0,1,\ldots$), asociada a la invariancia de traslaciones generada por los ${\cal P}_i$.

Por otro lado, la ecuación (\ref{JNN}) revela la diferencia esencial entre las dos fases no críticas. En la fase subcrítica, el momento angular toma valores semi-enteros de cualquier signo, mientras que en la fase supercrítica valores semi-enteros de sólo un signo (el signo del campo magnético).

Es útil notar que la diferencia entre las dos fases no-críticas se revela también en otro aspecto. Procediendo de la estructura cuántica (\ref{conmutadores1}), uno puede construir las variables \emph{vectoriales} $q_i$ y $p_i$ con relaciones de conmutación canónicas $[q_i,q_j]=[p_i,p_j]=0$, $[q_i,p_j]=i\delta_{ij}$. Ellos pueden ser presentados, hasta una transformación unitaria, en una forma simple en términos de los operadores mutuamente conmutantes ${\cal P}_i$ y $P_i$,
\begin{equation}
    q_i=\frac{1}{B}\epsilon_{ij}\left({\cal P}_j-\sqrt{1-\beta}\,P_j\right),
    \qquad
    p_i=\frac{1}{2}\left({\cal P}_i+\sqrt{1-\beta}\, P_i\right).
    \label{qp}
\end{equation}
En el límite $B\rightarrow 0$, $q_i$ se vuelve la coordenada canónica para una partícula libre en el plano no conmutativo, $q_i=x_i+\frac{\theta}{2}\epsilon_{ij}P_j$, y $p_i=P_i$, ver referencias \cite{Duval:2000xr,Duval:2001hu,Horvathy:2002vt,Horvathy:2004fw}. La ecuación (\ref{qp}) nos provee con coordenadas canónicas y momentos en ambas fases sub y supercríticas. Sin embargo, en el caso supercrítico, a diferencia del subcrítico, los operadores (\ref{qp}) son \emph{no hermíticos}.

Consistentemente con (\ref{H}), en el límite $\beta\rightarrow 1$ la frecuencia tiende a infinito, $|\omega|\rightarrow \infty$. La fase crítica $\beta=1$ puede ser obtenida por reducción del sistema al nivel de Landau más bajo $E=0$, donde $n_a=n_f=0$. En esta fase el sistema está descrito por las variables oscilatorias $b^\pm$ (que son las complexificaciones de los generadores de traslaciones no conmutantes ${\cal P}_i$), y $H=0$.
Tomando en cuenta las ecuaciones (\ref{S3}) y (\ref{f+-}), encontramos que $S_3=\frac{1}{2}\varepsilon_B\varepsilon_{(1-\beta)}$, es decir, la proyección de spin esta fija. Curiosamente, su valor depende de la fase desde donde se hace la reducción. En la fase crítica los grados de libertad de spin, como también los asociados a las variables oscilatorias $a^\pm$, se congelan, y la supersimetría desaparece.

La algebra de Virasoro puede ser realizada en términos de las integrales bosónicas restantes $b^\pm$ \cite{Kumar}.
Restringiéndonos a las integrales de orden no mayor a $2$ en los operadores $b^\pm$, nos provee con la álgebra de simetría del grupo Euclideano planar, generada por el momento $J=\varepsilon_{B}(N_b+\frac{1}{2})$, y por los generadores de traslación no conmutantes ${\cal P}_i$, mirar ecuaciones (\ref{calPP}) y (\ref{crea-anhi}).

En lo que sigue supondremos $\beta\neq 1$.

\section{Simetrías}\label{sim NCLP chap}

Identificaremos otras simetrías del sistema, adicionales a las ya descritas en la sección previa.
Para esto, consideraremos las ecuaciones Hamiltonianas de movimiento,
\begin{equation}
    \label{xpsevol}
    \dot{x}_{i}=\frac{1}{m^*}P_{i},
    \qquad
    \dot{P}_{i}=\omega \epsilon _{ij}P_{j},
    \qquad
    \dot{S}_{i}=\omega \epsilon _{ij}S_{j}.
\end{equation}
En las fases sub- y super- críticas la evolución es de la misma forma, pero (suponiendo un signo dado para el campo  $B$) el signo de la masa efectiva, $m^*$, (y el de la frecuencia, $\omega$,) es opuesto para  $\beta<1$ y $\beta>1$. Sorprendentemente, se puede producir el mismo efecto mediante una reflexión temporal, $t\rightarrow -t$. La integración de las ecuaciones de movimiento entrega,
\begin{equation}
     x_{i}(t) = \frac{1}{B}\left(\epsilon_{ij}{\cal P}_j
    -\Delta _{jk}^{-1}(t) P_{k}(0)\right) ,\quad
    P_{i}(t) = \Delta _{ij}^{-1}(t) P_{j}(0) ,\quad
    S_{i}(t) = \Delta _{ij}^{-1}(t) S_{j}(0) .
\end{equation}
Donde $\Delta_{ij}(t)=\cos{\omega t} \, \delta_{ij}-\sin{\omega t} \, \epsilon_{ij}$ es una matriz de rotación, $\Delta_{ij}^{-1}(t)=\Delta_{ji}(t)=\Delta_{ij}(-t)$ es su inversa, y $\frac{1}{B}\epsilon_{ij}{\cal P}_j=\frac{1}{B}\epsilon_{ij}{\cal P}_j(0)=x_i(0)+\frac{1}{B}\epsilon_{ij}P_{j}(0)$.

Ahora, identificaremos los generadores de boost como las integrales que, cuando actúan sobre $x_i(0)$ y $\dot{x}_i(0)$, producen la forma necesaria de las transformaciones infinitesimales, $[x_{i}(0) ,\mathcal{K}_{j}] =0,$ $[ \dot{x}_{i}( 0) ,\mathcal{K} _{j}] =-i\delta _{ij}$. Esto da $\mathcal{K}_{i}=m^{*}\left( x_{i}\left( 0\right) +\theta \epsilon _{ij}P_{j}\left( 0\right) \right)$. Usando la solución de las ecuaciones de movimiento, los generadores pueden ser rescritos en términos de las variables $x_i$ y $P_i$,
\begin{equation}
    \mathcal{K}_{i}=m^{*}\left(x_{i}+\frac{1}{B}
    \epsilon_{ij}\rho_{jk}P_{k}\right),\qquad
    \rho_{jk}\equiv \left( \delta_{jk}-\left(1-\beta\right)
    \Delta _{jk}\left( t\right) \right).
    \label{K_i}
\end{equation}
Los generadores de boost (\ref{K_i}) son integrales de movimiento dinámicas, en el sentido que dependen explícitamente del tiempo, $\frac{d}{dt}{\cal K}_j=\frac{\partial}{\partial t}{\cal K}_j+\frac{1}{i}[{\cal K}_j,H]=0$.

Para $\theta=0$,   (\ref{K_i}) reproduce correctamente los generadores de boosts del problema de Landau usual \cite{OP}. En el caso libre $B=0$, (\ref{K_i}) se reduce a los generadores de boost usuales de una partícula libre en el plano no-conmutativo.

Los conmutadores entre $\mathcal{P}_{i}$ y $\mathcal{K}_{i}$ están dados por
\begin{equation}
    \left[ \mathcal{K}_{i},\mathcal{K}_{j}\right] =-i\theta m^{\ast
    2}\epsilon _{ij},\qquad \left[
    \mathcal{K}_{i},\mathcal{P}_{j}\right] =im^{\ast }\delta
    _{ij},\qquad \left[ \mathcal{P}_{i},\mathcal{P}_{j}\right]
    =-im^*\omega\epsilon _{ij},  \label{K_P}
\end{equation}
y sus relaciones de conmutación con  $H$ son
\begin{equation}
    \left[ \mathcal{P}_{i},H\right] =0,\qquad \left[ \mathcal{K}_{i},H\right] =i\left(
    \mathcal{P}_{i}+\omega \epsilon _{ij}\mathcal{K}_{j}\right). \label{P_K_H}
\end{equation}
Las integrales bosónicas  $H$, $J$, $\mathcal{P}_{i}$ y $\mathcal{K}_{i}$ generan la álgebra de simetría de Newton-Hooke, en la cual $\omega$ es un parámetro, mientras $C=m^{\ast }$ y $\tilde{C}=\theta m^{\ast 2}$ son cargas centrales \cite{NH-NCL}. Los conmutadores con las super-cargas  $Q_a$ muestran que ellas son invariantes bajo traslaciones, pero no bajo boosts,
\begin{equation}
    \left[ \mathcal{P}_{i},Q_{a}\right] =
    0,\qquad \left[ \mathcal{K} _{i},Q_{a}\right]=i\left(
    \delta_{a1}\Sigma
    _{i}+\delta_{a2}\epsilon _{ij}\Sigma _{j}\right).
    \label{P,K-Q}
\end{equation}
Hemos identificado un nuevo generador vectorial fermiónico $\Sigma _{i}=\left( 1-\beta \right)
\sqrt{2m}S_{i}\left( 0\right)$. Esto nuevamente corresponde a una integral dinámica,
\begin{equation}
    \Sigma _{i}=\left( 1-\beta \right)
    \sqrt{2m}\Delta _{ij}\left( t\right) S_{j}.
\end{equation}
{}Las relaciones de anti-conmutación,
\begin{equation}
    \left\{ \Sigma _{i},\Sigma _{j}\right\} ={\cal C}\delta _{ij},\qquad
    {\cal C}=C-\omega \tilde{C}=m(1-\beta)^2>0,
    \label{sigma-sigma}
\end{equation}
implican que  $\Sigma_i$ es la raíz cuadrada de una adecuada combinación lineal de cargas centrales positivamente definidas.

Las relaciones de conmutación de $\Sigma _{i}$
 con $\mathcal{K}_{i}$, $\mathcal{P}_{i}$, $J$ y $H$ son
\begin{equation}
    \left[\Sigma _{i},\mathcal{K}_{j}\right] =0,
    \qquad \left[ \Sigma_{i}, \mathcal{P}_{j}\right] =0,\qquad
    [J,\Sigma_i]=i\epsilon_{ij}\Sigma_j,\qquad
    \left[ \Sigma _{i},H\right] =i\omega
    \epsilon _{ij}\Sigma _{j}.
    \label{Sigma-H}
\end{equation}
Como se sigue de, (\ref{P,K-Q}), los $\Sigma_i$ heredan la dependencia temporal explícita de los $\mathcal{K}_i$. Los anti-conmutadores con las super-cargas $Q_a$ son
\begin{equation}
    \left\{ \Sigma _{i},Q_{a}\right\}
    =\left(\delta_{a1}\delta_{ij}-\delta_{a2}\epsilon_{ij}\right)
    \left(
    \mathcal{P}_{j}+\omega \epsilon _{jk}\mathcal{K}_{k}\right) .
\end{equation}
Las integrales $H$, $J$, $\mathcal{P}_i$, $\mathcal{K}_i$, $Q_a$ y $\Sigma_i$ generan una
superálgebra tipo Lie cerrada, extendida centralmente por $C$ y $\tilde{C}$, en la cual la
frecuencia, $\omega$, juega el rol de parámetro.

Ahora nos preguntaremos por la simetría conforme. Para identificar sus generadores, primero consideraremos los análogos directos de los generadores de dilatación y simetría especial conforme (expansión) de una partícula libre \cite{LeivaPl},
\begin{equation}
    D=\frac{1}{4m^{\ast }}\left( \mathcal{K}_{i}\mathcal{P
    }_{i}+\mathcal{P}_{i}\mathcal{K}_{i}\right) ,\qquad
    K=\frac{1}{2m^{\ast } }\mathcal{K}_{i}^{2}.  \label{DK}
\end{equation}
Las relaciones de conmutación,
\begin{eqnarray}
    \left[K,H\right]= 2iD,\qquad
    \left[ D,H\right]=\frac{i}{2}
    \Big(\left( 2-\beta\right)H+\omega
    \left(J+\left( 1-\beta\right)S_{3}-\omega K\right) \Big),
    \label{D-H}
\\[6pt]
    \left[K,D\right]  = \frac{i}{2}
    \Big( \left( 2-\beta \right) K-m^*\theta
    \left( J-\left( 1-\beta \right) S_{3}+m^*\theta H\right) \Big),  \label{K-D}
\end{eqnarray}
generalizan las relaciones ``exóticas'' encontradas anteriormente para una campo ``Moyal'' de una partícula sin spin libre \cite{HMS}.

En contraste con el caso libre y sin spin usual, $\theta=B=0$, los conmutadores no cierran a un álgebra  $so(2,1)$. Pero, ya que  $J$ y $S_3$ conmutan con $H$, $D$ y $K$, uno podría concluir que tenemos una especie de extensión central de $so(2,1)$. Como veremos a continuación, esto es sólo cierto en la fase subcrítica, mientras en la fase supercrítica la álgebra no compacta $so(2,1)$ se cambia por la álgebra compacta  $so(3)$. Ya que $\omega$ juega el rol de parámetro en la simetría de Newton-Hooke en el caso conmutativo $\theta=0$ las relaciones (\ref{D-H}), (\ref{K-D}) corresponden a un álgebra $so(2,1)$, centralmente extendida por $J+S_3$. En cambio, en el caso no conmutativo $\theta\neq 0$, no tenemos una estructura algebraica de Lie, debido a la dependencia de los coeficientes en  (\ref{D-H}) y (\ref{K-D}) de las cargas centrales de la simetría de Newton-Hooke exótica. En la siguiente sección consideraremos una elección alternativa de los generadores que linealizan la estructura super-algebraica.

Para identificar la estructura super-algebraica completa, también necesitaremos los conmutadores de $D$ y $K$ con los otros generadores de la simetría super-extendida  de Newton-Hooke exótica. Los conmutadores con $\mathcal{P}_{i},\ \mathcal{K}_{i}$ y $\Sigma _{i}$ son
\begin{equation}\label{KPKSig}
    \left[K, \mathcal{P}_{i}\right]  =i\mathcal{K}_{i},
    \qquad
    \left[K, \mathcal{K}_{i}\right]  =im^*\theta\epsilon _{ij}\mathcal{K}_{j},\qquad
    \left[K, \Sigma _{i}\right]  =0,
\end{equation}
\begin{equation}\label{DPK1}
    \left[D, \mathcal{P}_{i}\right]  =\frac{i}{2 }
    \left( \mathcal{P}_{i}+\omega \epsilon _{ij}\mathcal{K}_{j}\right)
    ,\qquad
    \left[D, \mathcal{K}_{i}\right]
    =\frac{i}{2}m^*\theta\epsilon_{ij}
    \left(\mathcal{P}_{j}+\omega \epsilon _{jk}\mathcal{K}_{k}\right)
    ,\qquad
    \left[D, \Sigma _{i}\right]  =0,
\end{equation}
donde, nuevamente, está manifiesta la no-linealidad. Los conmutadores con las supercargas $Q_a$,
\begin{equation}\label{QKD}
    \left[K, Q_a\right] =iZ_a,\qquad
    \left[D, Q_a\right] =\frac{i}{2}
    \Big( (1-\beta)Q_a+\omega \epsilon
    _{ab}Z_b\Big),
\end{equation}
generan un nuevo conjunto de supercargas escalares $Z_a$,
\begin{equation}\label{Zdef}
    Z_{1} =\frac{1}{m^{\ast } }
    \mathcal{K}_{i}\Sigma_{i}, \qquad
    Z_{2} =\frac{1}{m^{\ast } }
    \epsilon _{ij}\mathcal{K} _{i}\Sigma _{j},
\end{equation}
con anticonmutadores
\begin{equation}
    \left\{ Z_{a },Z_{b }\right\} =2 (1-\beta)
    \delta _{ab}\left( K+m^*\theta
    S_{3}\right).
\end{equation}
Las relaciones de (anti)-conmutación de  $Z_a$ con otros generadores de simetría,
\begin{equation}
    \left[ Z_{a },H\right] =iQ_{a},\qquad
    \left[ Z_{a},K\right] =im^*\theta
    \epsilon _{ab }Z_{b },\qquad
    \left[ Z_{a },D\right] =\frac{i}{2}(1-\beta)\left( Z_{a}+m^*\theta
    \epsilon _{ab }Q_{b}\right),
\end{equation}
\begin{equation}
    \left[Z_a, \mathcal{P}_{i}\right]  = i\left(
    \delta_{a1}\delta_{ij}+\delta_{a2}\epsilon
    _{ij}\right)\Sigma _{j}, \qquad
    \left[Z_a, \mathcal{K}_{i}\right]  = im^*\theta \left(\delta_{a1}
    \epsilon _{ij}-\delta_{a2}\delta_{ij}\right)\Sigma _{j},
    \label{ZZ1}
\end{equation}
\begin{equation}
    \left\{Z_a, \Sigma _{i}\right\}  = (1-\beta)\left(\delta_{a1}
    \delta_{ij}-\delta_{a2}\epsilon _{ij}\right)
    \mathcal{K}_{j},
    \qquad
    \left\{Z_a, Q_{b}\right\} =2D\ \delta _{ab}-
    \left( J+\left( 1-\beta \right) S_{3}-\omega
    K +m^*\theta H\right)\epsilon _{ab},
    \label{Q,Z}
\end{equation}
muestran que se obtiene una estructura super-algebraica cerrada, y no se generan nuevas integrales
independientes. En el caso conmutativo $\theta=0$ esta estructura super-algebraica se reduce a la
super-álgebra de  Schr\"odinger estudiada en \cite{Duval:1993hs}.

\section{Base alternativa}\label{base alt nclp chap}

En esta sección mostraremos que, cambiando la base de los generadores de simetría conformes, la no linealidad debida a la presencia de las cargas centrales en los coeficientes en las relaciones de (anti)conmutación puede ser removida, y con eso obtenemos una extension superalgebraica \emph{de Lie} de la simetría conforme. El procedimiento de linealización puede ser extendido para incluir también los generadores de traslaciones y boosts, y las supercargas vectoriales $\Sigma_i$.

Consideraremos
\bea
    \mathcal{J}^0&=&\frac{1}{\omega}H+\frac{1}{2}(J+S_3),\label{J0}\\
    \mathcal{J}^1&=&\frac{\varepsilon_{1-\beta}}{2\sqrt{|1-\beta|}}
    \left(\frac{2-\beta}{\omega}H+J+(1-\beta)S_3-\omega
    K\right),\label{J1}\\
    \mathcal{J}^2&=&\frac{1}{\sqrt{|1-\beta|}}\,D,\label{J2}
\eea
en vez de $H$, $D$, $K$. Notar que ${\cal J}^1$ y ${\cal J}^2$ dependen no trivialmente del parámetro de no conmutatividad $\theta$ vía $\beta$. Todas las tres integrales (\ref{J0}-\ref{J2}) dependen sólo de las variables bosónicas $a^\pm$, $b^\pm$ pero no de los operadores fermiónicos $f^\pm$,
\begin{eqnarray}
    &\mathcal{J}^0=\frac{1}{2}\varepsilon_B(N_b+\varepsilon_{1-\beta}N_a)+
    \frac{1}{4}\varepsilon_B(1+\varepsilon_{(1-\beta)}),&
    \\[6pt]
    &\mathcal{J}^1_{sub}=\frac{1}{2}\varepsilon_B\left(a^+(0)b^++a^-(0)b^-\right),\qquad
    \mathcal{J}^2_{sub}=\frac{-i}{2}\left(a^+(0)b^+-a^-(0)b^-\right),&
    \\[6pt]
    &\mathcal{J}^1_{sup} =
    \frac{1}{2}\varepsilon_B\left(a^+(0)b^-+a^-(0)b^+\right),\qquad
    \mathcal{J}^2_{sup} =
    \frac{i}{2}\left(a^+(0)b^--a^-(0)b^+\right),&
\end{eqnarray}
donde los subíndices \emph{sub} y \emph{sup} se refieren a las fases sub- y super- críticas, y $a^\pm(0)=a^\pm e^{\mp i|\omega|t}$. Las relaciones de conmutación de $\cal{J}^\mu$, $\mu=0,1,2$, son dadas por
\begin{equation}\label{so}
    \left[ \mathcal{J}^{1},\mathcal{J}^{2}\right]
    =-i\varepsilon _{1-\beta } \mathcal{J}^{0},\qquad
    \left[ \mathcal{J}^{2},\mathcal{J}^{0}\right]
    =i \mathcal{J}^{1},\qquad \left[
    \mathcal{J}^{0},\mathcal{J}^{1}\right] =i \mathcal{J}^{2}.
\end{equation}
En el casi sub-crítico es la álgebra $so(2,1)\sim su(1,1)$, pero en el super-crítico los operadores (\ref{J0}-\ref{J2}) forman $so(3)\sim su(2)$.

Junto a la nueva base de operadores bosónicos, definimos también las siguientes combinaciones de las supercargas escalares $Q_a$ y $Z_a$,
\begin{eqnarray}
    {\cal Q}_a^+&=&\frac{1}{2|\omega|}
    \left((1+\sqrt{|1-\beta|})Q_a+\frac{\omega}{\sqrt{|1-\beta|}}\epsilon_{ab}Z_b\right),\label{comsusy1_1}\\
    {\cal Q}_a^-&=&\frac{1}{2|\omega|}\epsilon_{ab}\left((1-\sqrt{|1-\beta|})Q_b
    -\frac{\omega}{\sqrt{|1-\beta|}}\epsilon_{bc}Z_c\right).\label{comsusy1_2}
\end{eqnarray}
Entonces, en la fase subcrítica, obtenemos las relaciones de (anti-)conmutación
\begin{equation}\label{VV WW sub}
    \left\{ {\cal Q}_a^+,{\cal Q}_b^+\right\} =2\varepsilon_{B}\delta_{ab}
    \left(\mathcal{J} ^{0}+\mathcal{J}^{1}\right), \
    \qquad \left\{ {\cal Q}_a^-,{\cal Q}_b^-\right\}
    =2\varepsilon_{B}\delta_{ab}\left(\mathcal{J}^{0}-
    \mathcal{J}^{1}\right),
\end{equation}
\begin{equation}\label{VW sub}
    \left\{ {\cal Q}_a^+,{\cal Q}_b^-\right\} =
    2\varepsilon_{B}\left(\delta _{ab} \mathcal{J} ^{2}
    +\epsilon_{ab}\frac{1}{2}\left( J+S_{3}\right)\right),
\end{equation}
\begin{equation}\label{J,V}
    \left[\mathcal{J}^{1},{\cal Q}_a^\pm\right]
    =-\frac{i}{2}{\cal Q}_a^\mp,\qquad
    \left[\mathcal{J}^{2},{\cal Q}_a^\pm\right]=\pm\frac{i}{2}{\cal
    Q}_a^\pm.
\end{equation}
En cambio en el caso super-crítico encontramos
\begin{equation}\label{VV WW sup}
    \left\{ {\cal Q}_a^+,{\cal Q}_b^+\right\} =
    2\varepsilon _{B}\delta _{ab}
    \left( \frac{1}{2}\left(J+S_{3}\right)
    -\mathcal{J}^{1}\right),
    \qquad \left\{ {\cal Q}_a^-,{\cal Q}_b^-\right\} =
    2\varepsilon _{B}\delta _{ab} \left(
     \frac{1}{2}\left(J+S_{3}\right)
    +\mathcal{J}^{1}\right),
\end{equation}
\begin{equation}
\label{VWsup}
    \left\{ {\cal Q}_a^+,{\cal Q}_b^-\right\} =
    2\varepsilon_{B}\left(\epsilon_{ab}\mathcal{J}^{0}\ -
    \delta _{ab}\mathcal{J}^{2}\right),
\end{equation}
\begin{equation}
    \left[
    \mathcal{J}^{1},{\cal Q}_a^\pm\right]
    =\mp\frac{i}{2}\epsilon _{ab}{\cal Q}_b^\pm,
    \qquad \left[ \mathcal{J}^{2},{\cal Q}_a^\pm\right] =
    \frac{i}{2}\epsilon_{ab}{\cal Q}_b^\mp.
\end{equation}
En ambas fases, también tenemos
\begin{equation}
      \left[ \mathcal{J}^{0},{\cal Q}_a^\pm
    \right] =\pm\frac{i}{2}{\cal Q}_a^\mp,\qquad
    \left[ S_{3},{\cal Q}_a^\pm\right] =
    i\epsilon _{ab}{\cal Q}_b^\pm.
    \label{SVW}
\end{equation}

Queremos enfatizar que todas estas relaciones son lineales, como fue dicho antes.

Notar que los conmutadores de las super-cargas con ${\cal J}^0$ son los mismos para las fases sub- y super- críticas, pero no así las con ${\cal J}^1$ y ${\cal J}^2$. Las relaciones (\ref{VV WW sub}), (\ref{VW sub}) y (\ref{VV WW sup}), (\ref{VWsup}) son transformadas mutuamente haciendo el cambio ${\cal J}^0 \leftrightarrow \frac{1}{2}(J+S_3)$, ${\cal J}^1 \leftrightarrow -{\cal J}^1$ y ${\cal J}^2 \leftrightarrow -{\cal J}^2$.

Las relaciones (\ref{so}) y (\ref{VV WW sub}-\ref{SVW}) muestran que el sistema está descrito por la simetría super-conforme $osp(2|2)$ centralmente extendida en la fase subcrítica, y por la análoga extension super-algebraico de Lie de la simetría compacta $so(3)$ en la fase supercrítica. En ambos casos el momento angular $J$ juega el rol de carga central en esas superálgebras, mientras que ${\cal R}=2S_3$ es el generador de simetría-$R$.

Tomemos
\begin{equation}\label{defES}
    {\cal E}^+_i=\sqrt{\frac{|1-\beta|}{2|\omega|}}\, {\cal P}_i,\qquad
    {\cal E}^-_i=\frac{1}{\sqrt{2|\omega|}}
    \left({\cal P}_i+\omega\epsilon_{ij}{\cal K}_j\right)
\end{equation}
en vez de los generadores de traslaciones y boosts. Los (anti-)conmutadores no triviales de las integrales ${\cal E}^\pm_i$ y $\Sigma_i$ entre sí mismos son dados por (\ref{sigma-sigma}) y
\begin{equation}\label{EES}
    [{\cal E}^+_i,{\cal E}^+_j]=-\frac{i}{2}\varepsilon_{B}{\cal C} \epsilon_{ij},
    \qquad [{\cal E}^-_i,{\cal E}^-_j]=\frac{i}{2}
    \varepsilon_{B}\varepsilon_{(1-\beta)}{\cal C} \epsilon_{ij}.
\end{equation}
Sus relaciones de (anti-)conmutación con el resto de generadores,
${\cal J}^\mu$, $J$, $S_3$ y ${\cal Q}^\pm_\alpha$, son
\begin{eqnarray}\label{J0ES}
    &[{\cal J}^0,{\cal E}^\pm_i]=\pm\frac{i}{2}\epsilon_{ij}{\cal
    E}^\pm_j,\qquad [J,{\cal E}^\pm_i]=i\epsilon_{ij}{\cal E}^\pm_j,
    \qquad
     [S_3,{\cal E}^\pm_i]=0,&\\
    &[{\cal E}^+_i,{\cal J}^\mu]=-\frac{i}{2}\varepsilon_{(1-\beta)}
    \left(\delta^{\mu 1}\epsilon_{ij}+\delta^{\mu 2}\delta_{ij}\right){\cal
    E}^-_j,
    \quad
    [{\cal E}^-_i,{\cal J}^\mu]=\frac{i}{2}
    \left(\delta^{\mu 1}\epsilon_{ij}-\delta^{\mu 2}\delta_{ij}\right){\cal
    E}^+_j,\,\, \mu=1,2,&
     \label{EJ12}\\
     &[S_3,\Sigma_i]=i\epsilon_{ij}\Sigma_j,
     \qquad
    [J,\Sigma_i]=i\epsilon_{ij}\Sigma_j,\qquad
    [{\cal J}^\mu,\Sigma_i]=0,&\label{SJmu}
\end{eqnarray}
\begin{eqnarray}
    &[{\cal E}^+_i,{\cal Q}^\pm_a]=\frac{i}{2}\varepsilon_{B}\left(
    -\delta_{a1}\epsilon_{ij}\pm \delta_{a2}\delta_{ij}\right)\Sigma_j,\quad
    [{\cal E}^-_i,{\cal Q}^\pm_a]=
    \frac{i}{2}\varepsilon_{B}\varepsilon_{(1-\beta)}\left(
    \pm\delta_{a1}\epsilon_{ij}\mp\delta_{a2}\delta_{ij}\right)\Sigma_j,&\label{comsusy2}\\
    &\qquad \{\Sigma_i,{\cal Q}^+_a\}=\left(\delta_{a1}\delta_{ij}-\delta_{a2}\epsilon_{ij}\right)({\cal
    E}^+_j+{\cal E}^-_j),\quad
    \{\Sigma_i,{\cal Q}^-_a\}=\left(\delta_{a1}\epsilon_{ij}+\delta_{a2}\delta_{ij}\right)({\cal
    E}^+_j-{\cal E}^-_j).\label{comsusy4}&
\end{eqnarray}
Los índices de generadores bosónicos en (\ref{defES}) corresponden a los signos en sus conmutadores con ${\cal J}^0$ en (\ref{J0ES}).

De acuerdo con (\ref{sigma-sigma}) y (\ref{EES}), las integrales pares, ${\cal E}^\pm_i$, y las impares, $\Sigma_i$, generan álgebras de Heisenberg dos dimensionales con ${\cal C}$ como carga central, la cual, en el caso conmutativo $\theta=0$, se vuelve la masa, $m$.
Esta es la única carga central de la álgebra de simetría completa resultante dada por (\ref{sigma-sigma}), (\ref{so}), (\ref{VV WW sub}-\ref{SVW}), (\ref{EES}-\ref{comsusy4}).  Para $\theta=0$ nos provee con una forma alternativa para la super-álgebra de Schrödinger \cite{Duval:1993hs}.

Permítanos enfatizar que la linealización de las super-extensiones de las simetrías de Newton-Hooke exótica y simetría conforme es lograda mediante la inclusión de la dependencia de $\theta$ en los coeficientes de cambio de base. El parámetro de no conmutatividad es una función de las cargas centrales de la simetría de Newton-Hooke exótica, $\theta=\tilde{C}/C^2$.

Notar que mientras las super-cargas escalares ${\cal Q}^\pm_a$ generan, vía relaciones de anti-conmutación, las integrales escalares pares ${\cal J}^\mu$ y $J+S_3$, sus relaciones de anti-conmutación con $\Sigma_i$ reproducen los generadores vectoriales pares ${\cal E}^\pm_i$. El generador ${\cal R}=2S_3$ de la simetría $R$ está relacionado a la supercarga vectorial $\Sigma_i$ vía uno de los operadores de Casimir de la álgebra, $i\epsilon_{ij}\Sigma_i\Sigma_j+{\cal R}{\cal C}$, que aquí toma valor cero.

\section{Conclusiones del capítulo}\label{conclu nclp chap}

Hemos observado que los niveles de energía son infinitamente degenerados en todas las tres fases, debido a la invariancia bajo las traslaciones magnéticas. En las fases sub- y super- críticas los niveles exitados de energía poseen una degeneración doble adicional, asociada con la super-simetría $N=2$. Gracias a la super-simetría la fase crítica de borde puede ser obtenida vía una reducción simple del sistema al estado básico. En esta fase un grado de libertad bosónico y los fermiónicos se congelan, y SUSY desaparece. La simetría asociada con las integrales, de orden no mayor a dos, en los grados de libertad bosónicos residuales forman el grupo Euclideano de transformaciones en dos dimensiones,  generado por las traslaciones no conmutativas y las rotaciones. El momento angular aquí toma valores de un signo, que coincide con el del campo magnético.

Las dos fases no críticas poseen propiedades esencialmente distintas. El momento angular toma valores semienteros de ambos signos en la subcrítica, y de un solo signo para la supercrítica. En ambas fases las coordenadas y momentos canónicos \emph{vectoriales} pueden ser construidos de las coordenadas y momentos no conmutativos iniciales. En la fase subcrítica ellos son hermíticos, pero no para el caso supercrítico. En la fase subcrítica la parte bosónica de la simetría super-conforme está descrita por la álgebra no compacta $so(2,1)\sim su(1,1)$, mientras que en el caso supercrítico se cambia a $so(3)\sim su(2)$.

Cuando tratamos de unificar la extensión central doble de la simetría super-extendida de Newton-Hooke con la simetría conforme, las constantes de estructura de la super-álgebra resultante se transforman en ciertas funciones dependientes de las cargas centrales. Una estructura lineal super-algebraica de Lie puede ser obtenida mediante el cambio de base con coeficientes dependientes del parámetro de no conmutatividad $\theta$. La super-álgebra de Lie resultante tiene sólo una carga central. 
\chapter{Conclusiones generales y problemas abiertos}

La simetría de Newton-Hooke oscilatorio aparece como límite no relativista de AdS. Los sistemas físicos que realizan esta simetría están asociados a representaciones proyectivas del grupo de Newton-Hooke. Las álgebras de simetría de estos grupos están centralmente extendidas por un generador asociado a la masa de la partícula.

La simetría exótica de Newton-Hooke esta descrita por la álgebra de Newton-Hooke extendida por dos cargas centrales, lo cual sólo es posible en 2+1 dimensiones de espacio tiempo. Como consecuencias de la estructura exótica aparecen coordenadas no conmutativas, representaciones proyectivas adicionales y una estructura de fases donde el sistema exhibe propiedades físicas esencialmente distintas. La fase subcrítica reproduce en el límite plano la simetría de Galileo exótica. En la fase crítica el sistema posee menos grados de libertad físicos. La fase supercrítica tiene algunas propiedades peculiares, la energía tiene un espectro no acotado y el momento angular uno acotado. Esto está relacionado con una simetría de dualidad que relaciona la fase sub con la supercrítica. En este sentido la simetría de Newton-Hooke exótica trigonométrica (oscilante) es más ``rica'' que la hiperbólica (expansiva), o incluso la de Galileo exótica, ya que estas dos últimas sólo viven en una fase subcrítica.

Esta estructura de fases y la no conmutatividad aparece de forma similar en el problema de Landau no conmutativo. La investigación de las simetrías del problema de Landau no conmutativo nos llevo a determinar que su álgebra de simetría es isomorfa a la de Newton-Hooke exótico. Más aún, es posible determinar un sistema uniparamétrico que interpola continuamente ambas simetrías, es el oscilador armónico anisótropo exótico. La gracia de este último es que la anisotropía y la estructura exótica se conjugan de tal modo que la isometría del espacio se preserva (simetría de rotaciones), con lo cual puede ser asociado a un sistema de partícula no relativista. El oscilador anisótropo también posee de forma genérica coordenadas no conmutativas y la estructura de fases.

Aunque la álgebra de simetría del problema de Landau es isomorfa a la de Newton-Hooke exótico, esto no quiere decir que ambos sistemas sean físicamente equivalentes. De hecho hay una propiedad muy simple del problema de Landau no conmutativo que destaca la diferencia entre ambos sistemas, la energía es acotada de abajo para cualquiera de sus fases subcrítica, crítica y supercrítica. En virtud de las aplicaciones que ha mostrado el problema de Landau no conmutativo en el efecto de Hall cuántico, aparece de forma natural el estudio de las simetrías extendidas por los generadores conformes de Schrödinger, y también superextendidas de tal forma de incorporar grados de libertad de spin. La álgebra de (super)simetría resultante contiene ciertas funciones no lineales de los generadores, por lo que su (super)estructura de Lie se rompe. Sin embargo, debido a que la no linealidad aparece en las cargas centrales (constantes), es posible reescalar los generadores de tal modo de recuperar la superestructura de álgebra de Lie. Como resultado queda sólo una carga central independiente.\\

\bigskip\noindent{\it {\Large Algunos problemas abiertos}}

\bigskip A continuación daremos un listado de problemas que pueden extender el trabajo de esta tesis, y de problemas abiertos relacionados

\begin{itemize}

\item Uno puede esperar que, mediante el método de las realizaciones no lineales, sea posible construir un Lagrangiano relativista que en cierto límite se reduzca al modelo la partícula con simetría de Newton-Hooke exótica. Entonces, una pregunta interesante es a qué corresponden, a nivel relativista, las fases que exhibe el sistema de Newton-Hooke exótico.

\item El Hamiltoniano del sistema de Newton-Hooke exótico en la fase supercrítica tiene formalmente la misma estructura que el Hamiltoniano correspondiente al sistema con altas derivadas de Pais-Uhlenbeck \cite{Pais:1950za}. Este sistema ha sido considerado durante muchos años como un ejemplo de un modelo con estados fantasmas, y por tanto físicamente inaceptable. Recientemente, se mostró que, usando las reglas de mecánica cuántica pseudohermítica, es posible obtener una teoría cuántica razonable \cite{Bender:2008vh}. Es interesante estudiar si es posible aplicar las ideas de mecánica cuántica pseudohermítica al sistema de Newton-Hooke exótico, y observar, por ejemplo, que nos dice sobre la estructura del espacio de Hilbert y del espectro del momento angular.

\item Otro problema interesante es el estudio de una posible relación entre la simetrías de Newton-Hooke exótica y algunas soluciones de relatividad general. Como candidato aparece como una posibilidad el agujero negro BTZ \cite{Banados:1992wn}, ya que este es localmente como un espacio AdS$_3$, y además porque el agujero BTZ posee distintas fases en dependencia de los valores de su masa y su momento angular. En esta linea, se ha visto que hay una relación entre una familia de soluciones de relatividad general de tipo Gödel con el problema de Landau \cite{Drukker:2003mg,Hik}. Si tal relación existiese, sería interesante clarificar a qué corresponde la simetría de dualidad del sistema con simetría de Newton-Hooke exótica en el agujero BTZ.

\item El problema de Landau no conmutativo ha sido relacionado con el efecto de Hall cuántico, mediante la ``substitución de Peierls''. En el efecto de Hall cuántico aparecen cuasi-partículas anyonicas que llevan cargas eléctricas fraccionarias. Este tipo de partículas son asociadas a una razón giromagnética anómala $g\ne 2$. En \cite{Klishevich:2001fu} se mostró que es posible generalizar la supersimetría del Hamiltoniano de Pauli para los casos $g=2k$, con $k$ un entero positivo, obteniéndose superálgebras no lineales. En este contexto es muy interesante el estudio de las extensiones supersimétricas no lineales \cite{AIS,Pl_Para,Klishevich:2000dp} del problema de Landau no conmutativo.

\item Una pregunta abierta es ¿qué esta detrás de la no linealidad que aparece en el sector conforme de la álgebra de simetría del problema de Landau no conmutativo?. Mediante un cambio de base se mostró que era posible linearizar la álgebra, y cómo esto nos conducía a álgebras de tipo $so(2,1)$ o $so(3)$ en las fases sub y supercrítica respectivamente. Es primera vez que se encuentra una álgebra conforme de tipo compacta, por lo que es interesante realizar un estudio más profundo.

\item La simetría de Schrödinger que posee el problema de Landau no conmutativo y el sistema con simetría de Newton-Hooke exótica nos permite situar estos modelos en el marco de la correspondencia AdS/CFT no relativista, en este contexto aparecen varios problemas interesantes. La identificación de geometrías aparece vinculada a la extension central de la álgebra de Schrödinger \cite{Duval:1984cj,Duval:2008jg}, la cual corresponde la masa y es la única admisible en dimensiones mayores o iguales a cuatro. Entonces la pregunta natural es si es posible identificar nuevas geometrías en 2+1 dimensiones, donde tenemos una extensión central adicional. De hecho la respuesta a esta interrogante quizás puede ayudar a esclarecer los problemas mencionados en los dos puntos anteriores.
\end{itemize}


\appendix
\renewcommand{\appendixname}{Apéndice}
\renewcommand{\appendixtocname}{Apéndices}
\noappendicestocpagenum
\addappheadtotoc




\chapter{SO(4) Inhomogéneo}\label{Inhso(4)}

La álgebra de $E(4)$ está dada por
\begin{equation}
\left[ M_{ab},M_{cd}\right] =\eta _{ac}M_{bd}-\eta _{ad}M_{bc}-\eta_{bc}M_{ad}+\eta _{bd}M_{ac},
\end{equation}
\begin{equation}
\left[ M_{ab},T_{c}\right] =\eta _{ac}T_{b}-\eta _{bc}T_{a},
\end{equation}
\begin{equation}
\left[ T_{a},T_{b}\right] =0,
\end{equation}
donde $\eta _{ab}=diag(+,+,+,+)$.

Es posible reescribir esta álgebra haciendo las identificaciones
\begin{equation}
M_{12}=J_{1},\qquad M_{23}=J_{2},\qquad M_{31}=J_{3},
\end{equation}
\begin{equation}
M_{43}=P_{1},\qquad M_{41}=P_{2},\qquad M_{42}=P_{3},
\end{equation}
\begin{equation}
\left\{ T_{1},T_{2},T_{3},T_{4}\right\} =\left\{
K_{2},K_{3},K_{1},-H\right\} ,
\end{equation}
donde $J_{i}$ son rotaciones, $P_{i}$ traslaciones, $K_{i}$ boosts y $H$ traslaciones en el tiempo de un espacio en 3+1 dimensiones que satisfacen
\begin{equation}
\left[ P_{i},K_{j}\right] =\delta _{ij}H,\qquad \left[ P_{i},H\right] =K_{i},
\end{equation}
\begin{equation}
\left[ J_{i},K_{j}\right] =\epsilon _{ijk}K_{k},\qquad \left[ J_{i},H\right]=0,
\end{equation}
\begin{equation}
\left[ J_{i},J_{j}\right] =\epsilon _{ijk}J_{k},
\end{equation}
\begin{equation}
\left[ J_{i},P_{j}\right] =\epsilon _{ijk}P_{k},
\end{equation}
\begin{equation}
\left[ P_{i},P_{j}\right] =\epsilon _{ijk}J_{k}.
\end{equation}
Esta álgebra corresponde a una de las álgebras de ``Para-Poincaré'', del tipo R3, ver la tabla \ref{possible kinematics tabla}.




\begin{thebibliography}{200}
\bibitem{AGKP}
  P.~D.~Alvarez, J.~Gomis, K.~Kamimura y M.~S.~Plyushchay,
  \emph{``(2+1)D exotic Newton-Hooke symmetry, duality and projective phase,''}
  Annals Phys. {\bf 322} (2007) 1556
  [arXiv:hep-th/0702014].

\bibitem{NH-NCL}
  P.~D.~Alvarez, J.~Gomis, K.~Kamimura y M.~S.~Plyushchay,
  \emph{ ``Anisotropic harmonic oscillator, non-commutative Landau problem and exotic
  Newton-Hooke symmetry,''}
  Phys.\ Lett.\  B {\bf 659} (2008) 906
  [arXiv:0711.2644 [hep-th]].

\bibitem{Alvarez:2009nz}
  P.~D.~Alvarez, J.~L.~Cortes, P.~A.~Horvathy y M.~S.~Plyushchay,
  \emph{``Super-extended noncommutative Landau problem and conformal symmetry,''}
  JHEP {\bf 0903} (2009) 034
  [arXiv:0901.1021 [hep-th]].

\bibitem{Leinaas:1977fm}
  J.~M.~Leinaas y J.~Myrheim,
  \emph{``On the theory of identical particles,''}
  Nuovo Cim.\  B {\bf 37} (1977) 1.

\bibitem{Goldin:1981sm}
  G.~A.~Goldin, R.~Menikoff y D.~H.~Sharp,
  \emph{``Representations of A local current algebra in nonsimply connected space and
  the Aharonov-Bohm effect,''}
  J.\ Math.\ Phys.\  {\bf 22}, 1664 (1981).

\bibitem{Wilczek:1981du}
  F.~Wilczek,
  \emph{``Magnetic flux, angular momentum, and statistics,''}
  Phys.\ Rev.\ Lett.\  {\bf 48} (1982) 1144.

\bibitem{Wilczek:1982wy}
  F.~Wilczek,
  \emph{``Quantum mechanics of fractional spin particles,''}
  Phys.\ Rev.\ Lett.\  {\bf 49}, 957 (1982).

\bibitem{Semenoff:1988jr}
  G.~W.~Semenoff,
  \emph{``Canonical quantum field theory with exotic statistics,''}
  Phys.\ Rev.\ Lett.\  {\bf 61} (1988) 517.

\bibitem{Plyushchay:1990cv}
  M.~S.~Plyushchay,
  \emph{``Relativistic model of the anyon,''}
  Phys.\ Lett.\  B {\bf 248} (1990) 107.

\bibitem{Jackiw:1990ka}
  R.~Jackiw y V.~P. Nair,
  \emph{``Relativistic wave equations for anyons,''}
  Phys.\ Rev.\  D {\bf 43}, (1991) 1933.

\bibitem{Pl1}
  M.~S.~Plyushchay,
  \emph{``Relativistic particle with torsion, Majorana equation and fractional spin,''}
  Phys.\ Lett.\  B {\bf 262} (1991) 71;

  \emph{``The Model of relativistic particle with torsion,''}
  Nucl.\ Phys.\  B {\bf 362} (1991) 54.

\bibitem{Pl2}
  M.~S.~Plyushchay,
  \emph{``Deformed Heisenberg algebra, fractional spin fields and supersymmetry without fermions,''}
  Annals Phys.\  {\bf 245} (1996) 339.

\bibitem{mariano1}
  J. Negro, M. A. del Olmo y J. Tosiek,
  \emph{``Anyons, group theory and planar physics,''}
  J. Math. Phys. \textbf{47} (2006) 033508
  [arXiv:math-ph/0512007].



\bibitem{Green:1952kp}
  H.~S.~Green,
  \emph{``A Generalized method of field quantization,''}
  Phys.\ Rev.\  {\bf 90} (1953) 270.

\bibitem{Messiah:1900zz}
  A.~M.~L.~Messiah and O.~W.~Greenberg,
  \emph{``Symmetrization postulate and its experimental foundation,''}
  Phys.\ Rev.\  B {\bf 136} (1964) 248.

\bibitem{Greenberg:1964pe}
  O.~W.~Greenberg,
  \emph{``Spin and unitary spin independence in a paraquark model of baryons and
  mesons,''}
  Phys.\ Rev.\ Lett.\  {\bf 13} (1964) 598.

\bibitem{Greenberg:1963kk}
  O.~W.~Greenberg and A.~M.~L.~Messiah,
  \emph{``Selection rules for parafields and the absence of para particles in
  nature,''}
  Phys.\ Rev.\ B {\bf 138} (1965) 1155.

\bibitem{Greenberg:1989ty}
  O.~W.~Greenberg,
  \emph{``Example of infinite statistics,''}
  Phys.\ Rev.\ Lett.\  {\bf 64} (1990) 705.

\bibitem{Polychronakos:1999sx}
  A.~P.~Polychronakos,
  \emph{``Generalized statistics in one dimension,''}
  Les Houches Lectures,
  [arXiv:hep-th/9902157].


\bibitem{Greenberg:1991ec}
  O.~W.~Greenberg,
  \emph{``Quon statistics,''}
  Phys.\ Rev.\  D {\bf 43} (1991) 4111.

\bibitem{LL1963}
  J.~Levy-Leblond,
  \emph{``Galilei group and nonrelativistic quantum mechanics,''}
  J.\ Math.\ Phys.\  {\bf 4}, (1963) 776.

\bibitem{Bacry:1968zf}
  H.~Bacry y J.~Levy-Leblond,
  \emph{``Possible kinematics,''}
  J.\ Math.\ Phys.\  {\bf 9} (1968) 1605.

\bibitem{LL}
  J.-M.~L\'evy-Leblond,
  \emph{``Galilei group and Galilean invariance,''}
  In: \emph{ Group theory and its applications} (E. M. Loebl Ed.), {\bf II},
  Acad. Press, New York, p. 222 (1972).

\bibitem{JackN}
  R.~Jackiw y V.~P. Nair,
  \emph{``Anyon spin and the exotic central extension of the planar
  Galilei  group,''}
  Phys.\ Lett.\ B {\bf 480} (2000) 237
  [arXiv:hep-th/0003130].

\bibitem{Snyder:1946qz}
  H.~S.~Snyder,
  \emph{``Quantized space-time,''}
  Phys.\ Rev.\  {\bf 71} (1947) 38.


\bibitem{Jackiw:2001dj}
  R.~Jackiw,
  \emph{``Physical instances of noncommuting coordinates,''}
  Nucl.\ Phys.\ Proc.\ Suppl.\  {\bf 108} (2002) 30
  [Phys.\ Part.\ Nucl.\  {\bf 33} (2002\ LNPHA,616,294-304.2003) S6]
  [arXiv:hep-th/0110057].

\bibitem{Connes:1997cr}
  A.~Connes, M.~R.~Douglas y A.~S.~Schwarz,
  \emph{``Noncommutative geometry and matrix theory: Compactification on tori,''}
  JHEP {\bf 9802} (1998) 003
  [arXiv:hep-th/9711162].

\bibitem{Seiberg:1999vs}
  N.~Seiberg y E.~Witten,
  \emph{``String theory and noncommutative geometry,''}
  JHEP {\bf 9909} (1999) 032
  [arXiv:hep-th/9908142].

\bibitem{Szabo:2006wx}
  R.~J.~Szabo,
  \emph{``Symmetry, gravity and noncommutativity,''}
  Class.\ Quant.\ Grav.\  {\bf 23}, R199 (2006)
  [arXiv:hep-th/0606233].

\bibitem{Harikumar:2006xf}
  E.~Harikumar y V.~O.~Rivelles,
  \emph{``Noncommutative gravity,''}
  Class.\ Quant.\ Grav.\  {\bf 23}, 7551 (2006)
  [arXiv:hep-th/0607115].

\bibitem{Chaichian:2001py}
  M.~Chaichian, P.~Presnajder, M.~M.~Sheikh-Jabbari y A.~Tureanu,
  \emph{``Noncommutative Standard Model: model building,''}
  Eur.\ Phys.\ J.\  C {\bf 29}, 413 (2003)
  [arXiv:hep-th/0107055].

\bibitem{Carroll:2001ws}
  S.~M.~Carroll, J.~A.~Harvey, V.~A.~Kostelecky, C.~D.~Lane y T.~Okamoto,
  \emph{``Noncommutative field theory and Lorentz violation,''}
  Phys.\ Rev.\ Lett.\  {\bf 87} (2001) 141601
  [arXiv:hep-th/0105082].

\bibitem{CCGM}
   J.~M.~Carmona, J.~L.~Cortes, J.~Gamboa y F.~Mendez,
  \emph{``Quantum theory of noncommutative fields,''}
  JHEP {\bf 0303} (2003) 058
  [arXiv:hep-th/0301248].

\bibitem{NCS1}
  M.~Chaichian, P.~P.~Kulish, K. Nishijima y A.~Tureanu,
  \emph{``On a Lorentz-invariant interpretation of noncommutative space-time and its
  implications on noncommutative QFT,''}
  \textsl{ Phys. Lett. }  {\bf B 604} (2004) 98
  [arXiv:hep-th/0408069];

  M.~Chaichian, P.~Presnajder y A.~Tureanu,
  \emph{``New concept of relativistic invariance in NC space-time: twisted  Poincare
  symmetry and its implications,''}
  \textsl{Phys. Rev. Lett.}  {\bf 94} (2005) 151602
  [arXiv:hep-th/0409096].

\bibitem{NCS2}
  B.~Chakraborty, S.~Gangopadhyay y A.~Saha,
  \emph{``Seiberg-Witten map and Galiliean symmetry violation in a non-commutative planar system,''}
  \textsl{Phys. Rev. }  {\bf D 70} (2004) 107707
  [arXiv:hep-th/0312292];

  R.~Banerjee, B.~Chakraborty y K.~Kumar,
  \emph{``Noncommutative gauge theories and Lorentz symmetry,''}
  \textsl{Phys. Rev. }  {\bf D 70} (2004) 125004
  [arXiv:hep-th/0408197].

\bibitem{NCS3}
  O.~Bertolami y L.~Guisado,
  \emph{``Noncommutative field theory and violation of translation
  invariance,''}
  \textsl{JHEP} {\bf 0312} (2003) 013
  [arXiv:hep-th/0306176];

  X.~Calmet,
  \emph{``Space-time symmetries of noncommutative spaces,''}
  \textsl{Phys. Rev. }  {\bf  D 71} (2005) 085012
  [arXiv:hep-th/0411147].

\bibitem{NCS4}
  C.~Gonera, P.~Kosinski, P.~Maslanka y S.~Giller,
  \emph{``Space-time symmetry of noncommutative field theory,''}
  \textsl{Phys. Lett.} {\bf B 622} (2005) 192
  [arXiv:hep-th/0504132];

  \emph{``Global symmetries of noncommutative space-time,''}
   \textsl{Phys. Rev.} {\bf D 72} (2005) 067702
  [arXiv:hep-th/0507054].


\bibitem{Luk}
  J.~Lukierski, H.~Ruegg, A. Nowicki y V. N.~Tolstoi,
  \emph{``Q deformation of Poincar\'e algebra,''}
  \textsl{Phys. Lett.} {\bf  B 264} (1991) 331;

  J.~Lukierski, A. Nowicki y H.~Ruegg,
  \emph{``New quantum Poincar\'e algebra and k deformed field theory,''}
  \textsl{  Phys. Lett.} {\bf B 293} (1992) 344.

\bibitem{Az}
  J.~A.~de Azcarraga y J.~C.~Perez Bueno,
  \emph{``Relativistic and Newtonian kappa space-times,''}
  \textsl{J. Math. Phys.}  {\bf 36} (1995) 6879
  [arXiv:q-alg/9505004];

  \emph{``Deformed and extended Galilei group Hopf algebras,''}
  \textsl{ J. Phys.}  {\bf A 29} (1996) 6353
  [arXiv:q-alg/9602032].



\bibitem{Bigatti:1999iz}
  D.~Bigatti y L.~Susskind,
  \emph{``Magnetic fields, branes and noncommutative geometry,''}
  Phys.\ Rev.\  D {\bf 62} (2000) 066004
  [arXiv:hep-th/9908056].

\bibitem{Gamboa:2000yq}
  J.~Gamboa, M.~Loewe y J.~C.~Rojas,
  \emph{``Non-Commutative quantum mechanics,''}
  Phys.\ Rev.\  D {\bf 64} (2001) 067901
  [arXiv:hep-th/0010220].

\bibitem{Chaichian:2000si}
  M.~Chaichian, M.~M.~Sheikh-Jabbari y A.~Tureanu,
  \emph{``Hydrogen atom spectrum and the Lamb shift in noncommutative QED,''}
  Phys.\ Rev.\ Lett.\  {\bf 86}, 2716 (2001)
  [arXiv:hep-th/0010175].

\bibitem{Nair:2000ct}
  V.~P. Nair,
  \emph{``Quantum mechanics on a noncommutative brane in M(atrix) theory,''}
  Phys.\ Lett.\  B {\bf 505}, (2001) 249
  [arXiv:hep-th/0008027].

\bibitem{Nair:2000ii}
  V.~P. Nair y A.~P.~Polychronakos,
  \emph{``Quantum mechanics on the noncommutative plane and sphere,''}
  Phys.\ Lett.\  B {\bf 505}, (2001) 267
  [arXiv:hep-th/0011172].




\bibitem{Chaichian:2000hy}
  M.~Chaichian, A.~Demichev, P.~Presnajder, M.~M.~Sheikh-Jabbari y A.~Tureanu,
  \emph{``Aharonov-Bohm effect in noncommutative spaces,''}
  Phys.\ Lett.\  B {\bf 527}, (2002) 149
  [arXiv:hep-th/0012175].

\bibitem{Gamboa:2001fg}
  J.~Gamboa, M.~Loewe, F.~Mendez y J.~C.~Rojas,
  \emph{``Estimating noncommutative effects from the quantum Hall effect,''}
  Mod.\ Phys.\ Lett.\  A {\bf 16}, (2001) 2075
  [arXiv:hep-th/0104224].

\bibitem{Morariu:2001dv}
  B.~Morariu y A.~P.~Polychronakos,
  \emph{``Quantum mechanics on the noncommutative torus,''}
  Nucl.\ Phys.\  B {\bf 610}, (2001) 531
  [arXiv:hep-th/0102157].

\bibitem{Gamboa:2001qa}
  J.~Gamboa, M.~Loewe, F.~Mendez y J.~C.~Rojas,
  \emph{``Noncommutative quantum mechanics: The two-dimensional central field,''}
  Int.\ J.\ Mod.\ Phys.\  A {\bf 17}, (2002) 2555
  [arXiv:hep-th/0106125].

\bibitem{Falomir:2002ih}
  H.~Falomir, J.~Gamboa, M.~Loewe, F.~Mendez y J.~C.~Rojas,
  \emph{``Testing spatial noncommutativity via the Aharonov-Bohm effect,''}
  Phys.\ Rev.\  D {\bf 66}, (2002) 045018
  [arXiv:hep-th/0203260].

\bibitem{Horv}
  P.~A.~Horvathy,
  \emph{  ``Non-commutative mechanics, in mathematical and in condensed matter physics,''}
  [arXiv:cond-mat/0609571].


\bibitem{Horvathy:2002wc}
  P.~A.~Horvathy,
  \emph{ ``The non-commutative Landau problem,''}
  Annals Phys.\  {\bf 299} (2002) 128
  [arXiv:hep-th/0201007].

\bibitem{Hellerman:2001rj}
  S.~Hellerman y M.~Van Raamsdonk,
  \emph{``Quantum Hall physics equals noncommutative field theory,''}
  JHEP {\bf 0110}, (2001) 039
  [arXiv:hep-th/0103179].

\bibitem{Dunne:1992ew}
  G.~V.~Dunne y R.~Jackiw,
  \emph{``'Peierls substitution' and Chern-Simons quantum mechanics,''}
  Nucl.\ Phys.\ Proc.\ Suppl.\  {\bf 33C} (1993) 114
  [arXiv:hep-th/9204057].

\bibitem{Duval:2000xr}
  C.~Duval y P.~A.~Horvathy,
  \emph{``The exotic Galilei group and the 'Peierls substitution',''}
  Phys.\ Lett.\  B {\bf 479}, 284 (2000)
  [arXiv:hep-th/0002233].

\bibitem{Duval:2001hu}
  C.~Duval y P.~A.~Horvathy,
  \emph{``Exotic Galilean symmetry in the non-commutative plane, and the Hall
  effect,''}
  J.\ Phys.\ A  {\bf 34} (2001) 10097
  [arXiv:hep-th/0106089].

\bibitem{DH2}
  C.~Duval, Z.~Horvath y P.~A.~Horvathy,
  \emph{``Exotic plasma as classical Hall liquid,''}
  Int.\ J.\ Mod.\ Phys.\ B {\bf 15} (2001) 3397
  [arXiv:cond-mat/0101449].

\bibitem{Schonfeld:1980kb}
  J.~F.~Schonfeld,
  \emph{``A Mass Term For Three-Dimensional Gauge Fields,''}
  Nucl.\ Phys.\  B {\bf 185} (1981) 157.

\bibitem{anyonnc}
  M.~S.~Plyushchay,
  \emph{``The model of a free relativistic particle with fractional spin,''}
  Int.\ J.\ Mod.\ Phys.\  A {\bf 7} (1992) 7045;

  J.~L.~Cortes y M.~S.~Plyushchay,
  \emph{``Anyons as spinning particles,''}
  Int.\ J.\ Mod.\ Phys.\  A {\bf 11} (1996) 3331
  [arXiv:hep-th/9505117].


\bibitem{bargmann}
 V.~Bargmann,
 \emph{``On unitary ray representations of continuous groups,''}
  Annals Math.\  {\bf 59} (1954) 1.

\bibitem{azcarragabook}
  J.~A.~de Azcarraga y J.~M.~Izquierdo,
  \emph{``Lie groups, Lie algebras, cohomology and some applications in physics,''}
  Cambridge Univ. Press, 1995,   \href{http://www.slac.stanford.edu/spires/find/hep/www?irn=3285570}{SPIRES entry}.

\bibitem{BGO}
  A. Ballesteros, M. Gadella y M.~del Olmo,
  \emph{``Moyal quantization of 2+1 dimensional Galilean systems,''}
  Journ. Math. Phys. {\bf 33} (1992) 3379;

\bibitem{Grigore:1993fz}
  D.~R.~Grigore,
  \emph{``The Projective unitary irreducible representations of the Galilei group in
  (1+2)-dimensions,''}
  J.\ Math.\ Phys.\  {\bf 37} (1996) 460
  [arXiv:hep-th/9312048].

\bibitem{Bose:1994sj}
  S.~K.~Bose,
  \emph{``The Galilean group in (2+1) space-times and its central extension,''}
  Commun.\ Math.\ Phys.\  {\bf 169}, (1995) 385.

\bibitem{Bose:1994si}
  S.~K.~Bose,
  \emph{``Representations of the (2+1)-dimensional Galilean group,''}
  J.\ Math.\ Phys.\  {\bf 36}, (1995) 875.

\bibitem{Brihaye:1995nv}
  Y.~Brihaye, C.~Gonera, S.~Giller y P.~Kosinski,
  \emph{``Galilean invariance in (2+1)-dimensions,''}
  [arXiv:hep-th/9503046];

  \emph{``Remarks on central extensions of the Galilei group in 2+1 dimensions,''} [arXiv:q-alg/9508020].

\bibitem{Hagen:2002pg}
  C.~R.~Hagen,
  \emph{``Second central extension in Galilean covariant field theory,''}
  Phys.\ Lett.\  B {\bf 539} (2002) 168
  [arXiv:quant-ph/0203109].

\bibitem{Horvathy:2002vt}
  P.~A.~Horvathy y M.~S.~Plyushchay,
  \emph{``Non-relativistic anyons,  exotic Galilean symmetry and
  noncommutative plane,''}
  JHEP {\bf 0206} (2002) 033
  [arXiv:hep-th/0201228].

\bibitem{LSZ}
  J.~Lukierski, P.~C.~Stichel y W.~J.~Zakrzewski,
  \emph{``Galilean-invariant (2+1)-dimensional models with a Chern-Simons-like term
  and D = 2 noncommutative geometry,''}
  Annals Phys.\  {\bf 260} (1997) 224
  [arXiv:hep-th/9612017].

\bibitem{OP}
 M.~A.~del Olmo y M.~S.~Plyushchay,
  \emph{``Electric Chern-Simons term, enlarged exotic Galilei symmetry and
  noncommutative plane,''}
  Annals Phys.\  {\bf 321} (2006) 2830
  [arXiv:hep-th/0508020].

\bibitem{Dunne:1989hv}
  G.~V.~Dunne, R.~Jackiw y C.~A.~Trugenberger,
  \emph{``Topological (Chern-Simons) quantum mechanics,''}
  Phys.\ Rev.\  D {\bf 41} (1990) 661.

\bibitem{olmo nh}
  O. Arratia, M. A. Martín y M. A. del Olmo,
  \emph{``Classical systems and representations of (2+1) Newton-Hooke symmetries,''}
  [arXiv:math-ph/9903013].

\bibitem{Maldacena:1997re}
  J.~M.~Maldacena,
  \emph{ ``The large N limit of superconformal field theories and
  supergravity,''}
  Adv.\ Theor.\ Math.\ Phys.\  {\bf 2} (1998) 231
  [Int.\ J.\ Theor.\ Phys.\  {\bf 38} (1999) 1113]
  [arXiv:hep-th/9711200].

\bibitem{Gubser:1998bc}
  S.~S.~Gubser, I.~R.~Klebanov y A.~M.~Polyakov,
  \emph{ ``Gauge theory correlators from non-critical string
  theory,''}
  Phys.\ Lett.\  B {\bf 428} (1998) 105
  [arXiv:hep-th/9802109].

\bibitem{Witten:1998qj}
  E.~Witten,
  \emph{``Anti-de Sitter space and holography,''}
  Adv.\ Theor.\ Math.\ Phys.\  {\bf 2} (1998) 253
  [arXiv:hep-th/9802150].

\bibitem{Aharony:1999ti}
  O.~Aharony, S.~S.~Gubser, J.~M.~Maldacena, H.~Ooguri y Y.~Oz,
  \emph{``Large N field theories, string theory and gravity,''}
  Phys.\ Rept.\  {\bf 323}, (2000) 183
  [arXiv:hep-th/9905111].

\bibitem{LeivaPl}
  C.~Leiva y M.~S.~Plyushchay,
  \emph{``Conformal symmetry of relativistic and nonrelativistic systems and  AdS/CFT
  correspondence,''}
  Annals Phys.\  {\bf 307}, (2003) 372
  [arXiv:hep-th/0301244].

\bibitem{Son:2008ye}
  D.~T.~Son,
  \emph{``Toward an AdS/cold atoms correspondence: a geometric realization of the
  Schroedinger symmetry,''}
  Phys.\ Rev.\  D {\bf 78} (2008) 046003
  [arXiv:0804.3972 [hep-th]].

\bibitem{CJPAdS}
  F.~Correa, V.~Jakubsky y M.~S.~Plyushchay,
  \emph{``Aharonov-Bohm effect on AdS$_2$ and nonlinear supersymmetry of reflectionless Poschl-Teller system,''}
  Annals Phys.\  {\bf 324} (2009) 1078
  [arXiv:0809.2854 [hep-th]].


\bibitem{Balasubramanian:2008dm}
  K.~Balasubramanian y J.~McGreevy,
  \emph{ ``Gravity duals for non-relativistic CFTs,''}
  Phys.\ Rev.\ Lett.\  {\bf 101} (2008) 061601
  [arXiv:0804.4053 [hep-th]].

\bibitem{Nishida:2007pj}
  Y. Nishida y D.~T.~Son,
  \emph{ ``Nonrelativistic conformal field theories,''}
  Phys.\ Rev.\  D {\bf 76} (2007) 086004
  [arXiv:0706.3746 [hep-th]].

\bibitem{NRConf}
  R.~Jackiw,
  \emph{``Introducing scale symmetry,''}
  Phys.\ Today {\bf 25}, (1972) 23.

  U.~Niederer,
  \emph{``The maximal kinematical invariance group of the free Schrodinger
  equation,''}
  Helv.\ Phys.\ Acta {\bf 45}, (1972) 802.

  C. R. Hagen,
  \emph{``Scale and conformal transformations in Galilean-covariant field theory,''}
  Phys. Rev. D {\bf 5}  (1972) 377;

  V.~de Alfaro, S.~Fubini y G.~Furlan,
  \emph{``Conformal invariance in quantum mechanics,''}
  Nuovo Cim.\  A {\bf 34}, (1976) 569.

  R.~Jackiw,
  \emph{``Dynamical symmetry of the magnetic monopole,''}
  Annals Phys.\  {\bf 129} (1980) 183;

  R.~Jackiw,
  \emph{``Dynamical symmetry of the magnetic vortex,''}
  Annals Phys.\  {\bf 201} (1990) 83;

C. Duval, G. W. Gibbons, y P. A. Horvathy,
  \emph{``Celestial mechanics, conformal structures and
  gravitational waves,''}
  Phys. Rev. D {\bf 43}   (1991) 3907;

  M.~Henkel y J.~Unterberger,
  \emph{``Schroedinger invariance and space-time symmetries,''} {\sl
  Nucl.\ Phys.}\  B {\bf 660}  (2003) 407
  [arXiv:hep-th/0302187].

\bibitem{Duval:1984cj}
  C.~Duval, G.~Burdet, H.~P.~Kunzle y M.~Perrin,
  \emph{``Bargmann structures and Newton-Cartan theory,''}
  Phys.\ Rev.\ D {\bf 31} (1985) 1841.

\bibitem{Duval:2008jg}
  C.~Duval, M.~Hassaine and P.~A.~Horvathy,
  \emph{``The geometry of Schr\'odinger symmetry in gravity background/non-relativistic CFT,''}
  Annals Phys.\  {\bf 324}, (2009) 1158
  [arXiv:0809.3128 [hep-th]].

\bibitem{Banados:1992wn}
  M.~Bañados, C.~Teitelboim y J.~Zanelli,
  \emph{``The black hole in three-dimensional space-time,''}
  Phys.\ Rev.\ Lett.\  {\bf 69} (1992) 1849
  [arXiv:hep-th/9204099];

  M.~Bañados, M.~Henneaux, C.~Teitelboim y J.~Zanelli,
  \emph{``Geometry of the (2+1) black hole,''}
  Phys.\ Rev.\ D {\bf 48} (1993) 1506
  [arXiv:gr-qc/9302012].

\bibitem{AchTown}
  A.~Achucarro y P.~K.~Townsend,
  \emph{``A Chern-Simons action for three-dimensional anti-de Sitter
  supergravity,"}
  Phys.\ Lett.\ B {\bf 180} (1986) 89.

\bibitem{Witten}
  E.~Witten,
  \emph{``(2+1)-dimensional gravity as an exactly soluble system,"}
  Nucl.\ Phys.\ B {\bf 311} (1988) 46;

  \emph{  ``Quantization of Chern-Simons gauge theory with complex
  gauge group,"}
  Commun.\ Math.\ Phys.\  {\bf 137} (1991) 29.

\bibitem{BacNuy}
  H.~Bacry y J. Nuyts,
  \emph{``Classification of ten-dimensional kinematical groups with
  space isotropy,"}
  J.\ Math.\ Phys.\  {\bf 27} (1986) 2455.

\bibitem{Mukunda:1974dr}
  N.~Mukunda y G.~Sudarshan,
  \emph{``Classical dynamics: a modern perspective,''}
  \emph{  Wiley, New York, 1974}, \href{http://www.slac.stanford.edu/spires/find/hep/www?irn=6999280}{SPIRES entry}

\bibitem{AzHer}
  J.~A.~de Azcarraga, F.~J.~Herranz, J.~C.~Perez Bueno y M.~Santander,
  \emph{``Central extensions of the quasiorthogonal Lie algebras,''}
  J.\ Phys.\ A  {\bf 31} (1998) 1373
  [arXiv:q-alg/9612021].

\bibitem{Segal}
  I. Segal, Duke,
  Math. J. {\bf 18}, (1951) 221.

\bibitem{IW1}
  E. Inönü y E. P. Wigner,
  Proc. Natl. Acad. Sci. U.S. {\bf 39}, 510 (1953); {\bf 40}, 119 (1954).

\bibitem{Saletan}
  E. Saletan,
  J. Math. Phys. {\bf 2}, (1961) 1.

\bibitem{IW2}
  E. Inönü y E. P. Wigner,
  Nuovo Cimento {\bf 9}, (1952) 705.

\bibitem{Wightman1}
  S. Wightman,
  Rev. Mod. Phys. {\bf 34}, (1962) 845.

\bibitem{Hamer}
  M. Hamermesh,
  Ann. Phys. {\bf 9}, (1960) 518.

\bibitem{Weyl}
  H. Weyl,
  \emph{``The theory groups and quantum mechanics,''}
  London 1931.

\bibitem{Hoogland:1978qs}
  H.~Hoogland,
  \emph{``Gauge equivalence of representations of symmetry groups in quantum
  mechanics,''}
  J.\ Phys.\ A  {\bf 11} (1978) 1557.

\bibitem{Dyson:1972sd}
  F. J. Dyson,
  \emph{``Missed opportunities,''}
  Bull.\ Am.\ Math.\ Soc.\  {\bf 78} (1972) 635.

\bibitem{AldBar}
  R.~Aldrovandi, A.~L.~Barbosa, L.~C.~B.~Crispino y J.~G.~Pereira,
  \emph{  ``Non-Relativistic spacetimes with cosmological constant,''}
  Class.\ Quant.\ Grav.\  {\bf 16} (1999) 495
  [arXiv:gr-qc/9801100].

\bibitem{Mariano}
  O. Arratia, M.~A.~Martin y M.~A.~Olmo,
  \emph{``Classical systems and representation of (2+1) Newton-Hooke symmetries,''}
  [arXiv:math-ph/9903013].

\bibitem{Gao}
  Y.~h.~Gao,
  \emph{ ``Symmetries, matrices, and de Sitter gravity,''}
  [arXiv:hep-th/0107067].

\bibitem{GibPat}
  G.~W.~Gibbons y C.~E.~Patricot,
  \emph{``Newton-Hooke space-times, Hpp-waves and the cosmological constant,''}
  Class.\ Quant.\ Grav.\  {\bf 20} (2003) 5225
  [arXiv:hep-th/0308200].

\bibitem{BGK}
  J.~Brugues, J.~Gomis y K.~Kamimura,
  \emph{``Newton-Hooke algebras, non-relativistic branes and generalized pp-wave metrics,''}
  Phys.\ Rev.\ D {\bf 73} (2006) 085011
  [arXiv:hep-th/0603023].

\bibitem{levyleblond69}
  J.-M. L\'evy-Leblond,
  \emph{``Group-theoretical foundations of classical mechanics: the Lagrangian gauge problem,"}
  Comm. Math. Phys. {\bf 12} (1969) 64.

\bibitem{Marmo:1987rv}
  G.~Marmo, G.~Morandi, A.~Simoni y E.~C.~G.~Sudarshan,
  \emph{``Quasiinvariance and central extensions,''}
  Phys.\ Rev.\  D {\bf 37} (1988) 2196.


\bibitem{Coleman}
  S.~R.~Coleman, J.~Wess y B.~Zumino,
  \emph{``Structure of phenomenological Lagrangians. 1,''}
  Phys.\ Rev.\  {\bf 177} (1969) 2239;

  C.~G.~.~Callan, S.~R.~Coleman, J.~Wess y B.~Zumino,
  \emph{``Structure of phenomenological Lagrangians. 2,''}
  Phys.\ Rev.\  {\bf 177} (1969) 2247.

\bibitem{West}
  P.~C.~West,
  \emph{``Automorphisms, non-linear realizations and branes,''}
  JHEP {\bf 0002} (2000) 024
  [arXiv:hep-th/0001216].

\bibitem{GomisKW}
  J.~Gomis, K.~Kamimura y P.~West,
  \emph{``The construction of brane and superbrane actions using non-linear
  realisations,''}
  Class.\ Quant.\ Grav.\  {\bf 23} (2006) 7369
  [arXiv:hep-th/0607057].

\bibitem{HeTe}
  Para la discusión sobre variables auxiliares en un marco de trabajo generico en formalismo Lagrangiano mirar, por ejemplo, Marc Henneaux y Claudio Teitelboim, \emph{``Quantization of gauge systems,''}
  Princeton University Press, 1992.

\bibitem{Horvathy:2004fw}
  P.~A.~Horvathy y M.~S.~Plyushchay,
  \emph{``Nonrelativistic anyons in external electromagnetic field,''}
  Nucl.\ Phys.\ B {\bf 714} (2005) 269
  [arXiv:hep-th/0502040].

\bibitem{PHMP}
  P.~A.~Horvathy y M.~S.~Plyushchay,
  \emph{``Anyon wave equations and the noncommutative plane,''}
  Phys.\ Lett.\ B {\bf 595} (2004) 547
  [arXiv:hep-th/0404137].

\bibitem{Gozzi}
  E.~Gozzi y D.~Mauro,
  \emph{``Mechanical similarity as a generalization of scale symmetry,''}
  J.\ Phys.\ A {\bf 39} (2006) 3411
  [arXiv:quant-ph/0508199].

\bibitem{BarSL2R}
  V.~Bargmann,
  \emph{``Irreducible unitary representations of the Lorentz group,''}
  Annals Math.\  {\bf 48} (1947) 568;

M.~S.~Plyushchay,
  \emph{``Quantization of the classical SL(2,R) system and representations of
  $\overline{SL(2,R)}$ group,''}
  J.\ Math.\ Phys.\  {\bf 34} (1993) 3954.

\bibitem{GQ}
  N.~M.~J.~Woodhouse,
  \emph{``Geometric quantization,''}
  \emph{New York, USA: Clarendon (1992) 307 p. (Oxford mathematical monographs),}
  \href{http://www.slac.stanford.edu/spires/find/hep/www?irn=2650371}{SPIRES entry};


A. A. Kirillov,
  \emph{``Lectures on the orbit method,''}
  Graduate studies in Mathematics, 64, AMS, Rhode Island, 2004.

\bibitem{TeiZan}
  C. Teitelboim  y J. Zanelli,
  \emph{``Dimensionally continued topological gravitation theory in Hamiltonian form,'',} Class. and Quant.Grav. {\bf 4} (1987) L125;

M. Henneaux, C. Teitelboim y J. Zanelli,
  \emph{``Quantum mechanics of multivalued Hamiltonians,''},
  Phys.Rev. A {\bf 36} (1987) 4417.

\bibitem{BanHen}
  M.~Bañados, L.~J.~Garay y M.~Henneaux,
  \emph{``The local degrees of freedom of higher dimensional pure Chern-Simons theories,''}
  Phys.\ Rev.\ D {\bf 53} (1996) 593
  [arXiv:hep-th/9506187];

  \emph{``The dynamical structure of higher dimensional Chern-Simons theory,''}
  Nucl.\ Phys.\ B {\bf 476} (1996) 611
  [arXiv:hep-th/9605159].

\bibitem{Jauch}
  J.~M.~Jauch y E.~L.~Hill,
  \emph{``On the problem of degeneracy in quantum mechanics,''}
  Phys. Rev.  {\bf 57} (1940)  641.

\bibitem{Louck}
  J.~D.~Louck, M.~Moshinsky y K.~B.~Wolf,
  \emph{``Canonical transformations and accidental degeneracy. 1. the anisotropic
  oscillator,''}
  J.\ Math.\ Phys.\  {\bf 14} (1973) 692.

\bibitem{Boer}
  J.~de Boer, F.~Harmsze y T.~Tjin,
  \emph{``Nonlinear finite W symmetries and applications in elementary systems,''}
  Phys.\ Rept.\  {\bf 272} (1996) 139
  [arXiv:hep-th/9503161].

\bibitem{Kij}
  A.~Kijanka y P.~Kosinski,
  \emph{``On noncommutative isotropic harmonic oscillator,''}
  Phys.\ Rev.\  D {\bf 70} (2004) 127702
  [arXiv:hep-th/0407246].

\bibitem{Sudar}
  E.~C.~G.~Sudarshan y N.~Mukunda,
  \emph{``Quantum theory of the infinite-component Majorana field and the relation of
  spin and statistics,''}
  Phys.\ Rev.\  D {\bf 1} (1970) 571.

\bibitem{Horvathy:2007pm}
  P.~A.~Horvathy, M.~S.~Plyushchay and M.~Valenzuela,
  \emph{``Bosonized supersymmetry from the Majorana-Dirac-Staunton theory and massive
  higher-spin fields,''}
  Phys.\ Rev.\  D {\bf 77} (2008) 025017
  [arXiv:0710.1394 [hep-th]].

\bibitem{GomisKamimura}
  J.~Gomis, J.~Herrero, K.~Kamimura y J.~Roca,
  \emph{``Diffeomorphisms, nonlinear W symmetries and particle models,''}
  Annals Phys.\  {\bf 244} (1995) 67
  [arXiv:hep-th/9309048].


\bibitem{BNS}
  S.~Bellucci, A. Nersessian y C.~Sochichiu,
  \emph{``Two phases of the non-commutative quantum mechanics,''}
  Phys.\ Lett.\  B {\bf 522} (2001) 345
  [arXiv:hep-th/0106138];

S.~Bellucci y A. Nersessian,
  \emph{``Phases in noncommutative quantum mechanics on (pseudo)sphere,''}
  Phys.\ Lett.\  B {\bf 542} (2002) 295
  [arXiv:hep-th/0205024].

\bibitem{HMS}
  P.~A.~Horvathy, L.~Martina y P.~C.~Stichel,
  \emph{``Galilean symmetry in noncommutative field theory,''}
  Phys.\ Lett.\  B {\bf 564} (2003) 149
  [arXiv:hep-th/0304215].

\bibitem{Duval:1993hs}
  C.~Duval y P.~A.~Horvathy,
 \emph{ ``On Schrodinger superalgebras,''}
  J.\ Math.\ Phys.\  {\bf 35} (1994) 2516
  [arXiv:hep-th/0508079].

\bibitem{LandauSUSY}
  R.~J.~Hughes, V.~A.~Kostelecky y M.~M. Nieto,
  \emph{``Supersymmetric quantum mechanics in a first order Dirac equation,''}
  Phys.\ Rev.\  D {\bf 34}   (1986) 1100.

\bibitem{CIMT}
  T.~Curtright, E.~Ivanov, L.~Mezincescu y P.~K.~Townsend,
  \emph{ ``Planar super-Landau models revisited,''}
  JHEP {\bf 0704} (2007) 020
  [arXiv:hep-th/0612300].

\bibitem{AnPl}
  A.~Anabalon y M.~S.~Plyushchay,
  \emph{ ``Interaction via reduction and nonlinear superconformal symmetry,''}
  Phys.\ Lett.\  B {\bf 572}  (2003) 202
  [arXiv:hep-th/0306210].

\bibitem{Harikumar:2004qc}
  E.~Harikumar, V.~S.~Kumar y A.~Khare,
  \emph{``Supersymmetric quantum mechanics on non-commutative plane,''}
  Phys.\ Lett.\  B {\bf 589}, 155 (2004)
  [arXiv:hep-th/0402064].

\bibitem{Ghosh:2004ye}
  P.~K.~Ghosh,
  \emph{``Supersymmetric quantum mechanics on noncommutative space,''}
  Eur.\ Phys.\ J.\  C {\bf 42}, 355 (2005)
  [arXiv:hep-th/0403083].

\bibitem{Lapointe:2004py}
  L.~Lapointe, H.~Ujino y L.~Vinet,
  \emph{``Supersymmetry in the non-commutative plane,''}
  Annals Phys.\  {\bf 314}, 464 (2004)
  [arXiv:hep-th/0405271].

\bibitem{SUSYGal2}
  J.~Lukierski, I.~Prochnicka, P.~C.~Stichel y W.~J.~Zakrzewski,
  \emph{``Galilean exotic planar supersymmetries and nonrelativistic supersymmetric wave equations,''}
  {\sl  Phys.\ Lett.}\ B {\bf 639} (2006)  389
  [arXiv:hep-th/0602198].

\bibitem{SUSYGal}
  G.~S.~Lozano, O.~Piguet, F.~A.~Schaposnik y L.~Sourrouille,
  \emph{ ``Nonrelativistic supersymmetry in noncommutative space,''}
  Phys.\ Lett. {\bf B630}  (2005) 108
  [arXiv:hep-th/0508009];

  \emph{``On 1+1 dimensional Galilean supersymmetry in ultracold quantum gases,''} [ArXiv:cond-mat/0609553].

\bibitem{DFGM}
  A.~Das, H.~Falomir, J.~Gamboa y F.~Mendez,
  \emph{``Non-commutative supersymmetric quantum mechanics,''}
  Phys.\ Lett.\  B {\bf 670} (2009) 407
  [arXiv:0809.1405 [hep-th]].

\bibitem{Geloun}
  J.~B.~Geloun y F.~G.~Scholtz,
  \emph{``Noncommutative supersymmetric quantum mechanics,''}
  [arXiv:0812.3289 [hep-th]].

\bibitem{KhareSukhatme}
  F.~Cooper, A.~Khare y U.~Sukhatme,
  \emph{``Supersymmetry and quantum mechanics,''}
  Phys.\ Rept.\  {\bf 251}  (1995) 267
  [arXiv:hep-th/9405029].

\bibitem{Kumar}
  J.~Kumar,
  \emph{``Conformal mechanics and the Virasoro algebra,''}
  JHEP {\bf 9904} (1999) 006
  [arXiv:hep-th/9901139].

\bibitem{Pais:1950za}
  A.~Pais y G.~E.~Uhlenbeck,
  \emph{``On Field theories with nonlocalized action,''}
  Phys.\ Rev.\  {\bf 79}, 145 (1950).

\bibitem{Bender:2008vh}
  C.~M.~Bender y P.~D.~Mannheim,
  \emph{``Giving up the ghost,''}
  J.\ Phys.\ A {\bf 41} (2008) 304018
  [arXiv:0807.2607 [hep-th]].

\bibitem{Drukker:2003mg}
  N.~Drukker, B.~Fiol y J.~Simon,
  \emph{``Goedel-type universes and the Landau problem,''}
  JCAP {\bf 0410} (2004) 012
  [arXiv:hep-th/0309199].

\bibitem{Hik}
  Y.~Hikida y S.~J.~Rey,
  \emph{  ``Can branes travel beyond CTC horizon in Goedel universe?,''}
  Nucl.\ Phys.\  B {\bf 669} (2003) 57
  [arXiv:hep-th/0306148].

\bibitem{Klishevich:2001fu}
  S.~M.~Klishevich y M.~S.~Plyushchay,
  \emph{``Nonlinear supersymmetry on the plane in magnetic field and  quasi-exactly
  solvable systems,''}
  Nucl.\ Phys.\  B {\bf 616} (2001) 403
  [arXiv:hep-th/0105135].

\bibitem{AIS}
  A.~A.~Andrianov, M.~V.~Ioffe y V.~P.~Spiridonov,
  \emph{``Higher derivative supersymmetry and the Witten index,''}
  Phys.\ Lett.\  A {\bf 174} (1993) 273
  [arXiv:hep-th/9303005].

\bibitem{Pl_Para}
  M.~Plyushchay,
  \emph{``Hidden nonlinear supersymmetries in pure parabosonic systems,''}
  Int.\ J.\ Mod.\ Phys.\  A {\bf 15} (2000) 3679
  [arXiv:hep-th/9903130].

\bibitem{Klishevich:2000dp}
  S.~M.~Klishevich y M.~S.~Plyushchay,
  \emph{``Nonlinear supersymmetry, quantum anomaly and quasi-exactly solvable
  systems,''}
  Nucl.\ Phys.\  B {\bf 606} (2001) 583
  [arXiv:hep-th/0012023].






\end{thebibliography}
\end{document}